\documentclass[cleveref, autoref]{paper}
 
\usepackage{fullpage,float,amsthm,amsfonts,amsmath,amssymb,keytheorems,url,tikz,circuitikz,tabularx,multirow,multicol,hhline,tablefootnote,xcolor,etoolbox,clipboard,authblk,titlesec,setspace,titletoc,tocloft,enumitem,xparse,xspace,pgfplots,stmaryrd,longtable,cellspace}
\usepackage[sortcites=true,maxnames=1000]{biblatex}
\usepackage[toc,page]{appendix}
\usepackage[margin=0pt,font={small},labelfont={bf},labelsep=colon,format=plain]{caption}
\usepackage[left]{lineno}
\usepackage[mathscr]{eucal}
\usepackage[ruled,vlined]{algorithm2e}
\usepackage[hidelinks]{hyperref}
\makeatletter 
\let\original@algocf@latexcaption\algocf@latexcaption
\long\def\algocf@latexcaption#1[#2]{%
  \@ifundefined{NR@gettitle}{%
    \def\@currentlabelname{#2}%
  }{%
    \NR@gettitle{#2}%
  }%
  \original@algocf@latexcaption{#1}[{#2}]%
}
\makeatother 
\usepackage[T1]{fontenc}

\usetikzlibrary{fadings}
\usetikzlibrary{patterns}
\usetikzlibrary{shadows.blur}
\usetikzlibrary{shapes}
\usetikzlibrary{positioning}

\pgfplotsset{width=10cm,compat=1.9}
\usepgfplotslibrary{external}

\def\titlename{}
\def\authorsnames{}

\def\squiggylinescale{0.08}
\def\squiggylinedist{0.6}
\def\squiggylinerange{3.2}

\def\dontcare{\rule{2mm}{0.4pt}}

\definecolor{turquoise}{RGB}{64,224,208}
\definecolor{salmon}{RGB}{250,128,114}

\newcommand{\floor}[1]{\left\lfloor #1 \right\rfloor}
\newcommand{\ceil}[1]{\left\lceil #1 \right\rceil}
\newcommand{\paren}[1]{\left( #1 \right)}
\newcommand{\bracke}[1]{\left[ #1 \right]}
\newcommand{\bracce}[1]{\left\{ #1 \right\}}
\newcommand{\pair}[1]{\left< #1 \right>}
\newcommand{\abs}[1]{\left| #1 \right|}

\newcommand{\tabref}[1]{\autoref{tab:#1}}
\newcommand{\secref}[1]{\hyperref[#1]{Section~\ref*{#1}}}
\newcommand{\appref}[1]{\hyperref[#1]{Appendix~\ref*{#1}}}
\newcommand{\thmref}[1]{~\protect\autoref{thm:#1}}
\newcommand{\lemref}[1]{\hyperref[lem:#1]{Lemma~\ref*{lem:#1}}}
\newcommand{\algref}[1]{\hyperref[alg:#1]{Algorithm~\ref*{alg:#1}}}
\NewDocumentCommand{\linesref}{mg}{%
    \IfNoValueTF{#2}{%
        \hyperref[algline:#1]{Line~\ref*{algline:#1}}%
    }{%
        \hyperref[algline:#1]{Lines~\ref*{algline:#1}}--%
        \hyperref[algline:#2]{\ref*{algline:#2}}%
    }%
}

\newcommand{\cncref}[1]{~\protect\hyperref[cnc:#1]{Conclusion~\ref*{cnc:#1}}}
\newcommand{\corref}[1]{~\protect\hyperref[cor:#1]{Corollary~\ref*{cor:#1}}}

\newcommand{\obvref}[1]{\hyperref[obv:#1]{Observation~\ref*{obv:#1}}}
\newcommand{\clmref}[1]{\hyperref[clm:#1]{Claim~\ref*{clm:#1}}}
\newcommand{\stpref}[3]{\hyperref[alg:#1_#2]{\textit{#3}}}
\newcommand{\exmref}[1]{\hyperref[exm:#1]{Example~\ref*{exm:#1}}}
\newcommand{\prbref}[1]{\hyperref[prb:#1]{Problem~\ref*{prb:#1}}}
\newcommand{\defref}[1]{\hyperref[def:#1]{Definition~\ref*{def:#1}}}
\newcommand{\queref}[1]{\hyperref[que:#1]{Question~\ref*{que:#1}}}
\ExplSyntaxOn
\NewDocumentCommand{\querefs}{m}{
  Questions~
  \clist_set:Nn \l_tmpa_clist {#1}
  \int_set:Nn \l_tmpa_int {0}
  \clist_map_inline:Nn \l_tmpa_clist {
    \int_incr:N \l_tmpa_int
    \hyperref[que:##1]{\ref*{que:##1}}
    \int_compare:nNnT { \l_tmpa_int } < { \clist_count:N \l_tmpa_clist } { ,~ }
  }
}
\ExplSyntaxOff
\newcommand{\eqnref}[1]{\hyperref[eqn:#1]{Equation~(\ref*{eqn:#1})}}

\NewDocumentCommand\mmin{mgg}{\displaystyle \min\IfNoValueTF{#2}{}{_{#2}}\IfNoValueTF{#3}{}{^{#3}}\bracce{#1}}
\NewDocumentCommand\mmax{mgg}{\displaystyle \max\IfNoValueTF{#2}{}{_{#2}}\IfNoValueTF{#3}{}{^{#3}}\bracce{#1}}
\NewDocumentCommand\mprod{mgg}{\displaystyle \prod\IfNoValueTF{#2}{}{_{#2}}\IfNoValueTF{#3}{}{^{#3}}{#1}}
\NewDocumentCommand\msum{mgg}{\displaystyle \sum\IfNoValueTF{#2}{}{_{#2}}\IfNoValueTF{#3}{}{^{#3}}{#1}}
\NewDocumentCommand\mcup{mgg}{\displaystyle \bigcup\IfNoValueTF{#2}{}{_{#2}}\IfNoValueTF{#3}{}{^{#3}}{#1}}
\NewDocumentCommand\mcap{mgg}{\displaystyle \bigcap\IfNoValueTF{#2}{}{_{#2}}\IfNoValueTF{#3}{}{^{#3}}{#1}}
\newcommand{\textoverline}[1]{\={#1}}
\NewDocumentCommand\case{mg}
    {%
    \ensuremath{\textbf{Case (#1\IfNoValueTF{#2}{}{$_#2$})}}%
    }
\NewDocumentCommand\notcase{mg}
    {%
    \ensuremath{\textbf{Case (\textoverline{#1}\IfNoValueTF{#2}{}{$_#2$})}}%
    }
\newcommand{\midline}{\,\middle| \,\,}

\newcommand{\qoute}[1]{``#1''}
\newcommand{\squiggly}{
\begin{tikzpicture} 
    {
    \draw[domain=-\squiggylinedist:0,smooth,variable=\x,scale=\squiggylinescale] plot({\x},{0.75+sin(4*0 r)});
    \draw[domain=0:\squiggylinerange,smooth,variable=\x,scale=\squiggylinescale] plot({\x},{0.75+sin(4*\x r)});
    \draw[domain=\squiggylinerange:\squiggylinerange+\squiggylinedist,smooth,variable=\x,scale=\squiggylinescale] plot({\x},{0.75+sin(4*\squiggylinerange r)});
    }
\end{tikzpicture}
}
\newcommand{\straight}{
\begin{tikzpicture} 
    {
    \draw[domain=-\squiggylinedist:-\squiggylinedist+0.0001,smooth,variable=\x,scale=\squiggylinescale] plot({\x},{-0.25+sin(4*\squiggylinerange r)});
    \draw[domain=-\squiggylinedist+0.0001:\squiggylinerange+\squiggylinedist,smooth,variable=\x,scale=\squiggylinescale] plot({\x},{0.75+sin(4*\squiggylinerange r)});
    }
\end{tikzpicture}
}
\newcommand{\fatrightarrow}
{
    \begin{tikzpicture}
        \draw[-{Triangle[width=6pt,length=6pt]}, line width=2pt](0,0) -- (0.3, 0);
    \end{tikzpicture}
}
\newcommand{\hollowrightarrow}
{
    \begin{tikzpicture}
        \node[draw, single arrow,
              minimum height=1pt, minimum width=1pt,
              single arrow head extend=4pt,
              anchor=west, rotate=0] at (0,-2) {};
    \end{tikzpicture}
}

\newcommand{\rightabovearrow}[2]{\overset{#2}{\overrightarrow{#1}}}
\newcommand{\leftabovearrow}[2]{\overset{#2}{\overleftarrow{#1}}}
\newcommand{\sizeindex}[2]{\abs{#1}_{#2}}
\ifdefined\socgtabular\else
  \let\socgtabular\tabular
\fi

\newcommand*{\settitlename}[1]{\title{#1}\def\titlename{#1}}
\newcommand*{\addauthor}[2]{\author[#2]{#1} 
\IfNoValueTF{\authorsnames}{\def\authorsnames{#1}}{\expandafter\def\expandafter\authorsnames\expandafter{\authorsnames, #1}}}
\renewcommand{\keywords}[1]{\def\keywordnames{#1}}
\newcommand{\printkeywords}{%
    \par
    \vspace{0.5em}
    \noindent
    \textbf{Keywords: }
    \keywordnames
    \par
}

\newcommand{\dashuline}[1]
{
  \tikz[baseline=(X.base)]
  {
    \node[inner sep=0pt] (X) {#1};
    \draw[dash pattern=on 2pt off 2pt] (X.south west) -- (X.south east);
  }
}
\NewDocumentCommand\specball{mmg}
{
    ball\paren{#1 , #2 \IfNoValueTF{#3}{}{, #3}}
}
\ExplSyntaxOn
\cs_new:Npn \ballformatvar:n #1
 {
  \tl_clear:N \l_tmpa_tl
  \tl_map_inline:nn {#1}
   {
    \str_if_in:nnTF {ijkl\ell} {##1} 
      { \tl_put_right:Nn \l_tmpa_tl {\textit{##1}} }
      { \tl_put_right:Nn \l_tmpa_tl {##1} } 
   }
  \l_tmpa_tl
}
\cs_new:Npn \ballformatarg:n #1
 {
  \regex_match:nnTF { \A (?:[0-9]|i|j|k|\\ell)(?:[0-9ij k\\ell\+\-]*?) \Z } { #1 }
       {S\textsubscript{\ballformatvar:n{#1}}} 
       { #1 }
 }

\NewDocumentCommand \ball {mmg}
 {
    ball\paren{#1 ,
               \ballformatarg:n {#2}
               \IfNoValueTF{#3}
                           {}
                           {, \ballformatarg:n {#3}}
              }
 }
\ExplSyntaxOff
\newcommand{\B}{{\cal B}}
\newcommand{\Y}{{\cal Y}}
\renewcommand{\P}{{\cal P}}
\newcommand{\daft}{d_{\mathrm{aft}}}
\newcommand{\dbef}{d_{\mathrm{bef}}}
\newcommand{\dom}{\mathbin{\olessthan}}
\newcommand{\minplus}{\mathbin{\star}}
\newcommand{\maxmin}{\mathbin{\ovee}}
\newcommand{\minmax}{\mathbin{\owedge}}
\newcommand{\rbal}[1]{\operatorname{rb}\paren{#1}}
\newcommand{\cbal}[1]{\operatorname{cb}\paren{#1}}
\newcommand{\NFR}[1]{\operatorname{NF}^{\mathrm{row}}_{#1}}
 \newcommand{\NFC}[1]{\operatorname{NF}^{\mathrm{col}}_{#1}}
\newcommand{\mmname}{\ensuremath{\paren{+,\cdot}}\xspace}
\newcommand{\domname}{\ensuremath{\paren{\operatorname{count},\leq}}\xspace}
\newcommand{\minplusname}{\ensuremath{\paren{\min,+}}\xspace}
\newcommand{\maxminname}{\ensuremath{\paren{\max,\min}}\xspace}
\newcommand{\minmaxname}{\ensuremath{\paren{\min,\max}}\xspace}
\makeatletter
\@ifundefined{ess@box}{\newsavebox{\ess@box}}{}
\providecommand{\ess}{}
\renewcommand{\ess}{%
  \mathord{\mathpalette\ess@draw\relax}%
}
\providecommand{\ess@draw}[2]{}
\renewcommand{\ess@draw}[2]{%
  \begingroup%
  \sbox{\ess@box}{$\m@th#1s$}%
  \resizebox{!}{\ht\ess@box}{%
    \begin{tikzpicture}[
      x=1ex,
      y=1ex,
      xslant=0.08,
      baseline=(current bounding box.south)
    ]
      \draw[
        line width=0.135ex,
        line cap=round,
        line join=round
      ]
      (1.000,1.590)
      .. controls (1.180,1.530) and (1.190,1.800)
        .. (0.970,1.780)
      .. controls (0.660,1.750) and (0.690,1.350)
        .. (1.000,1.350)
      .. controls (1.350,1.350) and (1.470,2.000)
        .. (1.120,2.000)
      .. controls (0.750,2.000) and (0.260,1.600)
        .. (0.280,1.290)
      .. controls (0.300,1.000) and (0.900,0.820)
        .. (0.910,0.540)
      .. controls (0.940,0.230) and (0.540,0.000)
        .. (0.250,0.000)
      .. controls (-0.160,0.000) and (-0.130,0.600)
        .. (0.230,0.600)
      .. controls (0.580,0.600) and (0.710,0.200)
        .. (0.360,0.200)
      .. controls (0.150,0.200) and (0.100,0.400)
        .. (0.280,0.400);
    \end{tikzpicture}%
  }%
  \endgroup%
}
\makeatother

\newtheorem{theorem}{Theorem}[section]
\newtheorem{lemma}[theorem]{Lemma}

\newtheorem{corollary}[theorem]{Corollary}
\newtheorem{conclusion}[theorem]{Conclusion}
\newtheorem{observation}[theorem]{Observation}

\newtheorem{question}{Question}
\newtheorem{definition}[theorem]{Definition}
\newtheorem{example2}[theorem]{Example}
\AtBeginEnvironment{example2}{\pushQED{\qed}}
\AtEndEnvironment{example2}{\popQED}

\settitlename{Smallest Multiplicative Stretch for an  $\alpha$-APASP Below $\widetilde O \left(n^{\omega}\right)$ Runtime} 

\addauthor{Lakshay Saggi}{1,$\S$}
\addauthor{Ariel Sapir}{2,$\propto$}

\affil[1]{Department of \href{https://homecse.iitd.ac.in/}{Computer Science and Engineering}, IIT Delhi}
\affil[2]{Department of \href{http://cs.biu.ac.il/}{Computer Science}, Bar-Ilan University}
\affil[$\S$]{csz228231@iitd.ac.in}
\affil[$\propto$]{sapirar@biu.ac.il}

\date{\today}


\makeatletter
\def\@maketitle{%
  \newpage
  \null
  \vskip 2em%
  \begin{center}%
  \let \footnote \thanks
    {\huge \textsf{\textbf{\@title}} \par}%
    \vskip 1.5em%
    {\large
      \lineskip .5em%
      \begin{tabular}[t]{c}%
        \@author
      \end{tabular}\par}%
    \vskip 1em
    {\large \@date}
  \end{center}
  \par
  \vskip 1.5em}
\makeatother

\makeatletter
\renewenvironment{abstract}{
    \begin{center}%
    {\bfseries\sffamily \Large\abstractname\vspace{\z@}}
      \end{center}%
      \quotation
    }
    \endquotation
\makeatother

\titleformat*{\section}{\Large\bfseries\sffamily}
\titleformat*{\subsection}{\large\bfseries\sffamily}

\dottedcontents{section}[1.5em]{}{1.3em}{.6em}

\setlist[itemize,1]{label=\scalebox{1.6}[1.0]{$\blacklozenge$}}
\setlist[itemize,2]{label=\scalebox{0.9}[1.3]{$\blacktriangleright$}}
\setlist[itemize,3]{label=\textbf{\scalebox{0.9}[1.2]{\fatrightarrow}}}
\setlist[itemize,4]{label=\textbf{\scalebox{0.666}[0.47]{\hollowrightarrow}}}
\setlist[itemize,5]{label=\textbf{\scalebox{0.9}[1.2]{$>$}}}

\begin{document}

\let\originalleft\left
\let\originalright\right
\renewcommand{\left}{\mathopen{}\mathclose\bgroup\originalleft}
\renewcommand{\right}{\aftergroup\egroup\originalright}

\setlength{\textfloatsep}{8pt}


\pagenumbering{gobble}

\maketitle

\begin{abstract}
\small 

Dor, Halperin and Zwick [FOCS'96 and SICOMP'00] showed that, for every $\varepsilon>0$, computing a
$\left(2-\varepsilon\right)$-APASP is at least as hard as BMM. Consequently, any  multiplicative APASP algorithm with a running time exponent below $\omega$ must have stretch at least $2$.  
Zwick [FOCS'98 and JACM'02] gave an algebraic
$2$-APASP algorithm with 
 $\widetilde O\left(
 n^{\omega+o(1)}
\cdot\log W_{\max}\right)$ runtime, where  $W_{\max}$ is the weight of the heaviest edge in the graph. Bringmann, K{\"u}nnemann and W{\k{e}}grzycki
[STOC'19] removed the logarithmic dependence on $W_{\max}$. Using OpenAI's recent bound $\omega\leq\frac{9}{4}$
therefore gives
$\widetilde O\left(n^{\frac{9}{4}+o(1)}\right)$ time.

We present two algorithms for weighted $2$-APASP.
Our first algorithm uses $\left(\min,\max\right)$-products.
Using the earlier rectangular FMM bounds of
Vassilevska Williams, Yinzhan Xu, Zixuan Xu and
Zhou [SODA'24], and Alman, Duan, Vassilevska Williams,
Yinzhan Xu, Zixuan Xu and Zhou [SODA'25], it has
 $\widetilde O\left(n^{2.36999465}\right)$ runtime, which is below the previous  bound of 
$\omega \leq 2.371177$ of Dupont, Eisenberger, Kozlovskii,
Mehrabian, Ruiz, See, Zhou, Alman, Vassilevska Williams
and Balog [arXiv'26]. The current rectangular FMM bounds improve the
runtime to
$\widetilde O\left(n^{2.34549151}\right)$, with no  dependence on $\log W_{\max}$. 

Our second algorithm assumes integer edge weights
and improves the exponent further, but has a dependence
on $\log W_{\max}$. It has a runtime of $\widetilde O\left(
n^{\omega-\frac{(\omega-2)^2}{\omega-1}+o(1)}
\cdot \log^{O(1)}W_{\max}\right)$, whose exponent of $n$ is strictly below $\omega$
if  $\omega>2$. Using the current rectangular FMM bounds of
Uzna{\'n}ski [arXiv'26] or Amanov, van den Berg
and Zuiddam [arXiv'26], the runtime becomes
$\widetilde O\left(
n^{2.17157288}
\cdot\log^{O\left(1\right)}W_{\max}\right)$.

Our algorithms use two main technical contributions.
First, we combine the idea of $\left(\max,\min\right)$-matrix multiplication with distance estimations.
To do so,
extend the
$\left(\max,\min\right)$-matrix multiplication
framework of Duan and Pettie [SODA'09] to fully
rectangular matrices. For matrices of dimensions
$n^a\times n^b$ and $n^b\times n^c$, we compute their
$\left(\max,\min\right)$-product in
$\widetilde O\left(
n^{\frac{a+b+c+\omega(a,b,c)}{2}+o(1)}
\right)$ time. To the best of our knowledge, this is the first time rectangular $\left(\max,\min\right)$-matrix multiplication is used for APASP.

Second, we introduce pivot-adjusted
$\left(\min,+\right)$-products.
We group vertices
according to their distances to their pivots, subtract
values determined by each weight scale, compute approximate  $\left(\min,+\right)$-products and restore the
corresponding corrections afterwards.

\end{abstract}

\printkeywords

\setstretch{1.3}

\newpage

\tableofcontents
\newpage

\pagenumbering{arabic}

\section{Introduction}~\label{intro}
Let $G=\paren{V,E,w}$ be an undirected graph with $\abs{V}=n$ vertices, $\abs{E}=m$ edges and a nonnegative 
weight function $w:E\rightarrow \mathbb{R}^{\geq 0}$. For a \textit{multiplicative stretch} $\alpha\geq 1$ and an \textit{additive stretch} $\beta\geq 0$, in the \textbf{A}ll-\textbf{P}airs \textbf{A}pproximate \textbf{S}hortest \textbf{P}aths (\emph{APASP}) problem, we compute an estimate $d\bracke{u,v}$ for every pair $u,v\in V$, such that:

$$d\bracke{u,v}\in\bracke{\delta\paren{u,v},\alpha\cdot\delta\paren{u,v}+\beta}\textnormal{,}$$
 where $\delta\paren{u,v}$ is the distance in $G$. If $\beta = 0$, the approximation is  \textit{multiplicative}, referred to
as an $\alpha$-APASP; if $\alpha = 1$, the approximation is  \textit{additive}, known as
$+\beta$-APASP; if $\alpha=1+\varepsilon$, the approximation is \textit{nearly additive}; otherwise, the approximation is \textit{mixed}.

Since any $\paren{\alpha,\beta}$-APASP algorithm  outputs $\Theta\paren{n^{2}}$ estimates, $\Omega\paren{n^{2}}$ time is unavoidable. A natural goal is therefore to obtain the smallest possible multiplicative stretch in $\widetilde O\paren{n^{2}}$ time. Cohen and Zwick \cite{CohZwi1997} and Baswana and Kavitha \cite{BasKav2010} achieve this running time for computing $3$-APASP in weighted graphs.
For unweighted graphs, Saha and Ye \cite{SahYe2023} obtain a $\frac{7}{3}$-APASP in $\widetilde O\paren{n^{2}}$ time.  These results lead to the following question:

\begin{question}\label{que:n^2}
What is the smallest constant $\alpha\geq 1$ for which an $\alpha$-APASP in weighted undirected graphs can be computed in $\widetilde O\paren{n^{2}}$ time?
\end{question}

Achieving a $2$-APASP is a
particularly important target: Dor, Halperin and
Zwick~\cite{DorHalZwi2000} show that, for every fixed
$0<\varepsilon<1$, computing a
$\paren{2-\varepsilon}$-APASP is at least as hard as
\textbf{B}oolean \textbf{M}atrix \textbf{M}ultiplication (\emph{BMM}), even for unweighted
undirected graphs. Thus a nearly quadratic algorithm
with stretch below $2$ would also give a nearly quadratic
BMM algorithm.

We distinguish between \emph{combinatorial} algorithms,
which do not use \textbf{F}ast \textbf{M}atrix
\textbf{M}ultiplication (\emph{FMM}), and
\emph{algebraic} algorithms, which do.
For weighted $2$-APASP, Cohen and Zwick~\cite{CohZwi1997}
give a combinatorial algorithm with running time
$\widetilde O\paren{n^2+n^{\frac{3}{2}}m^{\frac{1}{2}}}$.
Baswana and Kavitha~\cite{BasKav2010} improve this to
$\widetilde O\paren{n^2+mn^{\frac{1}{2}}}$.
Both bounds become $\widetilde O\paren{n^{\frac{5}{2}}}$
for dense graphs.

Algebraic algorithms give a smaller exponent.
Let $\omega\paren{a,b,c}$ denote the exponent of
multiplying matrices of dimensions $n^a\times n^b$
and $n^b\times n^c$, and write
$\omega=\omega\paren{1,1,1}$ for square matrices.
For every fixed $\varepsilon>0$,
Zwick~\cite{Zwick2002} computes a
$\paren{1+\varepsilon}$-APASP in
$\widetilde O\paren{\frac{1}{\varepsilon}\cdot n^{\omega+o\paren{1}}\cdot \log{W_{\max}}}$, where $W_{\max}$ is the weight of the heaviest edge in $G$. Bringmann, K{\"u}nnemann and
W{\k{e}}grzycki~\cite{BriKunWun2019} remove this
logarithmic factor, obtaining: $\widetilde O\paren{\frac{1}{\varepsilon}\cdot n^{\omega+o\paren{1}}}$. Setting $\varepsilon=1$ gives a $2$-APASP
in $\widetilde{O}\paren{n^{\omega+o\paren{1}}}$ runtime.

OpenAI~\cite{OpenAI2026} recently showed that
$\omega\leq\frac{9}{4}$, improving the previous  $\omega\leq 2.371177$
of Dupont, Eisenberger, Kozlovskii, Mehrabian, Ruiz,
See, Zhou, Alman, Vassilevska Williams and
Balog~\cite{DupEisKozMehRuiSeeZhoAlmVasBal2026}.
For unweighted graphs, current rectangular FMM bounds
already yield $2$-APASP in $n^{2+o\paren{1}}$ time (See: \appref{unweighted2}). For weighted graphs, however, obtaining  $2$-APASP in
$\widetilde O\paren{n^2}$ time remains open. With the current $\omega$ values, the
$2$-APASP algorithm of  Bringmann, K{\"u}nnemann and
W{\k{e}}grzycki~\cite{BriKunWun2019} takes  
$\widetilde O\paren{n^{\frac{9}{4}+o\paren{1}}}$ runtime.  
As a step toward \queref{n^2}, we ask:

\begin{question}\label{que:exact2}
If $\omega>2$, can $2$-APASP for weighted graphs  be computed in
$\widetilde O\paren{n^{\omega-\eta}}$ 
for some constant $\eta>0$?
\end{question}

We answer \queref{exact2} in the affirmative. For integer edge weights bounded by $W_{\max}$,
our algorithm computes a $2$-APASP in $\widetilde O\paren{
n^{\omega-\frac{\paren{\omega-2}^{2}}{\omega-1}+o\paren{1}}
\cdot\log^{O\paren{1}}W_{\max}}$ runtime, whose exponent is $<\omega$ if $\omega>2$. Substituting the value of $\omega\leq \frac{9}{4}$ \cite{OpenAI2026} and  
the rectangular FMM bounds of
Uzna{\'n}ski~\cite{Uznanski2026} or those of 
Amanov, van den Berg and
Zuiddam~\cite{AmaBerZui2026}, our runtime becomes $\widetilde O\paren{
n^{5-2\sqrt{2}+o\paren{1}}
\cdot\log^{O\paren{1}}W_{\max}}\approx \widetilde O\paren{
n^{2.17157288}
\cdot\log^{O\paren{1}}W_{\max}}$. 

While previous algorithms for weighted undirected graphs achieved
$\alpha$-APASP with a runtime exponent below $\omega$ for
$\alpha\in\left(2,3\right]$, our algorithm is the first to attain precise 
$\alpha=2$. Thus, $2$ is the
smallest  multiplicative stretch achievable with a
runtime exponent strictly below $\omega$, assuming $\omega>2$.

\subsection{Our Results}\label{ours}

\begin{table}[t]
\centering
\small
\setlength{\tabcolsep}{6pt}
\renewcommand{\arraystretch}{1.5}

\begin{tabular}{|c|c|c|c|}
\hline
\multirow{2}{*}{\textbf{Stretch}}
& \multirow{2}{*}{\textbf{Algorithm}}
& \multicolumn{2}{c|}{$\omega$ value}
\\
\cline{3-4}
& 
& $\omega\leq 2.371177$~\cite{DupEisKozMehRuiSeeZhoAlmVasBal2026,VasXuXuZho2024,AlmDuaVasXuXuZho2024}
& $\omega\leq \frac{9}{4}$~\cite{OpenAI2026,Uznanski2026,AmaBerZui2026}
\\
\hhline{|=|=|=|=|}
\multirow{2}{*}{$2$}
& \algref{2-APASP-slower-main}
& $\widetilde O\paren{n^{2.36999465}}$
& $\widetilde O\paren{n^{2.34549151}}$
\\
\cline{2-4}
& \algref{2-APASP-faster-main}
& $\widetilde O\paren{n^{2.24186294}}$
& $\widetilde O\paren{n^{2.17157288}}$
\\
\hline
\end{tabular}

\caption{Runtimes of both our algorithms, depending on different $\omega$ values.
\algref{2-APASP-faster-main}  assumes integer weights and
includes an additional factor of 
$\log^{O\paren{1}}W_{\max}$.}
\label{tab:runtimes}
\end{table}

We present two algebraic constructions for weighted
$2$-APASP, using different matrix products. The stated runtime bounds for our randomized
APASP algorithms hold with high probability.
Our first algorithm, \algref{2-APASP-slower-main},
reduces the runtime to $
\widetilde O\paren{
n^{3-b-c}+n^{2+b}+n^{2+c}
+n^{\frac{3-b+\omega\paren{1,1-b,1}}{2} + o\paren{1}}
}
$, by using \minmaxname-products. This runtime is at most 
$
\widetilde O\paren{
n^{2+\frac{\omega-1}{\omega+1}+o\paren{1}}
}$, for $b=\frac{\omega-1}{\omega+1}$ and $c=\frac{3-\omega}{\omega+1}$. Using the earlier rectangular FMM bounds of
Vassilevska Williams, Xu, Xu and
Zhou~\cite{VasXuXuZho2024} and Alman, Duan,
Vassilevska Williams, Xu, Xu and
Zhou~\cite{AlmDuaVasXuXuZho2024}, we can obtain an even slightly better runtime of 
$\widetilde O\paren{n^{2.36999465}}$.
The  exponent of $n$ is below the previous bound
$\omega\leq 2.371177$ of Dupont, Eisenberger,
Kozlovskii, Mehrabian, Ruiz, See, Zhou, Alman,
Vassilevska Williams and
Balog~\cite{DupEisKozMehRuiSeeZhoAlmVasBal2026}.

Following OpenAI's bound
$\omega\leq \frac{9}{4}$~\cite{OpenAI2026},
the rectangular bounds of
Uzna{\'n}ski~\cite{Uznanski2026} or of 
Amanov, van den Berg and
Zuiddam~\cite{AmaBerZui2026}, we improve this runtime to
$
\widetilde O\paren{
n^{\frac{21-\sqrt{5}}{8}+o\paren{1}}
}\approx \widetilde O\paren{n^{2.34549151}}$. Sadly, this exponent  exceeds $\frac{9}{4}$. The limitation comes from the three runtime terms:
$\mmin{
\mmax{3-b-c,2+b,2+c}
}{\substack{b,c\geq 0\\b+c\leq 1}}
=\frac{7}{3}$. 
Thus, while it has no dependence on $W_{\max}$, this runtime analysis cannot give an exponent
below $\frac{7}{3}$, even if the \minmaxname-product
were to be computed for free. 

To overcome this and to go below any value of $\omega>2$, our second algorithm, \algref{2-APASP-faster-main}, 
uses approximate \minplusname-products with corrections
determined by each weight scale. For integer edge weights bounded by $W_{\max}$
and a parameter $a\in\bracke{0,1}$, it  requires $\widetilde O\paren{
n^{2+a}+n^{a+\omega\paren{1-a,1,1-a}+o\paren{1}} \cdot\log^{O\paren{1}} W_{\max}
}$ runtime. This is at most  $\widetilde O\paren{
n^{\omega-\frac{\paren{\omega-2}^{2}}{\omega-1}
+o\paren{1}}
\cdot\log^{O\paren{1}}W_{\max}
}$ runtime by choosing $a=\frac{\omega-2}{\omega-1}$. If $\omega>2$ the exponent of $n$ is strictly below $\omega$, therefore answering \queref{exact2}. 

Using the earlier rectangular FMM bounds
\cite{VasXuXuZho2024,AlmDuaVasXuXuZho2024}, the 
 runtime is
$\widetilde O\paren{
n^{2.24186294}
\cdot\log^{O\paren{1}}W_{\max}}$.
The current rectangular bounds
\cite{Uznanski2026,AmaBerZui2026} give
$\omega\paren{1,1-a,1}
\leq 2-a+\frac{1}{4\paren{1-a}}$ and
$a+\omega\paren{1-a,1,1-a}
\leq 2+\frac{\paren{1-a}^{2}}{4}$,
for $ a\in\bracke{0,\frac{1}{2}}$.
Choosing $a=3-2\sqrt{2}$ therefore gives
$\widetilde O\paren{
n^{5-2\sqrt{2}+o\paren{1}}
\cdot\log^{O\paren{1}}W_{\max}}$ runtime.
Assuming the weights are polynomial in $n$, this expression becomes: $\widetilde O\paren{
n^{2.17157288}}$. We summarize the runtimes of  \algref{2-APASP-slower-main} and \algref{2-APASP-faster-main} under the earlier and current
FMM bounds in \tabref{runtimes}.

We also extend \maxminname-products to fully rectangular
matrices. Duan and Pettie~\cite{DuaPet2009} compute
the \maxminname-product of two $n\times n$ matrices in
$\widetilde O\paren{n^{\frac{3+\omega}{2}+o\paren{1}}}$
time. Grandoni, Italiano, {\L}ukasiewicz, Parotsidis and
Uzna{\'n}ski~\cite{GraItaLukParUzn2021} extend this
bound to matrices of dimensions $n\times n^b$ and
$n^b\times n$.
In \secref{rect_(max,min)}, we allow all three
dimensions to differ: for matrices of dimensions
$n^a\times n^b$ and $n^b\times n^c$, their
\maxminname-product can be computed in
$\widetilde O\paren{
n^{\frac{a+b+c+\omega\paren{a,b,c}}{2}+o\paren{1}}
}$.
The same bound holds for the \minmaxname-product
(See: \corref{rect_minmax}).

\subsection{Related Results}\label{related}

For weighted graphs, Dory, Forster, Kirkpatrick, Nazari, Vassilevska Williams and de Vos~\cite{DorForKirNazVasVos2023} compute a $\paren{2+\varepsilon}$-APASP, for any $\varepsilon>0$, in $\widetilde O\paren{\paren{\frac{1}{\varepsilon}}^{O\paren{1}}\cdot \paren{n^{2+\alpha}+n^{\omega\paren{1,1-\alpha,1}+o\paren{1}}}\cdot\log W_{\max}}$ time, for any $\alpha\in\bracke{0,1}$. With the current rectangular FMM bounds
\cite{Uznanski2026,AmaBerZui2026}, this becomes
$\widetilde O\paren{
\paren{\frac{1}{\varepsilon}}^{O\paren{1}}
\cdot n^{2.14644661}
\cdot\log W_{\max}
}$ runtime. 
While their 
exponent is indeed below $\omega$, it has multiplicative stretch $2+\varepsilon$, for an $\varepsilon>0$.
It should be observed that even
 an arbitrarily small positive $\varepsilon$
still permits estimates larger than
$2\delta\paren{u,v}$.
The challenge is not to make $\varepsilon$ arbitrarily small, but to eliminate it entirely:
we achieve stretch exactly $2$ for every pair while keeping the running-time
exponent below $\omega$.

For unweighted graphs, Roditty~\cite{Roditty2023} and Dory, Forster, Kirkpatrick, Nazari, Vassilevska Williams and de Vos~\cite{DorForKirNazVasVos2023} give combinatorial $2$-APASP algorithms in $\widetilde O\paren{n^{\frac{9}{4}}}$ time.
A related direction seeks $\widetilde O\paren{n^2}$ runtime by requiring stretch $2$ only for pairs whose distance exceeds a certain threshold. Gupta~\cite{Gupta2025} gives a combinatorial algorithm that covers all pairs at distance at least  $\Omega\paren{\log\log n}$. Using FMM, Gupta and Shandilya~\cite{GupSha2026} reduce this threshold to a constant $c\approx 906$. 

Going back to all vertex pairs, Dory, Forster, Kirkpatrick, Nazari, Vassilevska Williams and de Vos~\cite{DorForKirNazVasVos2023} give an algebraic $2$-APASP\footnote{Their reduction from stretch $2+\varepsilon$ to stretch $2$ sets $\varepsilon=\frac{1}{\log n}$, rounds down, and takes the minimum with a $+\log n$-APASP of Dor, Halperin and Zwick~\cite{DorHalZwi2000}. The additional runtime is absorbed into polylogarithmic factors.}  algorithm for unweighted graphs in $\widetilde O\paren{n^{\frac{3}{2}+\alpha}+n^{\omega\paren{1,1-\alpha,1}+o\paren{1}}}$ time, for any $\alpha\in\bracke{0,1}$. Independently, Saha and Ye~\cite{SahYe2023} give an algebraic $2$-APASP 
algorithm with the same runtime. The rectangular FMM bounds of Uzna{\'n}ski~\cite{Uznanski2026} or of Amanov, van den Berg and Zuiddam~\cite{AmaBerZui2026} give $\omega\paren{1,\frac{1}{2},1}=2$. Choosing $\alpha=\frac{1}{2}$ therefore gives $n^{2+o\paren{1}}$ runtime for both algorithms. In \appref{unweighted2}, we present a simplified version of the algorithm of Dory, Forster, Kirkpatrick, Nazari, Vassilevska Williams and de Vos~\cite{DorForKirNazVasVos2023}.

Thus, unweighted algebraic $2$-APASP already admits a quadratic exponent. Obtaining $\widetilde O\paren{n^2}$ runtime remains open for weighted graphs and for combinatorial algorithms on unweighted graphs. \tabref{current2} compares these four settings.

\begin{table}[t]
\centering
\small
\setlength{\tabcolsep}{6pt}
\renewcommand{\arraystretch}{1.5}

\begin{tabular}{|c|c|c|}
\hline
\textbf{Graph}
& \textbf{Combinatorial}
& \textbf{Algebraic}
\\
\hhline{|=|=|=|}
Unweighted
& $\widetilde O\paren{n^{2.25}}$~\cite{Roditty2023,DorForKirNazVasVos2023}
& $n^{2+o\paren{1}}$ \cite[\appref{unweighted2}]{DorForKirNazVasVos2023}
\\
\hline
Weighted
& $\widetilde O\paren{n^{2.5}}$~\cite{BasKav2010}
& $\widetilde O\paren{n^{2.17157288}}$ [\secref{2-faster}]
\\
\hline
\end{tabular}

\caption{Runtimes for $2$-APASP on dense graphs.
The algebraic bounds use current rectangular FMM bounds.
Our weighted algorithm assumes integer weights and includes an additional factor
$\log^{O\paren{1}}W_{\max}$.}
\label{tab:current2}
\end{table}

\subsection{Organization}\label{organization}

\secref{toolkit} presents the notation and algorithmic tools,
and \secref{overview} explains the main ideas and technical
contributions behind our algorithms.
\secref{2-slower} presents our weighted $2$-APASP algorithm
using rectangular \minmaxname-products, and \secref{2-faster} presents
the faster algorithm using approximate \minplusname-products
with corrections determined by each weight scale.
\secref{rect_(max,min)} establishes the fully rectangular
\maxminname-product bound. \secref{conclusion} concludes with open problems.

\appref{balance} presents the row and column
balancing procedures used in
\secref{rect_(max,min)}. \appref{unweighted2} presents a simplified $2$-APASP
algorithm for unweighted graphs with
$n^{2+o\paren{1}}$ runtime.
\appref{2,W2} and \appref{2+eps,W2} present
$\paren{2,W_{2}}$-APASP and
$\paren{2+\varepsilon,W_{2}}$-APASP algorithms,
respectively.
\appref{app:allResults} summarizes known APASP results
and updates their runtimes by substituting the new
square and rectangular FMM bounds.

\section{Toolkit}\label{toolkit}
\subsection{Basic Graph Notions}\label{notions}
Unless stated otherwise, let $G=\paren{V,E,w}$ be an undirected graph with a non-negative weight function $w:E\rightarrow\mathbb{R}^{\geq 0}$. We denote $\abs{V}=n$,  $\abs{E}=m$, and the weight of an edge $\paren{u,v}\in E$ by $w\paren{u,v}$. For every integer $k\in \mathbb{N}\setminus\bracce{0}$, let $\bracke{k}=\bracce{1,2,\ldots,k}$.

A \textit{path} from $u$ to $v$ is a sequence
$P=\pair{y_{0},y_{1},\ldots,y_{\ell}}$ with $y_{0}=u$, $y_{\ell}=v$ and $\paren{y_{i-1},y_{i}}\in E$ for every $i\in\bracke{\ell}$. The number of edges in a path is denoted by $\abs{P}$ and the weight of a path is $w\paren{P}=\msum{w\paren{y_{i-1},y_{i}}}{i=1}{\ell}$. 
Furthermore, for any two vertices $x, y \in P$, $P\bracke{x, y}$ denotes the sub-path of $P$ lying between the vertices $x, y$.    

If the graph is unweighted, $w\paren{P}=\abs{P}$.  
A $(u,v)$-path of minimum weight is called a
\emph{shortest path} and is denoted by $u\squiggly v$.  The distance between $u$ and $v$ is denoted by $\delta_G\paren{u,v}=w\paren{u\squiggly v}$. If the graph $G$ is clear from the context, we just use $\delta\paren{u, v}$ for convenience. When there is only one edge in  $u\squiggly v$ we alternately use
$=u\straight v$ for $u\squiggly v$, in which case
$w\paren{u,v}=\delta\paren{u,v}$. For a vertex
$u\in V$ and a set $S\subseteq V$, we define
$\delta\paren{u,S}=\mmin{\delta\paren{u,s}}{s\in S}$.

For a path $P$, let $W_{i}\paren{P}$ denote the weight of its
$i^{\textnormal{th}}$ heaviest edge\footnote{We set $W_{i}\paren{P}=0$
for $i>\abs{P}$.}. In particular,
$W_{2}\paren{u\squiggly v}$ denotes the weight of the 
second heaviest edge on $u\squiggly v$. If the path $P = u\squiggly v$ is some fixed $\paren{u,v}$-shortest path clear from the context, we write $W_{i}\paren{P}$ as just $W_{i}$. Accordingly, every expression of the form
$\msum{W_{i}}{i\in I}$ is evaluated with respect to the same fixed path $u\squiggly v$.
When several shortest paths exist and no path is specified explicitly,
we choose one that minimizes the relevant upper bound.

Throughout our algorithms, $d\bracke{u,v}$ denotes the distance
estimate computed for the pair $u,v\in V$, whereas
$\delta\paren{u,v}$ denotes their actual distance in $G$. For an $\paren{\alpha,\beta}$-APASP algorithm, the goal is to prove that for every pair of vertices: 
$
d\bracke{u,v}
\in
\bracke{
\delta\paren{u,v},
\alpha\cdot\delta\paren{u,v}+\beta
}
$.

\subsection{Balls and Pivots}\label{nearest}

The following sampling quantities admit standard
bounds in expectation. In each of the following observations, we 
state high probability upper bounds. These will be used
in our runtimes analysis.
Here, high probability means probability 
$\geq 1-n^{-c}$ for any fixed constant $c>0$, where
the implicit constants may depend on $c$.
These bounds follow from standard Chernoff
bounds with the independent sampling indicators.
All sampling size bounds in this section hold
simultaneously for the vertices and sampled
levels used by our algorithms.

Let $q\in\paren{0,1}$ and let $S\subseteq V$ be obtained by sampling each vertex independently with probability $q$. We use the convention $\delta\paren{u,\varnothing}=\infty$. Whenever $S\neq\varnothing$, the \emph{pivot} of a vertex $u\in V$ with respect to $S$ is a vertex $p_{S}\paren{u}\in S$ satisfying
$\delta\paren{u,p_{S}\paren{u}}=\delta\paren{u,S}$. Ties are resolved by a consistent criteria. 
The \emph{ball} of $u$ with respect to $S$ consists of the vertices that are closer to $u$ than $p_{S}\paren{u}$: $\ball{u}{S} = \bracce{ v \in V \midline \delta\paren{u,v}<\delta\paren{u,S}}$. The sizes of the sampled set and the resulting balls satisfy the following standard bounds:

\begin{observation}[\cite{ThoZwi2005}]~\label{obv:hssize} 
$\abs{S} \in \widetilde O \paren{nq}$.
\end{observation}

\begin{observation}[\cite{ThoZwi2005}]~\label{obv:ballsize} 
$\abs{\ball{u}{S}} \in \widetilde O \paren{\frac{1}{q}}$.
\end{observation}

We next associate each vertex $u\in V$ with the set of incident edges whose weight is smaller than its distance to $S$: $
E_{S}\paren{u}=\bracce{\paren{u,v}\in E\midline w\paren{u,v}<\delta\paren{u,S}}$. We then define: $E_{S}=
\mcup{E_{S}\paren{u}}{u\in V}$.

\begin{observation}[\cite{ThoZwi2005}]~\label{obv:edgessize} 
$\abs{E_{S}} \in \widetilde O \paren{\frac{n}{q}}$.
\end{observation}

If $\paren{x,y}\in E_{S}$ then $x\in \ball{y}{S}$ or $y\in \ball{x}{S}$. Otherwise, if $\paren{x,y}\notin E_{S}$, then $\delta\paren{x,p_{S}\paren{x}} \leq w\paren{x,y}$ and $\delta\paren{y,p_{S}\paren{y}} \leq w\paren{x,y}$. 
That is,  edges in $E_{S}$  are within the balls, whereas the weight of edges outside of $E_{S}$ is an upper-bound for the distance to the pivots. This distinction will be used repeatedly in our distance estimates.

\begin{observation}[\cite{BasKav2010}]~\label{obv:pathinball} 
Let $u\in V$, $S\subseteq V$ and $v\in \ball{u}{S}$. Then: $u\squiggly v \subseteq E_{S}$.
\end{observation}

We remark that the pivots and their exact distances can be computed simultaneously by introducing an auxiliary source connected to every vertex of $S$ by a zero weight edge and a single SSSP computation.

\begin{observation}[\cite{AinCheIndMot1999}]~\label{obv:pivotsdistance} 
The pivots $p_{S}\paren{u}$ for all vertices $u\in V$ and the exact distances $d\bracke{u,p_{S}\paren{u}} = \delta\paren{u,p_{S}\paren{u}}$ can be computed together in total $\widetilde O \paren{m}$ time.
\end{observation}

Finally, we extend the ball definition. Let $A\subseteq V$ and let $S\subseteq A$ be obtained by sampling each vertex of $A$ independently with probability $q'\in\paren{0,1}$. We define:
$\ball{u}{A}{S} = \bracce{ a\in A \midline \delta\paren{u,a} < \delta\paren{u,S}}$. Equivalently, $\ball{u}{A}{S} =\ball{u}{S}\cap A$.

\begin{observation}[\cite{ThoZwi2005}]~\label{obv:bunchsize} 
$\abs{\ball{u}{A}{S}} \in \widetilde O \paren{\frac{1}{q'}}$.
\end{observation}

\subsection{Hierarchy of Hitting Sets}~\label{hierarchy}
We extend the preceding notions to a hierarchy of $\ell$ nested sampled sets $V=S_{0} \supseteq S_{1} \supseteq  S_{2}\supseteq  \ldots \supseteq S_{\ell} \supseteq S_{\ell+1}=\varnothing$. To do so, we consider $\ell$ probabilities $q_{1},q_{2},\ldots , q_{\ell} \in\paren{0,1}$. For every $i\in\bracke{\ell}$, we obtain $S_{i}$ by sampling each vertex of $S_{i-1}$ independently with probability $q_{i}$. For completeness, set $S_{0}=V$ and $S_{\ell+1}=\varnothing$. 

For any level $i\in \bracke{\ell}$, we denote $p_{S_{i}}\paren{u}$ as $p_{i}\paren{u}$ and $E_{S_{i}}$ as $E_{i}$ for convenience. Since $S_{i-1}\supseteq S_{i}$, we have that $\delta\paren{u,S_{i-1}}\leq\delta\paren{u,S_{i}}$. Consequently,
$\ball{u}{i-1}\subseteq\ball{u}{i}$ and $E_{i-1}\subseteq E_{i}$. Particularly: $\varnothing=E_{0}\subseteq E_{1}\subseteq\ldots\subseteq E_{\ell}\subseteq E_{\ell+1}=E$.

\begin{observation}[\cite{DorHalZwi2000, BasKav2010}]~\label{obv:ESiwithinESi+1}
For $0\leq i<j\leq\ell+1$ it holds that $S_{j}\subseteq S_{i}$ and  $ E_{i} \subseteq E_{j}$.
\end{observation}

Let $j\in \bracke{\ell}$. A vertex belongs to $S_{j}$ with probability $\mprod{q_{k}}{k=1}{j}$. Thus, by \obvref{hssize} and \obvref{ballsize}:

\begin{observation}[\cite{DorHalZwi2000, BasKav2010}]~\label{obv:hssize2}
$\abs{S_{j}} \in \widetilde O\paren{n\cdot \mprod{q_{k}}{k=1}{j}}$. 
\end{observation}

\begin{observation}[\cite{DorHalZwi2000, ThoZwi2005}]~\label{obv:ballsize2}
$\abs{\ball{u}{j}} \in \widetilde O\paren{\mprod{\frac{1}{q_{k}}}{k=1}{j}}$. 
\end{observation}

Additionally, we classify paths according to the levels in which their edges appear. Let $P$ be a path between $u$ and $v$ and let $i\in\bracke{\ell}$. We denote by $\sizeindex{P}{i}$ the number of edges of $P$ that do not belong to $E_{i}$. If $\sizeindex{P}{i}>0$, let $\rightabovearrow{P}{i}$ and $\leftabovearrow{P}{i}$ denote, respectively, the first and last such edges when traversing $P$ from $u$ to $v$. Recall from \secref{nearest} that the weight of an edge outside $E_{i}$ provides an upper-bound for the distance of either endpoint to its $i^{\textnormal{th}}$ pivot.

\begin{observation}[\cite{RodSap2026}]~\label{obv:Pi=0,1}
$\sizeindex{P}{i}=0$ iff $P\subseteq E_{i}$ and $\sizeindex{P}{i}=1$ iff $\rightabovearrow{P}{i} = \leftabovearrow{P}{i}$.
\end{observation}

We also consider auxiliary edges connecting vertices to their pivots. For every $i\in\bracke{\ell}$, define
$D_{i}=\bracce{\paren{u,p_{i}\paren{u}}\midline u\in V}$ and let
$D=\mcup{D_{i}}{i=1}{\ell}$.

\begin{observation}[\cite{DorHalZwi2000, CohZwi1997}]~\label{obv:edgesToPivots}
Let $i\in \bracke{\ell}$. Then: $\abs{D_{i}}\in O \paren{n}$. Additionally, $\abs{D}\in O\paren{n\cdot \ell}$. If $\ell \in \widetilde O \paren{1}$ then $\abs{D} \in \widetilde O\paren{n}$. 
\end{observation}

Finally, for $0\leq i<j\leq\ell$, we define the refined ball $\ball{u}{i}{j}=
\bracce{
s\in S_{i}
\midline
\delta\paren{u,s}<\delta\paren{u,S_{j}}
}$. Since each vertex of $S_{i}$ is sampled to $S_{j}$ with probability
$\mprod{q_{k}}{k=i+1}{j}$, \obvref{bunchsize} immediately gives:

\begin{observation}[\cite{ThoZwi2005}]~\label{obv:bunchsize2}
Let $i,j\in \bracce{0}\cup\bracke{\ell}$ such that $i<j$. Then: $\abs{\ball{u}{i}{j}} \in \widetilde O\paren{\mprod{\frac{1}{q_{k}}}{k=i+1}{j}}$. 
\end{observation}

\subsection{Matrix Multiplications}\label{mm}
We  define the matrix products used throughout this paper. Let
$A\in\mathbb{R}^{n^{a}\times n^{b}}$
and
$B\in\mathbb{R}^{n^{b}\times n^{c}}$ be two rectangular matrices, where $a,b,c\in\bracke{0,1}$. Each of the following products results in an $n^{a}\times n^{c}$ matrix. 

\begin{definition}\label{def:matrix_products}
For every
$i\in\bracke{n^{a}}$ and $j\in\bracke{n^{c}}$, define:
\begin{equation*}
\begin{alignedat}{2}
\paren{A\cdot B}\bracke{i,j}
&=
\msum{A\bracke{i,k}\cdot B\bracke{k,j}}
      {k\in\bracke{n^{b}}},
&\qquad
\paren{A\minplus B}\bracke{i,j}
&=
\mmin{A\bracke{i,k}+B\bracke{k,j}}
      {k\in\bracke{n^{b}}},
\\[0.5em]
\paren{A\maxmin B}\bracke{i,j}
&=
\mmax{
\min\bracce{A\bracke{i,k},B\bracke{k,j}}
}{k\in\bracke{n^{b}}},
&
\paren{A\minmax B}\bracke{i,j}
&=
\mmin{
\max\bracce{A\bracke{i,k},B\bracke{k,j}}
}{k\in\bracke{n^{b}}},
\\[0.5em]
\paren{A\dom B}\bracke{i,j}
&=
\abs{\bracce{
k\in\bracke{n^{b}}
\midline
A\bracke{i,k}\leq B\bracke{k,j}
}}.
\end{alignedat}
\end{equation*}

We refer to these products as the
\mmname-product,  the \minplusname-product, ,the \maxminname-product, the
\minmaxname-product and the \domname-product, respectively, from left to right.
\end{definition}

When needed, we extend the domain of the values of the matrices
$A$ and $B$ to include infinite values. For the
\minplusname-product, both matrices may contain $+\infty$; for the \maxminname-product and the \minmaxname-product, both
matrices may contain $\pm\infty$; for the \domname-product, the matrix $A$ may contain $+\infty$
and the matrix $B$ may contain $-\infty$. The values $\pm \infty$ uphold $-\infty < x < +\infty$ for any $x\in\mathbb{R}$.

We follow with the classical notation of $\omega\paren{a,b,c}$ as the infimum over all $\tau>0$ such that, for every constant $\varepsilon>0$,
the product $A\cdot B$ can be computed in 
$O\paren{n^{\tau+\varepsilon}}$ runtime. Likewise,
rectangular \mmname of matrices of these dimensions requires
$O\paren{n^{\omega\paren{a,b,c}+o\paren{1}}}$ time. We write
$\omega=\omega\paren{1,1,1}$ and
$\omega\paren{b}=\omega\paren{1,b,1}$.

\subsection{Useful Approximations}\label{useful}

We use the following two results as a black box for our algorithm. First, we mention the $\paren{1+\varepsilon}$-MSASP algorithm by Elkin and Neiman~\cite{ElkNei2020}.

\begin{theorem}\label{thm:ApproxMSSP} (Theorem 5 in \cite{ElkNei2020})
    Consider an undirected graph $G = \paren{V, E, w}$ and a set $S \subseteq V$ of sources such that $\abs{S} = O\paren{n^r}$ for some $r > 0$. Then, given any $\varepsilon > 0$, there exists a deterministic algorithm that computes $\paren{1+\varepsilon}$-MSASP for $S \times V$ in $\widetilde{O}\paren{m^{1+o\paren{1}} + \paren{\frac{1}{\varepsilon}}^{O\paren{1}} \cdot n^{\omega\paren{1,r,1}+o\paren{1}} }$ time.
\end{theorem}

Another well-known tool we use is the $\paren{1+\varepsilon}$-approximate \minplusname-product algorithm for integer weight matrices, which follows from the influential work of Zwick~\cite{Zwick2002}.

\begin{theorem}\label{thm:ApproxMinPlus}(Rectangular version of Lemma 8.1 in \cite{Zwick2002})
    Let $r\in\bracke{0,1}$ and  matrices $A, B$ of dimensions $n \times n^r  $ and $n^{r}\times n$, respectively, where each entry is from  $\bracce{0}\cup \bracke{ W_{\max}} \cup \bracce{\infty}$. For any $\varepsilon > 0$, there exists an algorithm which computes entrywise $\paren{1+\varepsilon}$-approximation of $A \minplus B$ in $\widetilde{O}\paren{\frac{1}{\varepsilon} \cdot n^{\omega\paren{1,r,1}+o\paren{1}}\cdot  \log {W_{\max}}}$ time.
\end{theorem}

\section{Overview}\label{overview}

\subsection{Sampling and Edge Sets}\label{samplingOV}

Both our  algorithms use a standard sampling framework
common to many  APASP algorithms (e.g.~\cite{RodSap2025,CohZwi1997,BasKav2010,
DorForKirNazVasVos2023,ThoZwi2005,Kavitha2012}). Such a framework considers a parameter $k\in\mathbb{N}$ which indicates the number of sampled sets. Starting with $S_{0}=V$, for each
$i\in\bracke{k-1}$, the idea is to sample independently every vertex from $S_{i-1}$
in $S_{i}$  with probability $q_{i}\in \bracke{0,1}$.  Setting $S_k=\varnothing$, obtaining: $V=S_{0}\supseteq S_1\supseteq\cdots\supseteq
S_{k-1}\supseteq S_{k}=\varnothing$. 
For every level $i$ and every vertex $u\in V$, a pivot   
$p_i\paren{u}$ is found and  $E_{i}$ is constructed (See: \secref{hierarchy}).  It follows that $\varnothing=E_{0}\subseteq E_{1}\subseteq\cdots
\subseteq E_{k}=E$. 
This hierarchy is used to balance the number of sources
against the size of the graph on which we compute
\textbf{S}ingle-\textbf{S}ource \textbf{S}hortest
\textbf{P}aths (\emph{SSSP}).  An SSSP from
a vertex $s\in S_{i-1}$ uses  $E_{i}$ and may also use auxiliary edges
$\bracce{s}\times V$, whose weights are given by previously
computed entries of the distance estimation matrix $d$. Since these auxiliary edges may differ between sources,
these are multiple SSSP invocations over different graphs. However, the searches from
$S_{k-1}$ use the entire edge set $E_{k}=E$, and can be described as
a single \textbf{M}ultiple-\textbf{S}ource \textbf{S}hortest
\textbf{P}aths (\emph{MSSP}) computation\footnote{For algebraic algorithms, we can  replace it by a $\paren{1+\varepsilon}$-MSASP from $S_{k-1}$.} from $S_{k-1}$. For our two algorithms, we use a simple hierarchy version where either $k=3$ or $k=2$. 

Apart from a runtime perspective, the  definition of $E_{i}$ has an important role in the
analysis. If $\paren{x,y}\notin E_{i}$, then
$\delta\paren{x,p_{i}\paren{x}}\leq w\paren{x,y}$
and
$\delta\paren{y,p_{i}\paren{y}}\leq w\paren{x,y}$. That is, an edge missing from $E_{i}$ gives
bounds on the distances from both endpoints to
their pivots. This allows an efficient use of the triangle inequality, to provide certain upper bounds and the estimates computed by the SSSP invocations.

\subsection{Using \minmaxname-products}\label{minmaxOV}

Our first algorithm,  \algref{2-APASP-slower-main}, uses two sampled sets. Each vertex of $V$ is included
in $S_{1}$ independently with probability $n^{-b}$.
Each member of $S_{1}$ is then included in $S_{2}$
independently with probability $n^{-c}$, and we set
$S_{3}=\varnothing$. Thus,
$\mathbb{E}\bracke{\abs{S_{1}}}=n^{1-b}$ and
$\mathbb{E}\bracke{\abs{S_{2}}}=n^{1-b-c}$.
By \obvref{edgessize}, the corresponding edge sets
satisfy
$\mathbb{E}\bracke{\abs{E_{1}}}\in O\paren{n^{1+b}}$
and
$\mathbb{E}\bracke{\abs{E_{2}}}\in O\paren{n^{1+b+c}}$.
The SSSP invocations from $V$ over $E_{1}$,
from $S_{1}$ over $E_{2}$ and from $S_{2}$ over
$E_{3}=E$ contribute
$\widetilde O\paren{n^{2+b}}$,
$\widetilde O\paren{n^{2+c}}$ and
$\widetilde O\paren{n^{3-b-c}}$ to the runtime,
respectively.

We then compute the estimation between any two vertices $u,v\in V$ by considering any member $s\in S_{2}$ and summing the estimations $d\bracke{u,v}\leftarrow d\bracke{u,s}+d\bracke{s,v}$. Considering every member of $S_{1}$
in this way would be too expensive. Instead, we use a rectangular \minmaxname-product. 
Let
$A$ be the $V\times S_{1}$ matrix whose entries are
$A\bracke{u,s}=d\bracke{u,s}$. The \minmaxname-product $C= A\minmax A^{\mathsf T}$ gives:
$$
C\bracke{u,v}
=
\mmin{
\mmax{d\bracke{u,s},d\bracke{s,v}}
}{s\in S_{1}}
$$

We update $d\bracke{u,v}$ with $2C\bracke{u,v}$. The uniqueness of this update is that it differs from the   common approach of summing
the estimates through a pivot. Although $C\bracke{u,v}$ may underestimate
$\delta\paren{u,v}$,  twice this value cannot. 

Indeed, for every $s\in S_{1}$, the triangle
inequality gives
$\delta\paren{u,v}\leq \delta\paren{u,s}+\delta\paren{s,v}\leq d\bracke{u,s}+d\bracke{s,v}
\leq 2\mmax{d\bracke{u,s},d\bracke{s,v}}$.
Taking the
minimum over $s\in S_{1}$ preserves the inequality.
Thus, $\delta\paren{u,v}\leq 2C\bracke{u,v}$.
Furthermore, if there exists $s\in S_{1}$ that satisfies  $d\bracke{u,s}\leq\delta\paren{u,v}$ and
$d\bracke{s,v}\leq\delta\paren{u,v}$, then
$2C\bracke{u,v}\leq 2\delta\paren{u,v}$.

The product considers all members of $S_{1}$
simultaneously for all vertex pairs. Its rectangular
dimensions allow us to use the smaller size of
$S_{1}$ to improve the runtime.
To compute it, we extend the \maxminname-product
framework of Duan and Pettie~\cite{DuaPet2009}
to fully rectangular matrices. 
For matrices of dimensions
$n^{a}\times n^{b}$ and $n^{b}\times n^{c}$, we obtain
a runtime of
$\widetilde O\paren{
n^{\frac{a+b+c+\omega\paren{a,b,c}}{2}+o\paren{1}}
}$.
The same bound applies to \minmaxname-products (See: \corref{rect_minmax}).

\subsection{Multiple Scales of \minplusname-products}\label{min+OV}

Our second algorithm, \algref{2-APASP-faster-main},
uses only one sampled set $S_{1}$. Each vertex of $V$
is included independently with probability $n^{-a}$,
and we set $S_{2}=\varnothing$.
The algorithm computes two SSSP invocations over $E_{1}$
with a $\paren{1+\varepsilon}$-MSASP computations from $S_{1}$
over $E_{2}=E$.  It then proceeds with multiple approximate \minplusname-products with corrections
determined by each weight scale.

Consider a shortest path $u\squiggly v$. 
If  $\sizeindex{u\squiggly v}{1}\leq 1$, two SSSP invocations from all $u\in V$ over $E_{1}\cup \paren{\bracce{u}\times V}$ would result in $d\bracke{u,v}=\delta\paren{u,v}$. 
Otherwise, let
$\paren{x,y}=\rightabovearrow{u\squiggly v}{1}$ and
$\paren{z,w}=\leftabovearrow{u\squiggly v}{1}$.
These are distinct edges, and the smaller of their
weights is at most $W_{2}$. 

This suggests seeking an
estimate of at most $\delta\paren{u,v}+2W_{2}$, which
would give multiplicative stretch $2$ since
$2W_{2}\leq\delta\paren{u,v}$. To do so, 
one would like to compute $A\minplus A^{\mathsf T}$, where
$A$ is the $V\times S_{1}$ matrix whose entries are
$A\bracke{u,s}=d\bracke{u,s}$. 
While an exact \minplusname-product is too expensive, an approximation may suffice. Yet, 
 an approximate \minplusname-product introduces
an additional multiplicative error and even an estimate
of $\paren{1+\varepsilon}\cdot\delta\paren{u,v}+2W_{2}$
may exceed $2\delta\paren{u,v}$.

A natural attempt is to choose $\varepsilon$ sufficiently
small that
$\varepsilon\cdot\delta\paren{u,v}\leq\delta\paren{u,v}-2W_{2}$. However, this choice
fails for two reasons. First, it
depends on a specific shortest path of a certain pair, while we want to compute a single approximate \minplusname-product for all vertices. 
Second, it  may require $\varepsilon=0$, while we need a  $\varepsilon>0$.

An improved attempt would be to choose a single
$\varepsilon>0$ depending on $W_{\max}$.
For integer weights, choosing
$\varepsilon=\frac{1}{4W_{\max}}$ and rounding down
would suffice. Indeed, it holds that 
$\paren{1+\varepsilon}\cdot \delta\paren{u,v}+2W_{2}
=\paren{1+\varepsilon}\cdot\delta\paren{u,v}
+\paren{1-\varepsilon}\cdot 2W_{2}+2\varepsilon W_{2}\leq 2\delta\paren{u,v}+2\varepsilon W_{2}
\leq 2\delta\paren{u,v}+\frac{1}{2}$. Rounding down therefore gives an estimate of at most
$2\delta\paren{u,v}$.
However, the polynomial runtime dependence on
$\frac{1}{\varepsilon}$ would therefore  be a polynomial
dependence on $W_{\max}$.

We therefore keep $\varepsilon$ constant and subtract
part of the distance before applying the approximation.
First, consider the shortest path
$u\squiggly x\straight y\squiggly z\straight w\squiggly v$.
Its prefix $u\squiggly x$ and suffix $w\squiggly v$
contain only edges of $E_{1}$.
If we obtain $d\bracke{x,w}\leq 2\delta\paren{x,w}$,
two additional rounds of SSSP extend this estimate
to the original endpoints, giving
$d\bracke{u,v}\leq
\delta\paren{u,x}+2\delta\paren{x,w}+\delta\paren{w,v}
\leq 2\delta\paren{u,v}$.
We therefore focus on estimating $d\bracke{x,w}$. 

To do so, we consider $\log W_{\max}$ weight scales. At the $j^{\textnormal{th}}$ scale, group $i$ contains the vertices $x$ with $\delta\paren{x,p_{1}\paren{x}}\in\bracke{i\cdot 2^{j},\paren{i+2}\cdot 2^{j}}$. We consider only nonempty groups. These ranges determine the corrections we subtract before computing the approximate \minplusname-product and restore afterwards.

We fix a sufficiently small constant $\varepsilon = \frac{1}{20} >0$ for  the MSASP computations and the approximate \minplusname-products. After restoring the corrections, we update $d\bracke{x,w}$ with the smallest estimate obtained over all groups $i$ and scales $j$. For each pair $x,w$, it suffices that one of these estimates is at most $2\delta\paren{x,w}$. 

\section{Warm Up: $2$-APASP faster than $n^{2.371}$}\label{2-slower}

In this section, we describe our first result: An algorithm which computes 2-APASP faster than $n^{2.371}$ time. The algorithm relies on the fairly standard vertex sampling ideas, but with an additional component: The rectangular \minmaxname matrix product.

The algorithm is quite simple once the appropriate sub-routines are defined. Since many of the sub-routines invoked would be standard to a reader familiar with the APASP literature, we first describe the sub-routine which uses rectangular \minmaxname matrix multiplication (\algref{MinMax-update-through-pivot}) over the set $S \times V$, where $S$ will be a set of sampled pivots. Then, we describe the overall algorithm (\algref{2-APASP-slower-main}) and present an illustration of the ideas underlying its case analysis. After this, we define the other (standard) sub-routines and present detailed proof of approximation and analysis of time complexity.

\algref{MinMax-update-through-pivot} performs a straightforward rectangular \minmaxname-product between the submatrix $A$ and $A^T$, where $A$ is a rectangular matrix storing the $S \times V$ distance estimates for an input set $S$ of vertices. Then, it updates the distance estimate table using twice the values of the obtained \minmaxname-product matrix. 

\medskip

\begin{algorithm}[H]\label{alg:MinMax-update-through-pivot}
\setstretch{1.15}
\DontPrintSemicolon
\caption{\textsc{\minmaxname-Update\_through\_pivot}$\paren{V_H, S, d}$}
\textbf{Input:} A set of vertices $V_H$, a set $S \subseteq V_H$ of pivots, a table $d\bracke{\cdot, \cdot}$ storing symmetric distance estimates for all vertex pairs in $V_H \times V_H$.

\textbf{Output:} Performs in-place symmetric update of distance estimate table using a \minmaxname-product through pivots.

\medskip

$A^{|V_H| \times |S|} \gets$ Rectangular Matrix constructed by the columns of table $d$ corresponding to the set $S$.\;

$M^{|V_H| \times |V_H|} \gets A \minmax A^T$ (\corref{rect_minmax}).\;

\ForEach{$\paren{u,v} \in V_H \times V_H$}
{
    $d\bracke{u,v} \gets \mmin{2 M\bracke{u,v}, d\bracke{u,v} }$ \;

    $d\bracke{v,u} \gets \mmin{2 M\bracke{u,v}, d\bracke{v,u} }$ \;
}

\Return $Null$
\end{algorithm}

\medskip

Now, we present the overall algorithm (\algref{2-APASP-slower-main}), which does the following on a high-level: 

\begin{itemize}
    \item Sample 2 levels of pivots $S_1, S_2$ and compute the distances of $V \times V$ in the sparse graph $G_1$, using $n$ estimate-augmented-Dijkstra calls. Estimate-augmented-Dijkstra just means that while computing SSSP from a source $s$, we also add $s X V$ edges with weight equal to the current distance estimate between the relevant pair. Since these are just $O(n)$ edges, they dont affect the sparsity of the underlying graph. 

    \item Then, for $i \in \bracce{1,2}$, the algorithm computes distances $S_{i-1} \times V$ in the graph $G_i$ (again using estimate-augmented-Dijkstra calls), and updates the distance estimates for every pair of vertices $u, v$, by considering the alternate $u-v$ path from $u$ to $p_i\paren{u}$, to $p_i\paren{u}$ to $v$.

    \item Afterwards, for every pair of vertices $\paren{u, v}$, the algorithm updates distance estimate updates for pairs in $S_1 \times V$ by considering the pair $\paren{p_1\paren{u}, v}$, and updating its estimate through the alternate $p_1\paren{u}$ to $u$, to $u$ to $v$ path.

    \item Then, following another round of updation of $S_1 \times V$ estimates through the graph $G_2$, the algorithm concludes with utilising the \minmaxname-product subroutine (\algref{MinMax-update-through-pivot}) to update distance estimates for $S_1 \times V$ vertex pairs.

\end{itemize}

Following is the pseudocode for the algorithm:

\medskip

{ \LinesNumbered
\begin{algorithm}[H]\label{alg:2-APASP-slower-main}
\setstretch{1.15}
\DontPrintSemicolon
\caption{All-Pair 2-Approximate Shortest Paths}
\textbf{Input:} A weighted undirected graph $G = \paren{V, E, w}$.

\textbf{Output:} Distance estimate $d\bracke{u,v}$ such that $\delta\paren{u,v} \leq d\bracke{u,v} \leq 2\delta\paren{u,v}$ for all $\paren{u,v} \in V \times V$.

\medskip

Set $q_1 \gets n^{-b}, q_2 \gets n^{-c}$.\;

$\paren{d, S_1, S_2, E_1, E_2, G_1, G_2} \gets \textsc{Initialise\_Hierarchy}\paren{G, q_1, q_2}$\; \label{algline:2-APASP-slower-main-initialiseHierarchy}

Initialise $E_3 \gets E, G_3 \gets G$.\;

\tcc{All initialisations done, the main part begins.}

\medskip

Invoke the function \textsc{Estimate\_augmented\_MSSP}$\paren{G_1, V, d}$ twice.\; \label{algline:2-APASP-slower-main-estimateAPSP}

\medskip

\ForEach{$i \in \bracce{2,3}$}
{
    \label{algline:2-APASP-slower-main-pivots-iterationloop}

    \textsc{Estimate\_augmented\_MSSP}$\paren{G_i, \ S_{i-1}, d}$\; \label{algline:2-APASP-slower-main-pivotMSSP}

    \textsc{Update\_through\_designated\_pivot}$\paren{V,\  p_{i-1}\paren{\cdot},\ d}$\; \label{algline:2-APASP-slower-main-pivotUpdate}
}

\ForEach{$\paren{u,v} \in V \times V$}
{
    \label{algline:2-APASP-slower-main-distanceUpdate-start}
    $d\bracke{p_1\paren{u},v} \gets \mmin{\delta\paren{p_1\paren{u},u} + d\bracke{u,v}, \ d\bracke{p_1\paren{u},v}}$ \;
    $d\bracke{v,p_1\paren{u}} \gets \mmin{\delta\paren{p_1\paren{u},u} + d\bracke{u,v}, \ d\bracke{p_1\paren{u},v}}$ \;
    \label{algline:2-APASP-slower-main-distanceUpdate-end}
}

\medskip

Invoke the function \textsc{Estimate\_augmented\_MSSP}$\paren{G_2, \ S_{1}, d}$ again.\; \label{algline:2-APASP-slower-main-pivotMSSP-again}

\textsc{\minmaxname-Update\_through\_pivot}$\paren{V, S_1, d}$\; \label{algline:2-APASP-slower-main-minmax-update}

\Return $d$
\end{algorithm}
}

\medskip

Before defining the other (standard) sub-routines used by the algorithm, we first present an illustration to provide the intuition for the algorithm, and an eagle-eye view of case analysis for the proof of approximation, which also informs us which lines of \algref{2-APASP-slower-main} are responsible for achieving the desired approximation ratio and the final running time. 

\medskip

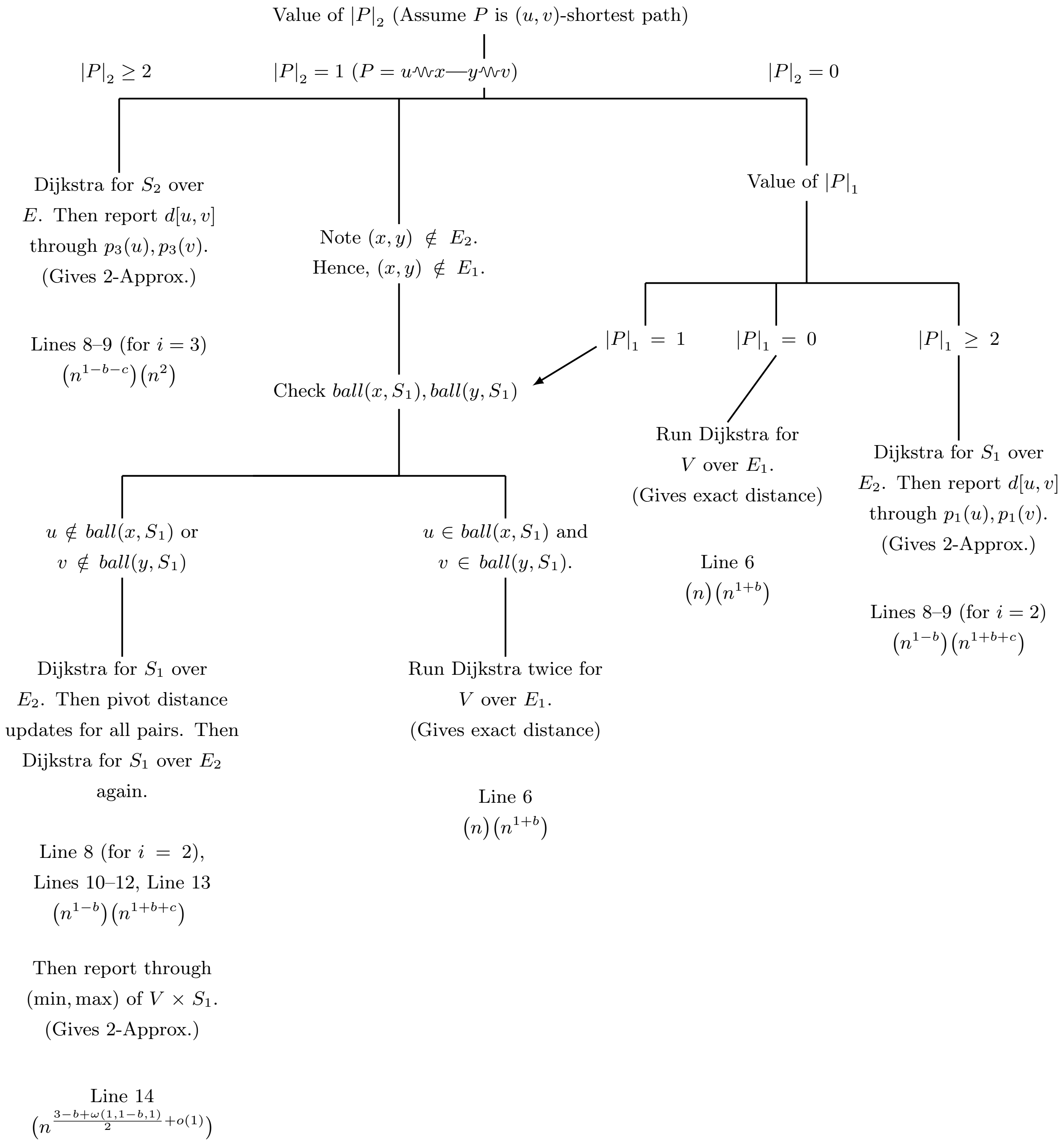
\usetikzlibrary{positioning,calc}

\begin{tikzpicture}[
  >=latex,
  every node/.style={
    font=\small,
    align=center
  },
  branch/.style={
    line width=0.8pt
  },
  caselabel/.style={
    font=\small,
    align=center,
    fill=white,
    inner sep=2pt
  },
  linelabel/.style={
    font=\small,
    align=center,
    fill=white,
    inner xsep=3pt,
    inner ysep=1pt
  },
  condition/.style={
    font=\small,
    align=center,
    text width=3.0cm,
    inner sep=2pt
  },
  action/.style={
    text width=3.2cm,
    align=center,
    inner sep=2pt
  },
  smallaction/.style={
    text width=2.7cm,
    align=center,
    inner sep=2pt
  }
]


\node (title)
{
  Value of $\sizeindex{P}{2}$
  (Assume $P$ is $(u,v)$-shortest path)
};

\coordinate (topLevel)
  at ([yshift=-1.35cm]title.center);

\coordinate (topMany)
  at ([xshift=-6.0cm]topLevel);

\coordinate (topOne)
  at ([xshift=-1.4cm]topLevel);

\coordinate (topZero)
  at ([xshift=5.3cm]topLevel);

\draw[branch]
  (title.south) -- (topLevel);

\draw[branch]
  (topMany) -- (topZero);

\node[caselabel, above=5pt of topMany]
{
  $\sizeindex{P}{2}\ge 2$
};

\node[caselabel, above=5pt of topOne]
{
  $\sizeindex{P}{2}=1$
  $(P=u\squiggly x\straight y\squiggly v)$
};

\node[caselabel, above=5pt of topZero]
{
  $\sizeindex{P}{2}=0$
};


\node[action] (reportP2)
  at ([yshift=-2.20cm]topMany)
{
  Dijkstra for $S_2$ over $E$. Then report $d[u,v]$
  through $p_3(u),p_3(v)$.\\
  (Gives 2-Approx.)
};

\draw[branch]
  (topMany) -- (reportP2.north);

\node (costP2)
  [below=6mm of reportP2]
{
  \linesref{2-APASP-slower-main-pivotMSSP}{2-APASP-slower-main-pivotUpdate} (for $i=3$)\\
  $\bigl(n^{1-b-c}\bigr)\bigl(n^2\bigr)$
};


\node[
  condition,
  text width=4cm
] (notE3)
  at ([yshift=-2.55cm]topOne)
{
  Note $(x,y)\notin E_2$. \\
  Hence, $(x,y) \notin E_1$.
};

\draw[branch]
  (topOne)
  -- (notE3.north);

\node (checkB1xy)
  [below=15mm of notE3]
{
  Check $\ball{x}{1}, \ball{y}{1}$
};

\draw[branch]
  (notE3.south)
  -- (checkB1xy.north);


\coordinate (xyDrop)
  at ([yshift=-1.10cm]checkB1xy.south);

\coordinate (xySplit)
  at ([xshift=-2.40cm]xyDrop);

\coordinate (xyBadPoint)
  at ([xshift=-2.15cm]xySplit);

\coordinate (xyGoodPoint)
  at ([xshift=4.15cm]xySplit);

\draw[branch]
  (checkB1xy.south)
  -- (xyDrop)
  -- (xySplit);

\draw[branch]
  (xyBadPoint)
  -- (xyGoodPoint);


\node[
  condition,
  below=7mm of xyBadPoint
] (xyBadCondition)
{
  $u\notin \ball{x}{1}$ or\\
  $v\notin \ball{y}{1}$
};

\node[
  condition,
  below=7mm of xyGoodPoint
] (xyGoodCondition)
{
  $u\in \ball{x}{1}$ and\\
  $v\in \ball{y}{1}$.\\
};

\draw[branch]
  (xyBadPoint)
  -- (xyBadCondition.north);

\draw[branch]
  (xyGoodPoint)
  -- (xyGoodCondition.north);


\node[action] (reportMinMax)
  [below=13mm of xyBadCondition, text width=4cm]
{
  Dijkstra for $S_1$ over $E_2$.
  Then pivot distance updates for all pairs.
  Then Dijkstra for $S_1$ over $E_2$ again.

  \vspace{5mm}
  \linesref{2-APASP-slower-main-pivotMSSP} (for $i=2$), \linesref{2-APASP-slower-main-distanceUpdate-start}{2-APASP-slower-main-distanceUpdate-end}, \linesref{2-APASP-slower-main-pivotMSSP-again} \\
  $\bigl(n^{1-b}\bigr)\bigl(n^{1+b+c}\bigr)$
  \vspace{4mm}

  Then report through $(\min,\max)$ of $V \times S_1$.\\
  (Gives 2-Approx.)

};

\node (costreportMinMax)
  [below=6mm of reportMinMax]
{
  \linesref{2-APASP-slower-main-minmax-update} \\
  $\bigl(n^{\frac{3-b+\omega(1,1-b,1)}{2} + o\paren{1}}\bigr)$
};

\draw[branch]
  (xyBadCondition.south)
  -- (reportMinMax.north);


\node[smallaction] (sampleXYGood)
  [below=13mm of xyGoodCondition, text width=4cm]
{
  Run Dijkstra twice for $V$ over $E_1$. \\
  (Gives exact distance)
};

\node (costsampleXYGood)
  [below=6mm of sampleXYGood]
{
  \linesref{2-APASP-slower-main-estimateAPSP}\\
  $\bigl(n\bigr)\bigl(n^{1+b}\bigr)$
};

\draw[branch]
  (xyGoodCondition.south)
  -- (sampleXYGood.north);


\node (valueE1)
  [below=11mm of topZero]
{
  Value of $\sizeindex{P}{1}$
};

\draw[branch]
  (topZero.south)
  -- (valueE1.north);


\coordinate (E1split)
  at ([yshift=-1.35cm]valueE1.south);

\coordinate (E1onePoint)
  at ([xshift=-2.65cm]E1split);

\coordinate (E1zeroPoint)
  at ([xshift=-0.5cm]E1split);

\coordinate (E1manyPoint)
  at ([xshift=2.5cm]E1split);

\draw[branch]
  (valueE1.south)
  -- (E1split);

\draw[branch]
  (E1onePoint)
  -- (E1manyPoint);


\node[
  condition,
  text width=2.45cm,
  below=7mm of E1onePoint
] (E1one)
{
  $\sizeindex{P}{1}=1$
};

\node[
  condition,
  text width=2.45cm,
  below=7mm of E1zeroPoint
] (E1zero)
{
  $\sizeindex{P}{1}=0$
};

\node[
  condition,
  text width=2.55cm,
  below=7mm of E1manyPoint
] (E1many)
{
  $\sizeindex{P}{1}\ge 2$
};

\draw[branch]
  (E1onePoint)
  -- (E1one.north);

\draw[branch]
  (E1zeroPoint)
  -- (E1zero.north);

\draw[branch]
  (E1manyPoint)
  -- (E1many.north);

%

\draw[
  branch,
  ->
]
  ([xshift=5mm, yshift=-1mm]E1one.west)
  -- ([xshift=0mm, yshift=1mm]checkB1xy.east);


\node[action] (apspE1Zero)
  [below=11mm of E1zero, xshift=-8mm]
{
  Run Dijkstra for $V$ over $E_1$. \\
  (Gives exact distance)
};

\draw[branch]
  (E1zero.south)
  -- (apspE1Zero.north);

\node (costE1Zero)
  [below=6mm of apspE1Zero]
{
  \linesref{2-APASP-slower-main-estimateAPSP} \\
  $\bigl(n\bigr)\bigl(n^{1+b}\bigr)$
};


\node[action] (reportP1)
  [below=14mm of E1many, text width=4cm]
{
  Dijkstra for $S_1$ over $E_2$. Then report $d[u,v]$
  through $p_1(u),p_1(v)$.\\
  (Gives 2-Approx.)
};

\draw[branch]
  (E1many.south)
  -- (reportP1.north);

\node (costP1)
  [below=6mm of reportP1]
{
  \linesref{2-APASP-slower-main-pivotMSSP}{2-APASP-slower-main-pivotUpdate} (for $i=2$)\\
  $\bigl(n^{1-b}\bigr) \bigl(n^{1+b+c}\bigr)$
};

\end{tikzpicture}

Important values to remember:
$$|S_1| = n^{1-b}, |S_2| = n^{1-b-c},
|E_1| = n^{1+b}, |E_2| = n^{1+b+c}$$

Also, $S_2 \subseteq S_1 \subseteq V$
and
$E_1 \subseteq E_2 \subseteq E$.

\medskip

Now, we complete the algorithm description by formally defining the other sub-routines utilised by \algref{2-APASP-slower-main}.

\medskip 

\begin{algorithm}[H]\label{alg:Sample}
\setstretch{1.15}
\DontPrintSemicolon
\caption{\textsc{Sample}$\paren{S, q}$}
\textbf{Input:} A set of vertices $S$, a parameter $q \in \bracke{0,1}$.

\medskip

$S_q \gets$ Random subset of $S$ obtained by sampling vertices of $S$ independently with probability $q$.\;

\Return $S_q$
\end{algorithm}

\medskip

The following \algref{InitialiseHierarchy} initialises the sampling set hierarchy according to the sample rates. 

\medskip

\begin{algorithm}[H]\label{alg:InitialiseHierarchy}
\setstretch{1.15}
\DontPrintSemicolon
\caption{\textsc{Initialise\_Hierarchy}$\paren{G, q_1, q_2}$}
\textbf{Input:} A weighted undirected graph $G = \paren{V, E, w}$, parameters $q_1, q_2$.

\textbf{Output:} Distance estimates table $d\bracke{\cdot, \cdot}$, Set of pivots $S_1, S_2$, Edge sets $E_{1}, E_{2}$, graphs $G_1, G_2$

\medskip

Initialise $d\bracke{u,v} \gets \infty$ for all $\paren{u,v} \in V \times V$, set $d\bracke{u,v}\gets w(u,v)$  for $\paren{u,v}\in E$.\;

$S_0 \gets V$, $S_1 \gets \textsc{Sample}\paren{S_0, q_1}$, $S_2 \gets \textsc{Sample}\paren{S_1, q_2}$\;

Compute the pivots $p_i\paren{u}$ and distances $\delta\paren{u, S_i}$ for all $u \in V$ and all $i \in \bracce{1,2}$  (\obvref{pivotsdistance}).\; \label{algline:initialise-computePivots}

\ForEach{$u \in V$}
{
    \label{algline:initialise-pivotDistances-start}
    $d\bracke{u,p_1\paren{u}} \gets \delta\paren{u, S_1}$, $d\bracke{p_1\paren{u}, u} \gets \delta\paren{u, S_1}$ \;

    $d\bracke{u,p_2\paren{u}} \gets \delta\paren{u, S_2}$, $d\bracke{p_2\paren{u}, u} \gets \delta\paren{u, S_2}$ \;
    \label{algline:initialise-pivotDistances-end}
} 

Compute the sets $E_1, E_2$ where $E_1 = E_{S_1}$, $E_2 = E_{S_2}$\; \label{algline:initialise-computeEi}

Initialise the graphs $G_1, G_2$ where $G_i = \paren{V, E_i, w}$ for $i \in \bracce{1,2}$.\; \label{algline:initialise-initialiseGi}

\Return $\paren{d, S_1, S_2, E_1, E_2, G_1, G_2}$
\end{algorithm}

\medskip

The \algref{Estimate-augmented-MSSP} does the following: For every source vertex $u \in S$ of the input graph $H$, first add to $H$ a set of edges from $u$ with weights corresponding to the currently stored distance estimates from $u$. Then, update the distances by invoking Dijkstra's algorithm to compute single source shortest path from $u$. Finally, update the distance estimate table. 

\medskip

\begin{algorithm}[H]\label{alg:Estimate-augmented-MSSP}
\setstretch{1.15}
\DontPrintSemicolon
\caption{\textsc{Estimate\_augmented\_MSSP}$\paren{H, S, d}$}
\textbf{Input:} A weighted undirected graph $H = \paren{V_H, E_H, w}$, a set $S \subseteq V_H$ of sources, a table $d\bracke{\cdot, \cdot}$ storing symmetric distance estimates for all vertex pairs in $V_H \times V_H$.

\textbf{Output:} Performs in-place symmetric update of distance estimate table using multiple estimate-augmented Single-Source Shortest Paths algorithm in $H$.

\medskip

\ForEach{$u \in S$}
{
    $E_u \gets$ Star on $V_H$ rooted at $u$, with the edge $\paren{u,v}$ having a weight of $d\bracke{u,v}$ for each $v \in V_H$.\;

    Compute $\delta_{H_u}\paren{u, v}$ for all $v \in V_H$ using Dijkstra's algorithm on the star-augmented graph $H_u = \paren{V_H, E_H \cup E_u}$.\;
}

\ForEach{$\paren{u,v} \in S \times V_H$}
{
    $d\bracke{u,v} \gets \mmin{\delta_{H_u}\paren{u, v}, d\bracke{u,v} }$ \;
    $d\bracke{v,u} \gets \mmin{\delta_{H_u}\paren{u, v}, d\bracke{v,u} }$ \;
}

\Return $Null$
\end{algorithm}

\medskip

We note the following property of \algref{Estimate-augmented-MSSP}:.

\begin{observation}\label{obv:Estimate-augmented-MSSP-property}
    Let $d, d_{end}$ be the distance estimates at the start and the end of \algref{Estimate-augmented-MSSP} respectively. Then, for any pair of vertices $\paren{u, v} \in \paren{S \times V_H} \cup \paren{V_H \times S}$, the following equality holds: 
    $$d_{end}\bracke{u, v} = \min \paren {d\bracke{u,v}, \min_{w \in V_H} d\bracke{u, w} + \delta_{H}\paren{w, v}}$$
\end{observation}

\medskip

Given a graph $H = \paren{V_H, E_H, w}$, a set $S \subseteq V_H$ of pivots, a designated-pivot mapping $p:V_H \rightarrow S$ and the symmetric distance estimates $d\bracke{\cdot, \cdot}$, \algref{Distance-update-through-designated-pivot} in-place updates the distance estimates by considering the vertex pair distances through their designated pivots.

\medskip

\begin{algorithm}[H]\label{alg:Distance-update-through-designated-pivot}
\setstretch{1.15}
\DontPrintSemicolon
\caption{\textsc{Update\_through\_designated\_pivot}$\paren{V_H, p\paren{\cdot}, d}$}
\textbf{Input:} A vertex set $V_H$, a designated-pivot mapping $p:V_H \rightarrow V_H$, a table $d\bracke{\cdot, \cdot}$ storing symmetric distance estimates for all pairs in $V_H \times V_H$.

\textbf{Output:} Symmetrically updates distance estimates in-place using estimates from designated pivots.

\medskip

\ForEach{$\paren{u,v} \in V_H \times V_H$}
{
    $d\bracke{u,v} \gets \mmin{d\bracke{u,p\paren{u}} + d\bracke{p\paren{u}, v}, d\bracke{u,v} }$ \;

    $d\bracke{v,u} \gets \mmin{d\bracke{u,p\paren{u}} + d\bracke{p\paren{u}, v}, d\bracke{v,u} }$ \;
}

\Return $Null$
\end{algorithm}

\subsection{Proof of Approximation}

First, we show that \algref{2-APASP-slower-main} never underestimates the distance between any pair of vertices.

\begin{lemma}\label{lem:lowerBound-proof}
    The output of \algref{2-APASP-slower-main} satisfies $\delta\paren{u,v} \leq d\bracke{u,v}$ for all $\paren{u,v} \in V \times V$.
\end{lemma}
\begin{proof}
    The claim is true initially, since $d\bracke{u, v}$ is initialised to $w\paren{u, v}$ if $\paren{u, v} \in E$, and $\infty$ otherwise. Now, we show that the claim continues to hold after every step of the algorithm modifying the $d$ values.

    In every step, we denote by $\dbef$ the distance estimates just before that step was executed and $\daft$ the distance estimates just after that step. Hence, at every step, we assume $\dbef\bracke{u, v} \geq \delta\paren{u, v}$ and assuming this, we show $\daft\bracke{u, v} \geq \delta\paren{u, v}$, for all $\paren{u, v} \in V \times V$.

    \begin{itemize}
        \item {\bf  \linesref{2-APASP-slower-main-initialiseHierarchy}}: For any $u \in V$ and $i \in \bracce{1,2}$, we know by definition $\delta\paren{u, p_i\paren{u}} = \delta\paren{u, S_i}$. Hence, setting $d\bracke{u,p_i\paren{u}}$ equal to $\delta\paren{u, p_i\paren{u}}$ does not violate the desired claim.
        
        \item {\bf \linesref{2-APASP-slower-main-estimateAPSP}}: Here the \algref{Estimate-augmented-MSSP} is invoked twice. Consider just one invocation for now, and define  $\dbef, \daft$ with respect to that invocation. Suppose there exists a pair $\paren{x,y}$ such that $\daft\bracke{x,y} < \dbef\bracke{x,y}$. Then, \obvref{Estimate-augmented-MSSP-property} implies that there exists a $w \in V$ such that $\daft\bracke{x,y} = \dbef\bracke{x, w} + \delta_{G_1}\paren{w, y}$. Since $\delta\paren{x, w} \leq \dbef\bracke{x, w}$ and $\delta\paren{w,y} \leq \delta_{G_1}\paren{w, y}$, we get that $\delta\paren{x, w} + \delta\paren{w,y} \leq \daft\bracke{x,y}$. By triangle inequality, this implies that $\delta\paren{x,y} \leq \daft\bracke{x,y}$. The same arguments hold for the next invocation as well. Hence, the desired claim holds at the end of \linesref{2-APASP-slower-main-estimateAPSP}.

        \item {\bf \linesref{2-APASP-slower-main-pivots-iterationloop}{2-APASP-slower-main-pivotUpdate}}: Cosnide any $u, v \in V$ and a fixed $i \in \bracce{2,3}$, it can be observed that \linesref{2-APASP-slower-main-pivotMSSP} never estimates the $S_{i-1} \times V$ distances. Then, define $\dbef, \daft$ with respect to execution of \linesref{2-APASP-slower-main-pivotUpdate}. If $\daft\bracke{u, v} < \dbef\bracke{u, v}$, this we can assume without loss of generality that $\daft\bracke{u, v} = \dbef\bracke{u, p_{i-1}\paren{u}} + \dbef\bracke{p_{i-1}\paren{u}, v}$. Thus, $\daft\bracke{u, v} \geq \delta\paren{u, p_{i-1}\paren{u}} + \delta\paren{p_{i-1}\paren{u}, v}$. From triangle inequality, we get that $\daft\bracke{u, v} \geq \delta\paren{u, v}$. So, we obtain that the execution of \linesref{2-APASP-slower-main-pivotMSSP}{2-APASP-slower-main-pivotUpdate} does not underestimate the distances.
        
        \item {\bf \linesref{2-APASP-slower-main-distanceUpdate-start}{2-APASP-slower-main-distanceUpdate-end}}: The desired claim holds trivially by triangle inequality.
        
        \item {\bf \linesref{2-APASP-slower-main-pivotMSSP-again}}: The proof is identical to the one discussed for the \linesref{2-APASP-slower-main-pivots-iterationloop}{2-APASP-slower-main-pivotUpdate}.

        \item {\bf \linesref{2-APASP-slower-main-minmax-update}}: Again, assume $\dbef$ is the distance estimate table just before the invocation of \algref{MinMax-update-through-pivot}, such that $\delta\paren{u,v} \leq \dbef\bracke{u, v}$ for all $u, v \in V$; and $\daft$ is the table just after that invocation. If $\daft\bracke{u, v} < \dbef\bracke{u, v}$, then there exists a $w \in V$ such that $\daft\bracke{u,v} = 2 \max \paren{\dbef\bracke{u,w}, \dbef\bracke{w,v}}$. This implies that $\daft\bracke{u,v} \geq \dbef\bracke{u,w} + \dbef\bracke{w,v}$. Since $\dbef$ does not underestimate $\delta\paren{\cdot}$, we get that $\daft\bracke{u,v} \geq \delta\paren{u,w} + \delta\paren{w,v}$. Therefore, $\daft\bracke{u,v} \geq \delta\paren{u,v}$ due to triangle inequality.
        
    \end{itemize}

    Hence, we do not underestimate the distances throughout \algref{2-APASP-slower-main}.
\end{proof}


Since we have established that \algref{2-APASP-slower-main} does not underestimate the distances, we now proceed to show that the distance estimates $d$ returned by \algref{2-APASP-slower-main} satisfy $d\bracke{u, v} \leq 2\delta\paren{u,v}$ for all $\paren{u, v} \in V \times V$. This will prove that the distance estimates returned by \algref{2-APASP-slower-main} are 2-Approximate.  

Note that whenever we attempt to update the distance estimate $d\bracke{u, v}$ for a pair of vertices $u, v$, we take its minimum with the existing estimate for the pair $u,v$. So, as the algorithm progresses, the value of an estimate can only decrease. Hence, it suffices to show that for every pair of vertices, there is some point in \algref{2-APASP-slower-main} by which the distance estimate is a 2-Approximation. 

The following observation is immediate:
\begin{observation}\label{obv:initialise-hierarchy}
    Let $d\bracke{\cdot, \cdot}$ be the distance estimates just after hierarchy is initialised. Then, for $u \in V$ and $i \in \bracce{1,2}$, we get that $d\bracke{u, p_i\paren{u}} = \delta\paren{u, p_i\paren{u}}$. Also, $d\bracke{u, v} \leq w\paren{u, v}$ for all $\paren{u, v} \in E$. 
\end{observation}

Now, we formally prove the approximation guarantee of \algref{2-APASP-slower-main}.

\begin{lemma}\label{lem:slower-mainAlgo-proof}
    The output of \algref{2-APASP-slower-main} satisfies $d\bracke{u,v} \leq 2\delta\paren{u,v}$ for all $\paren{u,v} \in V \times V$.
\end{lemma}
\begin{proof}
    Let $P = u \squiggly v$.
    After sampling the sets $S_1, S_2$, we first consider the following cases for value of $\sizeindex{P}{2}$:

    \begin{enumerate}
        \item \label{case:slower-P3-2} $\sizeindex{P}{2} \geq 2$: Suppose $\paren{x, y} = \leftabovearrow{P}{2}$ and $\paren{x', y'} = \rightabovearrow{P}{2}$. Since $\sizeindex{P}{2} \geq 2$, the edges $\paren{x, y}, \paren{x', y'}$ are distinct. Consider $P = u \squiggly x-y \squiggly y' - x' \squiggly v$.

        Note that the sub-paths $P\bracke{u, x}, P\bracke{v, x'}$ are a part of $G_2$, i.e.,  $E\paren{P\bracke{u, x}}, E\paren{P\bracke{v, x'}} \subseteq E_2$. Without loss of generality, assume that $\delta\paren{u, y} \leq \delta\paren{v, y'}$. Hence, $\delta\paren{u, y} \leq \frac{\delta\paren{u, v}}{2}$. Since $\paren{x, y} \notin E_2$, this implies that $\delta\paren{u, p_2\paren{u}} \leq \delta\paren{u, y}$. So, $\delta\paren{u, p_2\paren{u}} \leq \frac{\delta\paren{u, v}}{2}$.

        Furthermore, it can be seen that for $i = 3$, the \linesref{2-APASP-slower-main-pivotMSSP} corresponds to computing all the $S_2 \times V$ distances in the graph $G$. Hence, after execution of this line, the following equality holds: $d\bracke{p_2\paren{u}, v} = \delta\paren{p_2\paren{u}, v}$.

        Then, execution of \linesref{2-APASP-slower-main-pivotUpdate} (for $i = 3$) results in the following update to $d\bracke{u,v}$:
        
        \begin{align*}
            d\bracke{u, v} & \leq d\bracke{u, p_2\paren{u}} + d\bracke{p_2\paren{u}, v}\\
            & \leq \delta\paren{u, p_2\paren{u}} + \delta\paren{p_2\paren{u}, v} \\
            & \leq 2 \cdot \delta\paren{u, p_2\paren{u}} + \delta\paren{u, v} \quad \paren{\text{Triangle inequality}} \\
            & \leq \delta\paren{u, v} + \delta\paren{u, v} \quad \paren{ \text{As } \delta\paren{u, p_2\paren{u}} \leq \frac{\delta\paren{u, v}}{2}}
        \end{align*}
        
        Hence, $d\bracke{u, v} \leq 2 \cdot \delta\paren{u, v}$.

        \item \label{case:slower-P3-1} $\sizeindex{P}{2} = 1$: Consider the excluded edge to be $\paren{x,y}$, such that $P = u \squiggly x-y \squiggly v$. Divide into the following two cases:

        \begin{enumerate}
            \item $u \in \ball{x}{1}$ and $v \in \ball{y}{1}$: This case implies that the subpaths $P\bracke{u, x}, P\bracke{v, y}$ lie in $G_1$. So, after one invocation of \textsc{Estimate\_augmented\_MSSP}$\paren{G_1, V, d}$ in \linesref{2-APASP-slower-main-estimateAPSP}, we get that $d\bracke{x, y} = d\bracke{y, x} = \delta\paren{x, y}$. 
            
            Since $P\bracke{u, x}$ lies in $G_1$, another invocation of \textsc{Estimate\_augmented\_MSSP}$\paren{G_1, V, d}$ leads to the following guarantee: $d\bracke{v, u} = d\bracke{u, v} = \delta\paren{u, v}$.
            
            \item $u \notin \ball{x}{1}$ or $v \notin \ball{y}{1}$: Without loss of generality, assume that $u \notin \ball{x}{1}$. This implies that $\delta\paren{x, p_1(x)} \leq \delta\paren{u, x}$. Also, $\paren{x, y} \notin E_2$ implies $\paren{x, y} \notin E_1$. Hence, $\delta\paren{x, p_1(x)} \leq w\paren{x, y}$.
            
            So, the following bound will be useful for us:

            \begin{equation}\label{eqn:slower-analysis-helper-3}
                \delta\paren{x, p_1(x)} \leq \mmin{\delta\paren{u, x}, w\paren{x, y}}
            \end{equation}

            Since $P\bracke{u, x}, P\bracke{v, y}$ both lie in $E_2$, the following inequality holds after execution of \textsc{Estimate\_augmented\_MSSP}$\paren{G_2, \ S_{1}, d}$\; (corresponds to \linesref{2-APASP-slower-main-pivotMSSP} for $i=2$):

            \begin{align} 
                d\bracke{p_1\paren{x}, u} &\leq d\bracke{p_1\paren{x}, x} + \delta\paren{x, u} \notag \\
                &\leq \delta\paren{p_1\paren{x}, x} + \delta\paren{x, u} \label{eqn:slower-analysis-helper-1}
            \end{align}
            
            After the estimate updates in \linesref{2-APASP-slower-main-distanceUpdate-start}{2-APASP-slower-main-distanceUpdate-end}, it can be observed that:
            
            \begin{align*}
                d\bracke{p_1\paren{x}, y} &\leq \delta\paren{p_1\paren{x}, x} + d\bracke{x, y} \\
                &\leq \delta\paren{p_1\paren{x}, x} + w\paren{x, y}
            \end{align*}
            
            Since the path $P\bracke{y, v}$ lies in $G_2$, invocation of \textsc{Estimate\_augmented\_MSSP}$\paren{G_2, \ S_1, d}$ in \linesref{2-APASP-slower-main-pivotMSSP-again} results in the following bounds:

            \begin{align}
                d\bracke{p_1\paren{x}, v} &\leq d\bracke{p_1\paren{x}, y} + \delta\paren{y, v} \notag \\ 
                &\leq \delta\paren{p_1\paren{x}, x} + w\paren{x, y} + \delta\paren{y, v} \notag \\
                &\leq \delta\paren{p_1\paren{x}, x} + \delta\paren{x, v} \label{eqn:slower-analysis-helper-2}
            \end{align}

            Utilising \eqnref{slower-analysis-helper-1} and \eqnref{slower-analysis-helper-2}, we now show how the \minmaxname-product update (\linesref{2-APASP-slower-main-minmax-update}) gives us a 2-approximation. After \linesref{2-APASP-slower-main-minmax-update}, one can observe that:

            \begin{align*}
                d\bracke{u, v} &\leq 2 \cdot \mmax{d\bracke{u, p_1\paren{x}}, d\bracke{v, p_1\paren{x}}} \\
                &\leq 2 \cdot \mmax{\delta\paren{p_1\paren{x}, x} + \delta\paren{x, u}, \delta\paren{p_1\paren{x}, x} + \delta\paren{x, v}} \ \paren{\text{Using } \eqnref{slower-analysis-helper-1}, \eqnref{slower-analysis-helper-2}}  \\
                &\leq 2 \cdot \mmax{w\paren{x, y} + \delta\paren{x, u}, \delta\paren{x, u} + \delta\paren{x, v}} \ \paren{\text{Using } \eqnref{slower-analysis-helper-3}} \\
                &\leq 2 \cdot \mmax{\delta\paren{y, u}, \delta\paren{u, v}} \\
                &\leq 2 \cdot {\delta\paren{u, v}}
            \end{align*}
            
            Hence, the desired bound holds.

        \end{enumerate}
        
        \item $\sizeindex{P}{2} = 0$: In this case, the path $P$ lies in $G_2 = \paren{V, E_2}$ as well.
        
        Consider the following cases for the value of $\sizeindex{P}{1}$:

            \begin{enumerate}
                \item $\sizeindex{P}{1} \geq 2$: Its analysis and approximation guarantee are similar to the case~\ref{case:slower-P3-2}, i.e., the case when $\sizeindex{P}{2} \geq 2$. The only difference is here we consider the pivot $p_1(u)$ for analysis and the bounds follow from the $i=2$ iteration of the
                \linesref{2-APASP-slower-main-pivotMSSP}{2-APASP-slower-main-pivotUpdate}. 
                
                \item $\sizeindex{P}{1} = 0$: 
                Here, the complete path $P$ lies in the graph $G_1$. So, even after one invocation of \textsc{Estimate\_augmented\_MSSP}$\paren{G_1, V, d}$ in \linesref{2-APASP-slower-main-estimateAPSP}, we obtain that $d\bracke{u, v} = \delta\paren{u, v}$.

                \item $\sizeindex{P}{1} = 1$: Its analysis and approximation guarantee follow near-identically from the analysis of case~\ref{case:slower-P3-1}, i.e., the case when $\sizeindex{P}{2} = 1$. 
            \end{enumerate}

    \end{enumerate}
\end{proof}

\subsection{Running Time}

Throughout this subsection, we refer to the number of vertices and edges in $G$ as $n, m$ respectively. Note that in \algref{2-APASP-slower-main}, we sometimes deal with subgraphs of the form $H = (V, E_H)$. Hence, we denote the runtime of some subroutines in terms of size of $H$ instead. For ease of notation, we assume $|E_H| = m_H$. 

\begin{observation}\label{obv:designatedPivot-update-runtime}
    \algref{Distance-update-through-designated-pivot} runs in $O\paren{n^2}$ time.
\end{observation}

\begin{lemma}\label{lem:estimateMSSP-runtime}
    \algref{Estimate-augmented-MSSP} runs in time $\widetilde{O}\paren{|S| \ \paren{m_H + n}}$.
\end{lemma}
\begin{proof}
    The algorithm initially performs $|S|$ runs of Dijkstra's algorithm on a graph with $O\paren{m_H + n}$ edges. Since each run of Dijkstra takes $\widetilde{O}\paren{m_H + n}$ time, this makes the total time complexity over $|S|$ runs of Dijkstra as $\widetilde{O}\paren{|S| m_H + |S|n}$. A further $O(|S|n)$ time is spent on updating the distance estimate table. Hence, the overall time complexity of \algref{Estimate-augmented-MSSP} is $\widetilde{O}\paren{|S| \ \paren{m_H + n}}$.
\end{proof}

\begin{lemma}\label{lem:minmaxUpdate-runtime}
    If $|S| = n^{1-\gamma}$ for some $\gamma \in \bracke{0,1}$, \algref{MinMax-update-through-pivot} runs in total time $\widetilde{O}\paren{n ^ {\frac{3-\gamma+\omega\paren{1,1-\gamma,1}}{2} + o\paren{1}} + n^2}$.
\end{lemma}
\begin{proof}
    Note that the matrix $A$ is of dimension $n \times n^{1-\gamma}$. Hence, by \corref{rect_minmax}, the rectangular \minmaxname-product $A \minmax A^T$ can be computed in $\widetilde{O}\paren{n ^ {\frac{3-\gamma+\omega\paren{1,1-\gamma,1}}{2} + o\paren{1} }}$ time. Updating the distance estimate table takes further $O(n^2)$ time. This results in overall time complexity of $\widetilde{O}\paren{n ^ {\frac{3-\gamma+\omega\paren{1,1-\gamma,1}}{2}+ o\paren{1}} + n^2}$.
\end{proof}

\begin{lemma}\label{lem:mainAlgo-runtime}
    Assuming $|E| = O\paren{n^2}$, $q_1 = n^{-b}$ and $q_2 = n^{-c}$ for some $b,c, b+c \in \bracke{0,1}$, \algref{2-APASP-slower-main} runs in the following total time:

    $$
    \widetilde{O}\paren{n^{3-b-c} + n ^ {\frac{3-b+\omega\paren{1,1-b,1}}{2}+ o\paren{1}} + n^{2+b} + n^{2+c}}
    $$

Using the earlier  FMM bounds
\cite{DupEisKozMehRuiSeeZhoAlmVasBal2026,VasXuXuZho2024,AlmDuaVasXuXuZho2024},
the runtime is
$\widetilde O\paren{n^{2.36999465}}$.
Following new FMM bounds ~\cite{OpenAI2026, Uznanski2026, AmaBerZui2026} 
gives
$\widetilde O\paren{n^{2.34549151}}$ runtime.
\end{lemma}
\begin{proof}
    To bound the total running time, we analyse the time complexity of \algref{2-APASP-slower-main} piece by piece. Since $q_1 = n^{-b}$ and $q_2 = n^{-c}$, it will be helpful to keep in mind that $|S_1| = n^{1-b}, |S_2| = n^{1-b-c}, |E_1| = n^{1+b}, |E_2| = n^{1+b+c}$ :
    
    \begin{itemize}
        \item {\bf \linesref{2-APASP-slower-main-initialiseHierarchy}}: From \obvref{pivotsdistance}, this takes $\widetilde{O}\paren{|E|+n}$ total time, which can be written as $\widetilde{O}\paren{n^2}$.
        
        \item {\bf \linesref{2-APASP-slower-main-estimateAPSP}}: Here,  we invoke the \textsc{Estimate\_augmented\_MSSP} procedure twice, with set of sources as the whole vertex set, i.e., $S = V$. From \lemref{estimateMSSP-runtime}, this takes time $\widetilde{O}\paren{|E_1| n + n^2}$, which simplifies to $\widetilde{O}\paren{n^{2+b}}$.  

        \item {\bf \linesref{2-APASP-slower-main-pivotMSSP}{2-APASP-slower-main-pivotUpdate} (for $i = 3$)}: \lemref{estimateMSSP-runtime} and
        \obvref{designatedPivot-update-runtime} together imply that this iteration runs in overall $\widetilde{O}\paren{|S_2||E| + n^2}$ time. Substituting the bounds for $S_2, E$, we get the time complexity as $\widetilde{O}\paren{n^{3-b-c} + n^2}$. Since $b+c \leq 1$, this runtime can be simply written as $\widetilde{O}\paren{n^{3-b-c}}$.

        \item {\bf \linesref{2-APASP-slower-main-pivotMSSP}-\linesref{2-APASP-slower-main-pivotUpdate} (for $i = 2$)}: From \lemref{estimateMSSP-runtime} and
        \obvref{designatedPivot-update-runtime} we get that this iteration runs in total $\widetilde{O}\paren{|S_1||E_2| + n^2}$ time. Substituting the bounds for $S_1, E_2$, the time complexity simplifies to $\widetilde{O}\paren{n^{2+c}}$. 

        \item {\bf \linesref{2-APASP-slower-main-distanceUpdate-start}{2-APASP-slower-main-distanceUpdate-end}}: Since each line of the {\em for} loop takes only $O\paren{1}$ time, hence the loop overall takes $O\paren{n^2}$ time.

        \item {\bf \linesref{2-APASP-slower-main-pivotMSSP-again}}: From \lemref{estimateMSSP-runtime}, this line runs in $\widetilde{O}\paren{|S_1| \paren{|E_2| + n}}$ time. As discussed previously, the runtime can be simplified to $\widetilde{O}\paren{n^{2+c}}$.

        \item {\bf \linesref{2-APASP-slower-main-minmax-update}}: Since $|S_1| = n^{1-b}$, we get from \lemref{minmaxUpdate-runtime} that this line runs in time $\widetilde{O}\paren{n ^ {\frac{3-b+\omega\paren{1,1-b,1}}{2} + o\paren{1}} + n^2}$.
    \end{itemize}

    Hence, the total runtime incurred over the whole algorithm is $\widetilde{O}\paren{n^{3-b-c} + n ^ {\frac{3-b+\omega\paren{1,1-b,1}}{2} + o\paren{1}} + n^{2+b} + n^{2+c}}$.

For $b\in\bracke{\frac{1}{3},\frac{1}{2}}$,
set $c=1-2b$, so that
$\mmax{3-b-c,2+b,2+c}=2+b$.
The rectangular bounds
\cite{Uznanski2026,AmaBerZui2026} give:
$$
\omega\paren{1,1-b,1}
\leq 2-b+\frac{1}{4\paren{1-b}}
$$
Balancing the two remaining exponents gives:
$$
2+b
=
\frac{5-2b+\frac{1}{4\paren{1-b}}}{2}
\quad\Longleftrightarrow\quad
16b^2-20b+5=0
$$
Thus, choosing
$b=\frac{5-\sqrt{5}}{8}$ and
$c=\frac{\sqrt{5}-1}{4}$
gives
$\widetilde O\paren{
n^{\frac{21-\sqrt{5}}{8}+o\paren{1}}
}$
runtime, whose exponent is
$2.3454915028\ldots+o\paren{1}$.

Using the earlier rectangular FMM bounds
\cite{VasXuXuZho2024,AlmDuaVasXuXuZho2024}
instead, choosing
$b=0.36999465$ and $c=0.26001070$
gives
$\widetilde O\paren{n^{2.36999465}}$
runtime.
\end{proof}


\section{$2$-APASP faster than $n^{2.25}$}\label{2-faster}

Note that \algref{2-APASP-slower-main} did not require the edge weights to be integral and its runtime was not dependent on the weight range. In this section, we introduce a faster algorithm for 2-APASP. For the scope of this section, we will assume that the edges of the input graph $G$ are positive integers\footnote{We can also allow zero weight edges, but they can be contracted in the preprocessing stage.} and lie in the range $\bracke{W_{\max}} \cup \bracce{\infty}$.  

\medskip

\begin{algorithm}[H]\label{alg:ApproxMSSP-pivot}
\setstretch{1.15}
\DontPrintSemicolon
\caption{\textsc{Approximate\_MSSP}$\paren{H, S, d, \varepsilon}$}
\textbf{Input:} A weighted undirected graph $H = \paren{V_H, E_H, w}$, a set $S \subseteq V_H$ of sources, a table $d\bracke{\cdot, \cdot}$ storing symmetric distance estimates for all vertex pairs in $V_H \times V_H$, a parameter $\varepsilon > 0$.

\textbf{Output:} Computes $\paren{1+\varepsilon}$-approximate $S \times V_H$ distances in $H$, and then updates the distance estimate table symmetrically in-place.

\medskip

Compute $\bar{\delta} \paren{s,v}$ for each $\paren{s, v} \in S \times V_H$ where $\delta_H(s,v) \leq \bar{\delta} \paren{s,v} \leq \paren{1+\varepsilon}\delta_H(s,v)$ (\thmref{ApproxMSSP}). \;

\ForEach{$\paren{u,v} \in S \times V_H$}
{
    $d\bracke{u,v} \gets \mmin{\floor{\bar{\delta}\paren{u, v}}, d\bracke{u,v} }$ \;
    $d\bracke{v,u} \gets \mmin{\floor{\bar{\delta}\paren{u, v}}, d\bracke{v,u} }$ \;
}

\Return $Null$
\end{algorithm}

\medskip

First, since the edge weights are positive integers, taking the floor value of an approximate distance does not underestimate the true distance. So, the following observation is immediate:

\begin{observation}\label{obv:approx-mssp-lowerBound}
    \textsc{Approximate\_MSSP}$\paren{H, S, d, \varepsilon}$ does not underestimate the true distances of $S \times V$ in the graph $H$.
\end{observation}

\medskip

The following \algref{PivotDistance-bucketing} partitions the vertex set into buckets based on their distance to the pivots. The buckets correspond to the set of intervals $\bracce{[0, width), [width, 2 \cdot width), [2 \cdot width, 3 \cdot width), \dots}$.
That is, $\B\bracke{i}$ denotes the bucket corresponding to interval $[i \cdot width, \paren{i+1} \cdot width)$.
Since these buckets correspond to a partition of vertices in $V_H$, there are at most $n$ buckets at the end of the algorithm.

\medskip

\begin{algorithm}[H]\label{alg:PivotDistance-bucketing}
\setstretch{1.15}
\DontPrintSemicolon
\caption{\textsc{Bucket\_by\_Pivot\_Distance}$\paren{V_H, p\paren{\cdot}, d, width}$}
\textbf{Input:} A vertex set $V_H$, a designated-pivot mapping $p:V_H \rightarrow V_H$, a table $d\bracke{\cdot, \cdot}$ storing symmetric distance estimates for all pairs in $V_H \times V_H$, interval width for each bucket.

\textbf{Output:} Computes the vertex partition based on pivot distance buckets according to the interval width.

\medskip

Initialise the family of buckets $BucketFam \gets \varnothing$ as a dictionary.\;

\ForEach{$v \in V_H$}{
    Find the integer $i$ such that $d\bracke{v, p\paren{v}} \in [i \cdot width, \paren{i+1} \cdot width)$.\;
    
    \If{Bucket $\B\bracke{i}$ does not exist in $BucketFam$}{
        Initialise $\B\bracke{i} \gets \varnothing$.
    }
    
    Add $v$ to $\B\bracke{i}$.\;
}

\Return $BucketFam$\; 
\end{algorithm}

\medskip

We now describe the offset-shifted approximate \minplusname-matrix product, one of the main new algorithmic tools, which when combined with the bucketing idea, enables the faster 2-Approximate APSP.  

\medskip 

\begin{algorithm}[H]\label{alg:offset-shifted-ApproxminPlus}
\setstretch{1.15}
\DontPrintSemicolon
\caption{\textsc{Offset\_Shifted\_AMPMM}$\paren{M_1, M_2, \mathrm{offset}_1, \mathrm{offset}_2, \varepsilon}$}
\textbf{Input:} A $l_1 \times l_2$-sized matrix $M_1$, a $l_2 \times l_3$ sized matrix $M_2$, a value $\mathrm{offset}_1$ - with guarantee that each entry of $M_1$ is at least $\mathrm{offset}_1$, a value $\mathrm{offset}_2$ - with guarantee that each entry of $M_2$ is at least $\mathrm{offset}_2$, a parameter $\varepsilon > 0$.

\textbf{Output:} Computes approximation to $M_1 \minplus M_2$ by subtracting offsets, then computing $\paren{1+\varepsilon}$-approx. \minplusname-product and adding the offsets back.

\medskip

Compute $\bar{M}_i$ from $M_i$ by subtracting $\mathrm{offset}_i$ from each non-infinite entry of $M_i$, for $i \in \{1, 2\}$. \;

$\bar{M} \gets \paren{1+\varepsilon}$-Approximation of $\bar{M}_1 \minplus \bar{M}_{2}$. (\thmref{ApproxMinPlus}).\;  

Compute $M$ from $\bar{M}$ by adding $\mathrm{offset}_1 + \mathrm{offset}_2$ to each non-infinite entry of $\bar{M}$. \; 

\Return $M$; 
\end{algorithm}

\medskip

Since we assume that the input to \textsc{Offset\_Shifted\_AMPMM}$\paren{M_1, M_2, \mathrm{offset}_1, \mathrm{offset}_2, \varepsilon}$ is guaranteed to be such that an offset is never larger than the entries of its corresponding matrix, one can safely assume that \algref{offset-shifted-ApproxminPlus} does not underestimate any entry of the matrix product $M_1 \minplus M_2$.

\begin{observation}\label{obv:shifted-ApproxMinPlus-lowerBound}
    Assuming the input is correct, \algref{offset-shifted-ApproxminPlus} does not underestimate any entry of $M_1 \minplus M_2$. Concretely, if $M$ is the output of \textsc{Offset\_Shifted\_AMPMM}$\paren{M_1, M_2, \mathrm{offset}_1, \mathrm{offset}_2, \varepsilon}$, then $M\bracke{i, j} \geq \paren{M_1 \minplus M_2}\bracke{i, j}$ for any entry $i, j$. 
\end{observation}

We remark that in \algref{offset-shifted-ApproxminPlus}, we are only reducing the matrix entry values before multiplying the matrices, and later adding these values back. So this algorithm also follows the runtime bounds implied by dividing the input matrices into some block matrices and taking the block matrix products. This will be useful in our runtime analysis.

\begin{observation}\label{obv:shifted-ApproxMinPlus-blocks}
    Assuming offsets are fixed and compatible with input matrices $A$ and $B$, let $A \boxast B$ denote the offset-shifted $\paren{1+\varepsilon}$-approximate-\minplusname-product of \algref{offset-shifted-ApproxminPlus}. Then, the following block matrix identities hold (assuming the block matrices have compatible dimensions):
    \begin{itemize}
        \item $\begin{pmatrix}A_1 & A_2\end{pmatrix} \boxast \begin{pmatrix} B_1 \\ B_2 \end{pmatrix} = \begin{pmatrix}
        \mmin{A_1 \boxast A_2, B_1 \boxast B_2}
        \end{pmatrix}$

        \item $\begin{pmatrix}A_1 \\ A_2\end{pmatrix} \boxast \begin{pmatrix} B_1 & B_2 \end{pmatrix} = \begin{pmatrix}
        A_1 \boxast B_1 & A_1 \boxast B_2 \\
        A_2 \boxast B_1 & A_2 \boxast B_2
        \end{pmatrix}$
    \end{itemize}

    where $\mmin{\cdot}$ is an entrywise minimum.
\end{observation}

Now, we present the main \algref{2-APASP-faster-main}, which computes the All-Pair 2-Approximate Shortest Paths.

For ease of notation, hereon we denote $d[R, C]$ to be the submatrix of $d$ indexed by the row set $R$ and column set $C$. 
Also, we use the terms $\B_{i, j}$ and $\B_j\bracke{i}$ interchangeably in the \algref{2-APASP-faster-main}.

\medskip

{ \LinesNumbered
\begin{algorithm}[H]\label{alg:2-APASP-faster-main}
\setstretch{1.15}
\DontPrintSemicolon
\caption{\textsc{Weighted-2-APASP}$\paren{G}$}
\textbf{Input:} A weighted undirected graph
$G=\paren{V,E}$.

\textbf{Output:} A symmetric table $d\bracke{\cdot,\cdot}$
satisfying
$\delta\paren{u,v}\leq d\bracke{u,v}
\leq 2\delta\paren{u,v}$
for all $\paren{u,v}\in V\times V$.

\medskip

Set $q_1 \gets n^{-a}, q_2 \gets 0$.\;

$\paren{d, S_1, S_2, E_1, E_2, G_1, G_2} \gets \textsc{Initialise\_Hierarchy}\paren{G, q_1, q_2}$\; \label{algline:2-APASP-faster-main-initialiseHierarchy}

Define $p\paren{v} \gets p_1\paren{v}$ for all $v \in V$\;

\tcc{All initialisations done. No use of $S_2, E_2$ in this algorithm.}

\ForEach{$\paren{u,v} \in V \times V$}
{
    \label{algline:2-APASP-faster-main-distanceUpdate-start}
    $d\bracke{p\paren{u},v} \gets \mmin{\delta\paren{p\paren{u},u} + d\bracke{u,v}, \ d\bracke{p\paren{u},v}}$ \;
    $d\bracke{v,p\paren{u}} \gets \mmin{\delta\paren{p\paren{u},u} + d\bracke{u,v}, \ d\bracke{p\paren{u},v}}$ \;
    \label{algline:2-APASP-faster-main-distanceUpdate-end}
}

Invoke the function \textsc{Estimate\_augmented\_MSSP}$\paren{G_1, V, d}$ twice.\; \label{algline:2-APASP-faster-main-estimateAPSP}

Set $\varepsilon\gets 0.05$ and invoke \textsc{Approximate\_MSSP}$\paren{G, S_1, d, \varepsilon}$\; \label{algline:2-APASP-faster-main-approxMSSP}

$\bar{d} \gets$ $\paren{1+\varepsilon}$-Approximation of $d\bracke{V, S_1} \minplus d\bracke{S_1, V}$ (\thmref{ApproxMinPlus}). Symmetrically update $d$ using $\bar{d}$.\; \label{algline:2-APASP-faster-main-pivot-approxMinPlus}

\ForEach{$j \gets 0$ to $\ceil{\log_2 \paren{n W_{\max}}}$}{
    \label{algline:2-APASP-faster-main-bucketSizeLoop-start}

    Set $intervalWidth \gets 2^j$.\;
    
    $\B_{j} \gets$ \textsc{Bucket\_by\_Pivot\_Distance}$\paren{V, p\paren{\cdot}, d, intervalWidth}$\;\label{algline:2-APASP-faster-main-pivotDistance-bucketing}

    \ForEach{$\B_{i, j} \in \B_j$ such that $\B_j\bracke{i} = \B_{i, j}$} {
        \label{algline:2-APASP-faster-main-bucketIter-start}
        Initialise $\P_{i, j} \gets \bracce{p\paren{x} \ | \ x \in {\cal B}_{i, j} }$\;\label{algline:2-APASP-faster-main-Pij-initialise}


        Define  $\Y_{i, j} \gets \B_{i,j} \cup \B_j\bracke{i+1}$, if $\B_j\bracke{i+1}$ does not exist consider it empty.\;\label{algline:2-APASP-faster-main-Yij-initialise}

        Define two matrices $A_{i,j}^{|\P_{i, j}| \times |V|}, B_{i,j}^{|V| \times |\Y_{i, j}|}$ with each entry initialised to $\infty$.\;

        \ForEach{$\paren{x, y} \in {\cal B}_{i, j} \times V$ such that $\paren{x,y} \in E$ and $w\paren{x,y} \geq d\bracke{x, p\paren{x}}$}{
            \label{algline:2-APASP-faster-main-Aij-initialise}

            $A_{i,j}\bracke{p\paren{x}, y} \gets \mmin{A_{i,j}\bracke{p\paren{x}, y}, d\bracke{p\paren{x}, x} + w\paren{x,y} }$\;

        }

        \ForEach{$\paren{x', y} \in \Y_{i, j} \times V$ such that $d\bracke{x', y} \geq d\bracke{x', p\paren{x'}}$}{
            \label{algline:2-APASP-faster-main-Bij-initialise}

            $B_{i, j}\bracke{y, x'} \gets {d\bracke{x', y}} $\;   

        }

        Set $shift_A \gets 2 \cdot i \cdot 2^j, \ shift_B \gets i \cdot 2^j$\;\label{algline:2-APASP-faster-main-shift-values}

        $M_{i, j} \gets$ \textsc{Offset\_Shifted\_AMPMM}$\paren{A_{i,j}, B_{i, j}, shift_A, shift_B, \varepsilon}$\;\label{algline:2-APASP-faster-main-shifted-ApproxMinPlus}

        \ForEach{$\paren{x, x'} \in \B_{i, j} \times \Y_{i, j}$}{\label{algline:2-APASP-faster-main-updateEstimateAfterShift-start}
            $d\bracke{x, x'} \gets \mmin{d\bracke{x, x'}, d\bracke{x, p\paren{x}} + M_{i, j}\bracke{p\paren{x}, x'}}$\;
            $d\bracke{x', x} \gets \mmin{d\bracke{x, x'}, d\bracke{x', x} }$\; \label{algline:2-APASP-faster-main-updateEstimateAfterShift-end}
        }
        
    }
}

Invoke the function \textsc{Estimate\_augmented\_MSSP}$\paren{G_1, V, d}$ twice (again).\;\label{algline:2-APASP-faster-main-estimateAPSP-again}

\Return $d$
\end{algorithm}
}

\subsection{Proof of Approximation}

We first show that \algref{2-APASP-faster-main} never underestimates the distance between any pair of vertices.

\begin{lemma}\label{lem:faster-alg-lowerBound-proof}
    The output of \algref{2-APASP-faster-main} satisfies $\delta\paren{u,v} \leq d\bracke{u,v}$ for all $\paren{u,v} \in V \times V$.
\end{lemma}
\begin{proof}
    Its easy to see that the claim holds just after the hierarchy is initialised (\linesref{2-APASP-faster-main-initialiseHierarchy}). Now, we show that the claim continues to hold after every step of the algorithm which modifies the $d$ values. 
    In every step, we denote by $\dbef$ the distance estimates just before that step was executed and $\daft$ the distance estimates just after that step. Hence, at every step, we assume $\dbef\bracke{u, v} \geq \delta\paren{u, v}$ and assuming this, we show $\daft\bracke{u, v} \geq \delta\paren{u, v}$, for all $\paren{u, v} \in V \times V$.
    Also, we can safely assume moving forward that $\dbef\bracke{u, p\paren{u}} = \delta\paren{u, p\paren{u}}$.

    \begin{itemize}
        \item {\bf \linesref{2-APASP-faster-main-distanceUpdate-start}{2-APASP-faster-main-distanceUpdate-end}}: Observe that $ \delta\paren{p\paren{u}, u} + \dbef\bracke{u, v} \geq \delta\paren{p\paren{u}, u} + \delta\paren{u, v} \geq \delta\paren{p\paren{u}, v}$. So, $\daft\bracke{p\paren{u}, v} \geq \delta\paren{p\paren{u}, v}$.
        
        \item {\bf \linesref{2-APASP-faster-main-estimateAPSP}}: Here the \algref{Estimate-augmented-MSSP} is invoked twice. As shown previously in proof of \lemref{lowerBound-proof}, this invocation does not underestimate the distances.

        \item {\bf \linesref{2-APASP-faster-main-approxMSSP}}: \obvref{approx-mssp-lowerBound} implies that the desired claim holds here.
        
        \item {\bf \linesref{2-APASP-faster-main-pivot-approxMinPlus}}: The claim follows trivially from \thmref{ApproxMinPlus}.
        
        \item {\bf \linesref{2-APASP-faster-main-Aij-initialise}{2-APASP-faster-main-Bij-initialise}}: First, consider the {\em for} loop of \linesref{2-APASP-faster-main-Aij-initialise}. It can be observed that whenever value of $A_{i, j}\bracke{p\paren{x}, y}$ changes, it takes the updated value of form $\dbef\bracke{p\paren{x}, x} + w\paren{x, y} $, which by triangle inequality is at least $\delta\paren{p\paren{x}, y}$. Hence, $A_{i, j}$ never underestimates the distance between any vertex pair.
        
        Since $x \in \B_{i, j}$, it implies from definition of the buckets that $\delta\paren{x, p\paren{x}} \geq i \cdot 2^j$. Furthermore, $w\paren{x, y} \geq \delta\paren{x, p\paren{x}}$ implies $w\paren{x, y} \geq i \cdot 2^j$. Altogether, $\delta\paren{x, p\paren{x}} + w\paren{x, y} \geq 2 \cdot i \cdot 2^j$, resulting in a lower bound of $2 \cdot i \cdot 2^j$ for all entries of $A_{i, j}$.
        
        Similarly, one can show in \linesref{2-APASP-faster-main-Bij-initialise} that $B_{i, j}$ does not underestimate the distances and each entry of $B_{i, j}$ is lower bounded by $i \cdot 2^j$.
        
        \item {\bf \linesref{2-APASP-faster-main-shifted-ApproxMinPlus}}: As discussed above, $shift_A , shift_B$ are smaller than all entries of $A_{i, j}, B_{i, j}$ respectively. By \obvref{shifted-ApproxMinPlus-lowerBound}, the matrix $M_{i, j}$ never underestimates the entries of $A_{i, j} \minplus B_{i, j}$. Hence, for any $\paren{x, x'} \in \B_{i, j} \times \Y_{i, j}$, there exists a $y \in V$ such that the following inequality holds:
        
        $$M_{i, j}\bracke{p\paren{x}, x'} \geq A_{i, j}\bracke{p\paren{x}, y} + B_{i, j}\bracke{y, x'}$$
        Since $A_{i, j}, B_{i, j}$ do not underestimate true distances, we get that $M_{i, j}\bracke{p\paren{x}, x'} \geq \delta\paren{p\paren{x}, y} + \delta\paren{y, x'}$. By triangle inequality, we get that $M_{i, j}$ does not underestimate actual distances.

        \item {\bf \linesref{2-APASP-faster-main-updateEstimateAfterShift-start}{2-APASP-faster-main-updateEstimateAfterShift-end} }: We show that the claim holds for \linesref{2-APASP-faster-main-updateEstimateAfterShift-start}. Since the distance estimate table $d$ is symmetric at all times, then the claim follows directly for \linesref{2-APASP-faster-main-updateEstimateAfterShift-end} as well.
        
        Since $M_{i, j}\bracke{p\paren{x}, x'} \geq \delta\paren{p\paren{x}, x'}$, we get that if $\daft\bracke{x, x'} < \dbef\bracke{x, x'}$, then $\daft\bracke{x, x'} = \dbef\bracke{x, \paren{x}} + M_{i, j}\bracke{p\paren{x}, x'} \geq \delta\paren{x, \paren{x}} + \delta\paren{p\paren{x}, x'} \geq \delta\paren{x, x'}$.
        
        Hence, the bound holds.

        \item {\bf \linesref{2-APASP-faster-main-estimateAPSP-again}}: Again, the argument is similar to the one used in the proof of \lemref{lowerBound-proof}, i.e., \algref{Estimate-augmented-MSSP} does not underestimate the true distances.
        
    \end{itemize}

    So, we've proved that the distances reported by \algref{2-APASP-faster-main} are not an underestimation.
\end{proof}

By \lemref{faster-alg-lowerBound-proof}, the \algref{2-APASP-faster-main} does not underestimate the distances. To conclude the proof for 2-approximation of \algref{2-APASP-faster-main}, we now show that for every pair of vertices, there is some point in \algref{2-APASP-faster-main} by which the distance estimate is a 2-Approximation.

\begin{lemma}\label{lem:faster-mainAlgo-proof}
    The output of \algref{2-APASP-faster-main} satisfies $d\bracke{u,v} \leq 2\delta\paren{u,v}$ for all $\paren{u,v} \in V \times V$.
\end{lemma}
\begin{proof}
    Let $P = u \squiggly v$.
    After sampling the sets $S_1$, consider the following cases for value of $\sizeindex{P}{1}$:

    \begin{enumerate}
        \item $\sizeindex{P}{1} \leq 1$: Consider the following two cases:
        \begin{enumerate}
            \item If $\sizeindex{P}{1} = 0$, then the complete path $P$ lies in the graph $G_1$. After one invocation of \textsc{Estimate\_augmented\_MSSP} $\paren{G_1, V, d}$ in \linesref{2-APASP-faster-main-estimateAPSP}, we obtain that $d\bracke{u, v} = \delta\paren{u, v}$.

            \item If $\sizeindex{P}{1} = 1$, let $\paren{x, y} = \rightabovearrow{P}{1}$. So, $P = u \squiggly x - y \squiggly v$. Note that the subpaths $P\bracke{u, x}, P\bracke{v, y}$ lie in $G_1$. So, after one invocation of \textsc{Estimate\_augmented\_MSSP}$\paren{G_1, V, d}$ in \linesref{2-APASP-faster-main-estimateAPSP}, we get that $d\bracke{x, y} = d\bracke{y, x} = \delta\paren{x, y}$. 
                
            Since $P\bracke{u, x}$ lies in $G_1$, another invocation of \textsc{Estimate\_augmented\_MSSP}$\paren{G_1, V, d}$ leads to the following guarantee: $d\bracke{v, u} = d\bracke{u, v} = \delta\paren{u, v}$.
        \end{enumerate}

        \item $\sizeindex{P}{1} \geq 3$: \label{case:faster-case-3}
        In this case, there exist at least three edges in $P$ which do not lie in $E_1$. Consider three edges as $\paren{x_1, y_1}, \paren{x_2, y_2}, \paren{x_3, y_3}$ such that $w\paren{x_1, y_1} \leq w\paren{x_2, y_2} \leq w\paren{x_3, y_3}$. This implies that the weight of $\paren{x_1, y_1}$ is at most one-third the weight of $P$. That is,

        $$w\paren{x_1, y_1} \leq \frac{\delta\paren{u, v}}{3}$$

        Furthermore, since $\paren{x_1, y_1} \notin E_1$, this implies that $\delta\paren{x_1, p\paren{x_1}} \leq w\paren{x_1, y_1}$.

        From the above two bounds, we get that:
        
        \begin{equation}\label{eqn:faster-analysis-helper-1}
            \delta\paren{x_1, p\paren{x_1}} \leq w\paren{x_1, y_1} \leq \frac{\delta\paren{u, v}}{3}
        \end{equation} 

        Note that the \algref{2-APASP-faster-main} invokes $\paren{1+\varepsilon}$-Approximate-MSSP with $S_1$ as the source set (\linesref{2-APASP-faster-main-approxMSSP}) and then updates the $V \times V$ distance estimates by utilising $\paren{1+\varepsilon}$-\minplusname-product between the matrices $d\bracke{V, S_1}, d\bracke{S_1, V}$ (\linesref{2-APASP-faster-main-pivot-approxMinPlus}), one can write the following inequality:

        \begin{align*}
            d\bracke{u, v} &\leq \paren{1 + \varepsilon} \paren{\paren{1+\varepsilon}\cdot \delta\paren{p\paren{x_1}, u} + \paren{1+\varepsilon}\cdot \delta\paren{p\paren{x_1}, v} } \\
            &\leq \paren{1 + \varepsilon}^2 \cdot \paren{\delta\paren{p\paren{x_1}, u} + \delta\paren{p\paren{x_1}, v} } \\
            &\leq \paren{1 + \varepsilon}^2 \cdot \paren{ \delta\paren{p\paren{x_1}, x_1} + \delta\paren{x_1, u} + \delta\paren{p\paren{x_1}, x_1} + \delta\paren{x_1, v} } \quad \paren{\text{Using triangle inequality}} \\
            &\leq \paren{1 + \varepsilon}^2 \cdot \paren{ 2 \cdot \delta\paren{p\paren{x_1}, x_1} + \delta\paren{u, v} } \\
            &\leq \paren{1 + \varepsilon}^2 \cdot \paren{ \frac{2}{3}\delta\paren{u, v} + \delta\paren{u, v} } \quad \paren{\text{Using } \eqnref{faster-analysis-helper-1}} \\
            &\leq \paren{1 + \varepsilon}^2 \cdot { \frac{5}{3}\delta\paren{u, v} }
        \end{align*}
        
        Substituting $\varepsilon \gets 0.05$, we obtain $d\bracke{u, v} \leq \paren{1.05 \cdot 1.05 \cdot \frac{5}{3}} \delta\paren{u, v} \leq 2\delta\paren{u, v}$.

        \item $\sizeindex{P}{1} = 2$:
        
        Suppose $\paren{x, y} = \rightabovearrow{P}{1}$ and $\paren{x', y'} = \leftabovearrow{P}{1}$. Since $\sizeindex{P}{1} = 2$, the edges $\paren{x, y}, \paren{x', y'}$ are distinct (they may share endpoints). It can be seen that $P = u \squiggly x-y \squiggly y' - x' \squiggly v$. 
        
        The following Figure~\ref{fig:faster-2case} will be helpful to keep in mind:

        \usetikzlibrary{decorations.pathmorphing}

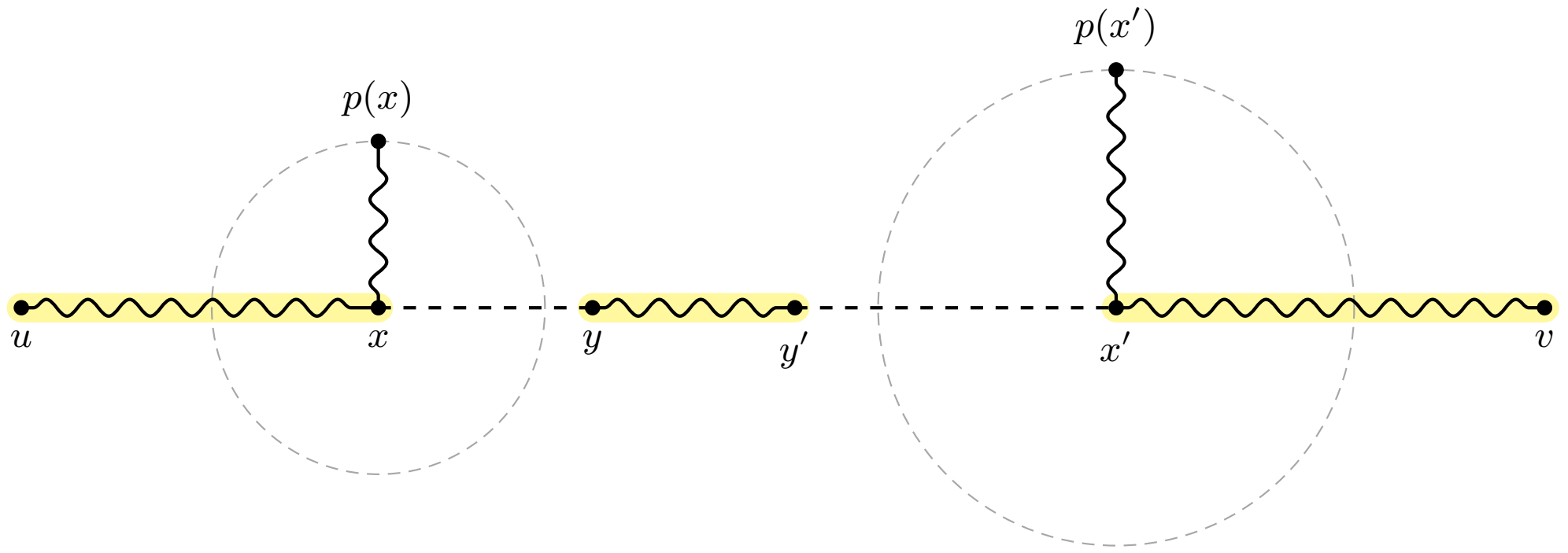
\begin{figure}[H]
\centering
\begin{tikzpicture}[
    every node/.style={font=\small},
    vtx/.style={circle, fill=black, inner sep=1.3pt},
    wave/.style={thick, decorate,
        decoration={snake, amplitude=0.7mm, segment length=3.5mm,
                    pre length=1mm, post length=1mm}},
    hl/.style={line width=7pt, yellow!80, opacity=0.55, line cap=round},
    nonE/.style={thick, dashed},
    ball/.style={densely dashed, gray!70}
]
    \coordinate (u)   at (0,0);
    \coordinate (x)   at (3.0,0);
    \coordinate (y)   at (4.8,0);
    \coordinate (yp)  at (6.5,0);
    \coordinate (xp)  at (9.2,0);
    \coordinate (v)   at (12.8,0);
    \coordinate (px)  at (3.0,1.4);
    \coordinate (pxp) at (9.2,2.0);

    \draw[hl] (u)  -- (x);
    \draw[hl] (y)  -- (yp);
    \draw[hl] (xp) -- (v);

    \draw[ball] (x)  circle (1.4);
    \draw[ball] (xp) circle (2.0);

    \draw[wave] (u)  -- (x);
    \draw[nonE] (x)  -- (y);
    \draw[wave] (y)  -- (yp);
    \draw[nonE] (yp) -- (xp);
    \draw[wave] (xp) -- (v);

    \draw[wave] (x)  -- (px);
    \draw[wave] (xp) -- (pxp);

    \node[vtx, label=below:{$u$}]           at (u)   {};
    \node[vtx, label=below:{$x$}]           at (x)   {};
    \node[vtx, label=below:{$y$}]           at (y)   {};
    \node[vtx, label=below:{$y'$}]          at (yp)  {};
    \node[vtx, label=below:{$x'$}]          at (xp)  {};
    \node[vtx, label=below:{$v$}]           at (v)   {};
    \node[vtx, label=above:{$p\paren{x}$}]  at (px)  {};
    \node[vtx, label=above:{$p\paren{x'}$}] at (pxp) {};
\end{tikzpicture}
\caption{The case $\abs{P}_{1}=2$, with $P = u \rightsquigarrow x - y \rightsquigarrow y' - x' \rightsquigarrow v$.
Dashed edges $\paren{x,y}$ and $\paren{x',y'}$ are the two edges of $P$ outside $E_{1}$.
Dashed circles are the balls of $x$ and $x'$, whose radii are $\delta\paren{x,p\paren{x}} \leq \delta\paren{x',p\paren{x'}}$.
The highlighted subpaths lie in $E_{1}$.}
\label{fig:faster-2case}
\end{figure}

        Without loss of generality, assume that $\delta\paren{x, p\paren{x}} \leq \delta\paren{x', p\paren{x'}}$. 
        
        As $\paren{x, y}, \paren{x', y'} \notin E_1$, the following inequalities also hold:

        \begin{equation}\label{eqn:faster-2case-helper-1}
            \delta\paren{x, p\paren{x}} \leq w\paren{x, y} \quad ; \quad \delta\paren{x, p\paren{x}} \leq \delta\paren{x', p\paren{x'}} \leq w\paren{x', y'}
        \end{equation}
        
        Note that $P\bracke{y, y'}$ lies in $G_1$ and hence, $d\bracke{x, y'} = \delta\paren{x, y'}, d\bracke{x', y} = \delta\paren{x', y}$ due to double SSSP invocation in $G_1$ in \linesref{2-APASP-faster-main-estimateAPSP}.
    
        \paragraph{Buckets are based on distance to nearest pivot}:
        First, we first show a simple observation that buckets generated in \linesref{2-APASP-faster-main-pivotDistance-bucketing} are based on the actual distances between vertices and their corresponding pivots. For this, recall that as soon as the sampling sets are initialised in the algorithm, the value $d\bracke{z, p\paren{z}}$ is set to $\delta\paren{z, p\paren{z}}$ for all $z \in V$.
        Since the bucketing procedure of \linesref{2-APASP-faster-main-pivotDistance-bucketing} maps any vertex $z$ into corresponding bucket based on $d\bracke{z, p\paren{z}}$ value and $d\bracke{z, p\paren{z}} = \delta\paren{z, p\paren{z}}$, it can be assumed that the $z$ is bucketed based on $\delta\paren{z, p\paren{z}}$.

        \paragraph{Proving $d\bracke{x, x'} \leq 2 \cdot \delta\paren{x, x'}$}:
        Now, we claim that after the completion of the {\em for} loop of \linesref{2-APASP-faster-main-bucketSizeLoop-start}, the value $d\bracke{x, x'}$ is at most $2 \cdot \delta\paren{x, x'}$. 
        First we recall some facts and notations: (i) As mentioned earlier, we  interchangeably use the notations $\B_{i, j}$ and $\B_j\bracke{i}$ for the buckets. Also, $\Y_{i, j} = \B_{i, j} \cup \B_{i+1, j}$.
        (ii) In any iteration $j$ of the {\em for} loop of \linesref{2-APASP-faster-main-bucketSizeLoop-start},
        recall from \algref{PivotDistance-bucketing} that any vertex $z$ lies in $\B_{i, j}$ if and only $\delta\paren{z, p\paren{z}} \in [i \cdot 2^j, \paren{i+1} \cdot 2^j)$.
        
        Now, we proceed with the proof. For a vertex $z$ in iteration $j$ of the {\em for} loop (\linesref{2-APASP-faster-main-bucketSizeLoop-start}), let $\B_j^{-1}\paren{z} = \floor{\frac{\delta\paren{z, p\paren{z}}}{2^j}}$ denote the value $i$ such that $z \in \B_{i, j}$. 

        Let $j^* \in \bracke{0, \ceil{\log_{2} \paren{n W_{\max}}}}$ be the minimum value $j$ such that $\delta\paren{x', p\paren{x'}} < \paren{\B_{j}^{-1}\paren{x} + 2} \cdot 2^j$. In other words, $\displaystyle j^* = \arg \min_j \paren{\exists i \  | \ x \in \B_{i, j} \land x' \in \Y_{i, j} }$. Note that such a $j$ always exists, because setting $j_{\max} = \ceil{\log_2 \paren{nW_{\max}}}$ implies that $0 \leq \delta\paren{x, p\paren{x}}, \delta\paren{x', p\paren{x'}} < 2^{j_{\max}}$. Hence, $\B_{j_{\max}}^{-1}\paren{x} = 0$ and $\delta\paren{x', p\paren{x'}} < 2^{j_{\max}}$. 
        For simplicity, we set $i^*=\B_{j^*}^{-1}\paren{x}$.
        Now, we consider 2 possible cases for $j^*$: 

        \begin{enumerate}

            \item $j^* > 0$: Since $j^*$ is the minimum chosen and not $j^* - 1$, this implies that  following: 
            
            \begin{align*}
                \delta\paren{x', p\paren{x'}} &\geq \paren{i^* + 2} \cdot 2^{j^* - 1} \\
                &\geq \paren{\floor{\frac{\delta\paren{x, p\paren{x}}}{2^{j^* - 1}}} + 2}\cdot 2^{j^* - 1} \\
                &> \paren{{\frac{\delta\paren{x, p\paren{x}}}{2^{j^* - 1}}} + 1}\cdot 2^{j^* - 1} \\
                &> {{{\delta\paren{x, p\paren{x}}}} + 2^{j^* - 1}} 
            \end{align*}
             
            Since $\delta\paren{x, p\paren{x}} \in \left[ i^* \cdot 2^{j^*}, {i^*\cdot 2^{j^*} + 2^{j^*}} \right)$, we get that:

            \begin{align} \label{eqn:faster-helper-5}
                \delta\paren{x, p\paren{x}} - i^* \cdot 2^{j^*} &\leq 2^{j^*}\notag \\  
                \delta\paren{x, p\paren{x}} - i^* \cdot 2^{j^*} &< 2 \cdot \paren{\delta\paren{x', p\paren{x'}} - \delta\paren{x, p\paren{x}}} \notag \\
                \implies 3\delta\paren{x, p\paren{x}} - 2 \cdot {\delta\paren{x', p\paren{x'}}} &< i^* \cdot 2^{j^*} 
            \end{align}

            Since $x \in \B_{i^*, j^*}, x' \in \Y_{i^*, j^*}$, the values of $A_{i^*, j} \bracke{p\paren{x}, y}, B_{i^*, j}\bracke{y, x'}$ will be modified in \linesref{2-APASP-faster-main-Aij-initialise}{2-APASP-faster-main-Bij-initialise} during the iteration $j^*$ of the {\em for} loop (\linesref{2-APASP-faster-main-bucketSizeLoop-start}). Hence, they satisfy the following:
            
            \begin{align} \label{eqn:faster-helper-6}
                A_{i^*, j^*} \bracke{p\paren{x}, y} &\leq \delta\paren{x, p\paren{x}} + w\paren{x,y}\\ 
                B_{i^*, j^*}\bracke{y, x'} &\leq d\bracke{x', y} \notag
            \end{align}
            
            Recall that $d\bracke{x', y} = \delta\paren{x', y}$ before the {\em for} loop starts (due to \linesref{2-APASP-faster-main-estimateAPSP}) and hence, we get that:
            
            \begin{align} \label{eqn:faster-helper-7}
                B_{i^*, j^*}\bracke{y, x'} \leq \delta\paren{x', y}
            \end{align}

            From \eqnref{faster-helper-6} and \eqnref{faster-helper-7}, we get that the value of $M_{i^*, j^*}\bracke{p\paren{x}, x'}$ after execution of \linesref{2-APASP-faster-main-shifted-ApproxMinPlus} will be set as the following:

            \begin{align*}
                M_{i^*, j^*}\bracke{p\paren{x}, x'}
            &\leq
            \paren{1+\varepsilon}
            \paren{
            \delta\paren{x,p\paren{x}}
            +w\paren{x,y}
            +\delta\paren{x',y}
            -3\cdot i^*\cdot 2^{j^*}
            }
            +3\cdot i^*\cdot 2^{j^*}\\
                &\leq \paren{1+\varepsilon} \paren{\delta\paren{x, p\paren{x}} +  \delta\paren{x, x'} - 3 \cdot i^* \cdot 2^{j^*}} + 3 \cdot i^* \cdot 2^{j^*}
            \end{align*}
            
            So, \linesref{2-APASP-faster-main-updateEstimateAfterShift-start} updates the value $d\bracke{x, x'}$ as the following:

            \begin{align}
                d\bracke{x,x'} &\leq \delta\paren{x, p\paren{x}} + \paren{1+\varepsilon} \paren{\delta\paren{x, p\paren{x}} + \delta\paren{x, x'} - 3 \cdot i^* \cdot 2^{j^*}} + 3 \cdot i^* \cdot 2^{j^*} \notag \\
                &\leq \delta\paren{x, p\paren{x}} + \paren{1+\varepsilon} \paren{\delta\paren{x, p\paren{x}} + \delta\paren{x, x'}} - 3 \cdot \varepsilon \cdot i^* \cdot 2^{j^*} \label{eqn:helper-12}
            \end{align}
            
           Using \eqnref{faster-helper-5}, the above bound inequality becomes:

           \begin{align}
                d\bracke{x,x'} &\leq \delta\paren{x, p\paren{x}} + \paren{1+\varepsilon} \paren{\delta\paren{x, p\paren{x}} + \delta\paren{x, x'}} - 9 \cdot \varepsilon \delta\paren{x, p\paren{x}} + 6 \cdot \varepsilon \cdot {\delta\paren{x', p\paren{x'}}} \notag \\
                &\leq \paren{2 - 8\varepsilon} \cdot \delta\paren{x, p\paren{x}} + \paren{6\varepsilon} \cdot \delta\paren{x', p\paren{x'}} + \paren{1+\varepsilon}\delta\paren{x, x'} \notag \\
                &\leq \paren{2 - 8\varepsilon} \cdot \delta\paren{x, p\paren{x}} + \paren{6\varepsilon} \cdot \delta\paren{x', p\paren{x'}} + \paren{1+\varepsilon}\delta\paren{x, x'} \notag\\
                &\leq \paren{2-14\varepsilon}\cdot \delta\paren{x,p\paren{x}} +6\varepsilon\cdot \paren{ \delta\paren{x',p\paren{x'}} +\delta\paren{x,p\paren{x}} } +\paren{1+\varepsilon}\delta\paren{x,x'} \notag\\
                &\leq \paren{2 - 14\varepsilon} \cdot \delta\paren{x, p\paren{x}} + \paren{6\varepsilon} \cdot \paren{w\paren{x', y'} + w\paren{x, y}} + \paren{1+\varepsilon}\delta\paren{x, x'} \quad \paren{\text{Using } \eqnref{faster-2case-helper-1}} \notag\\
                &\leq \paren{2 - 14\varepsilon} \cdot \delta\paren{x, p\paren{x}} + \paren{6\varepsilon} \cdot \delta\paren{x, x'} + \paren{1+\varepsilon}\delta\paren{x, x'} \quad \paren{\text{Using } \eqnref{faster-2case-helper-1}} \notag\\
                &\leq \paren{2 - 14\varepsilon} \cdot \delta\paren{x, p\paren{x}} + \paren{1+7\varepsilon}\delta\paren{x, x'}
                \label{eqn:helper-11}
            \end{align}

            Note that $\delta\paren{x, p\paren{x}}$ is bounded by both  $w\paren{x, y}$ and  $w\paren{x', y'}$ (\eqnref{faster-2case-helper-1}). Hence, $\delta\paren{x, p\paren{x}} \leq \frac{1}{2}\delta\paren{x, x'}$. Substituting this into \eqnref{helper-11}, we get that:
            
            \begin{align*}
                d\bracke{x,x'} &\leq \paren{1 - 7\varepsilon} \cdot \delta\paren{x, x'} + \paren{1+7\varepsilon}\delta\paren{x, x'} \\
                &\leq 2 \delta\paren{x, x'}
            \end{align*}

            Hence, the approximation bound holds in this case.
        
            \item $j^* = 0$: Assume that $x \in \B_0\bracke{i^*} \ \paren{i^* > 0}$  and hence, $x' \in \Y_0\bracke{i^*}$. Since weights are integers, we get that $\delta\paren{x, p\paren{x}} = i^*$ and $\delta\paren{x', p\paren{x'}} \in \bracce{i^*, i^*+1}$. 
            So, directly using \eqnref{helper-12} and substituting $j^* \gets 0$, we get that:
            
            \begin{align*}
                d\bracke{x,x'} &\leq \delta\paren{x, p\paren{x}} + \paren{1+\varepsilon} \paren{\delta\paren{x, p\paren{x}} + \delta\paren{x, x'}} - 3 \cdot \varepsilon \cdot i^*\\
                &\leq \paren{2+\varepsilon}\delta\paren{x, p\paren{x}} + \paren{1+\varepsilon}\delta\paren{x, x'} - 3 \cdot \varepsilon \cdot i^* \\
                &\leq \paren{2+\varepsilon}\delta\paren{x, p\paren{x}} + \paren{1+\varepsilon}\delta\paren{x, x'} - 3 \cdot \varepsilon \cdot \delta\paren{x, p\paren{x}} \\
                &\leq  \paren{2+\varepsilon}\delta\paren{x, p\paren{x}} + \paren{1+\varepsilon}\delta\paren{x, x'} - 3 \cdot \varepsilon \cdot \delta\paren{x, p\paren{x}}\\
                &\leq \paren{2-2\varepsilon}\delta\paren{x, p\paren{x}} + \paren{1+\varepsilon}\delta\paren{x, x'}\\
                &\leq \frac{1}{2}\paren{2-2\varepsilon}\delta\paren{x, x'} + \paren{1+\varepsilon}\delta\paren{x, x'}\\
                &\leq 2\delta\paren{x, x'}
            \end{align*}

            So, the 2-Approximation bound follows in this the $j^* = 0$ case as well.
        \end{enumerate}

        \paragraph{Proving $d\bracke{u, v} \leq 2 \cdot \delta\paren{u, v}$}: Since we have that $d\bracke{x, x'} \leq 2 \cdot \delta\paren{x, x'}$ just before \linesref{2-APASP-faster-main-estimateAPSP-again}, and $P\bracke{u, x}, P\bracke{v, x'}$ lie in $G_1$, invocation of one call to \textsc{Estimate\_augmented\_MSSP}$\paren{G_1, V, d}$ in \linesref{2-APASP-faster-main-estimateAPSP-again} gives $d\bracke{x, v} \leq 2\delta\paren{x, x'} + \delta\paren{x', v}$. Another invocation then gives $d\bracke{u, v} \leq \delta\paren{u, x} + 2\delta\paren{x, x'} + \delta\paren{x', v}$, thereby completing the proof for 2-Approximation. 

    \end{enumerate}
\end{proof}

\subsection{Running Time}

Throughout this subsection, we refer to the number of vertices and edges in $G$ as $n, m$ respectively. Note that like in previous section, we sometimes deal with subgraphs of the form $H = (V, E_H)$. Hence, we denote the runtime of some subroutines in terms of size of $H$ instead. For ease of notation, we assume $|E_H| = m_H$.

First, we note that the \algref{ApproxMSSP-pivot} just invokes the $\paren{1+\varepsilon}$-MSASP algorithm of \thmref{ApproxMSSP} and accordingly updates the distance estimates. So its runtime is comparable to the runtime of the algorithm in \thmref{thm:ApproxMSSP}. This leads to the following observation:

\begin{observation}\label{obv:ApproxMSSP-pivot-runtime}
    Assuming $|S| = n^{1-\gamma}$ for some $\gamma \in  [0, 1)$, and $\varepsilon > 0$ is a fixed constant, the \algref{ApproxMSSP-pivot} runs in $\widetilde{O}\paren{m_H^{1+o\paren{1}} + n^{\omega\paren{1, 1-\gamma, 1} + o\paren{1}}}$ time.
\end{observation}

\begin{lemma}\label{lem:PivotDistance-bucketing-runtime}
    \algref{PivotDistance-bucketing} runs in $O\paren{n}$ time.
\end{lemma}
\begin{proof}
    The \algref{PivotDistance-bucketing} just iterates over all vertices and puts them in their corresponding bucket (creates a new bucket if such a bucket not already present). Then it outputs this family of buckets. This just takes overall $O\paren{V_H} = O\paren{n}$ time.
\end{proof}

\begin{lemma}\label{lem:offset-shifted-ApproxminPlus-runtime}
    Suppose $M_1, M_2$ have entries in $\bracke{W_{\max}} \cup \bracce{\infty}$. Furthermore, assume their dimensions are $N^{\alpha} \times N^{\beta}$, $N^{\beta} \times N^{\gamma}$ respectively for some $N, \alpha, \beta, \gamma > 0$, and $\varepsilon > 0$ is a fixed constant. Then, \algref{offset-shifted-ApproxminPlus} runs in time $\widetilde{O}\paren{N^{\omega\paren{\alpha, \beta, \gamma} + o\paren{1}} \log W_{\max}}$. 
\end{lemma}
\begin{proof}
    Computing the matrices $\bar{M}_i$ can be done in time linear in input size. Since the entries are at least as large as the offsets, we observe that $\bar{M}_i$ will have entries in $\bracce{0}\cup\bracke{W_{\max}} \cup \bracce{\infty}$. Using \thmref{ApproxMinPlus}, for a fixed $\varepsilon > 0$, the matrix $\bar{M}$ (and subsequently $M$) can be computed in $\widetilde{O}\paren{N^{\omega\paren{\alpha, \beta, \gamma}+o\paren{1}} \log W_{\max}}$ time. This proves the desired claim. 
\end{proof}

We first note the following simple observation which we will need moving forward:

\begin{observation}\label{obv:omega-inequality-1}
    For any $a \in \bracke{0, 1}$, the following inequality holds: $\omega\paren{1-a, 1, 1-a} + a-1 \geq 1$.
\end{observation}
\begin{proof}
    The proof is quite simple. Any algorithm for multiplication of 2 rectangular matrices must at least read the whole input. Since the input has $n^{1-a} \times n$ sized matrices, the algorithm must take at least $n^{2-a}$ time and hence, $\omega\paren{1-a, 1, 1-a} \geq 2-a$. Adding $a-1$ to both sides of this inequality proves the desired claim.
\end{proof}

\begin{observation}\label{obv:omega-inequality-2}
    For any $a \in \bracke{0, 1}$, the following inequality holds: $\omega\paren{1-a, 1, 1-a} + a \geq \omega\paren{1, 1, 1-a}$.
\end{observation}
\begin{proof}
    Consider 2 rectangular matrices $M_1, M_2$ of dimensions $n \times n, \ n \times n^{1-a}$ respectively. Then, one way to compute the matrix product $M_1 \cdot M_2$ is to partition $M_1$ into $n^a$ blocks, with each block of dimension $n^{1-a} \times n$. Computing product of each block with $M_2$ takes $n^{\omega\paren{1-a, 1, 1-a} + o\paren{1}}$ time and hence, computation over all $n^a$ blocks takes $n^{\omega\paren{1-a, 1, 1-a} + a + o\paren{1}}$ time. Hence, $M_1 \cdot M_2$ can be computed in at most $n^{\omega\paren{1-a, 1, 1-a} + a + o\paren{1}}$ time, which implies that $\omega\paren{1, 1, 1-a} \leq \omega\paren{1-a, 1, 1-a} + a$.
\end{proof}

Following is a well-known cyclic invariance property of rectangular matrix multiplication exponent.

\begin{theorem}(\cite{Blaser13})\label{thm:omega-cyclicShift}
    For any $a,b > 0$, $\omega\paren{a,a,b} = \omega\paren{a,b,a} = \omega\paren{b,a,a}$. 
\end{theorem}

Recall that in a fixed iteration of the {\em for} loop starting at \linesref{2-APASP-faster-main-bucketIter-start} of \algref{2-APASP-faster-main}, $\Y_{i, j}$ for that specific iteration is defined as $\B_j\bracke{i} \cup \B_j\bracke{i+1}$ (or just $\B_j\bracke{i}$ if $\B_j\bracke{i+1}$ does not exist). Now, we state the runtime of this iterations in terms of the size of $\Y_{i, j}$, denoted by $|\Y_{i, j}|$.

\begin{lemma} \label{lem:2-APASP-faster-main-bucketIter-iter-runtime}
    For any $a < 1$, each iteration of the {\bf for} loop starting at \linesref{2-APASP-faster-main-bucketIter-start} of \algref{2-APASP-faster-main} runs in the following total time:

    $$ \widetilde{O} \paren{|\Y_{i, j}| n^{\omega\paren{1-a, 1, 1-a} + a-1 + o\paren{1}} \log W_{\max}}$$

    where $\Y_{i, j}$ is the set defined for that specific iteration at \linesref{2-APASP-faster-main-Yij-initialise}.
\end{lemma}
\begin{proof}
    We fix an iteration and hence, $\B_{i, j}, \P_{i, j}, \Y_{i, j}$ remain fixed here onwards in the proof. By definition, $\B_{i, j} \subseteq \Y_{i, j}; \P_{i, j} \subseteq S_1$ and hence, $|\B_{i, j}| \leq |\Y_{i, j}| \ ; \ |\P_{i, j}| \leq |S_1| = O\paren{n^{1-a}}$. Furthermore, each element of $\B_{i, j}$ contributes at most one element to $\P_{i, j}$ and hence, we also have that $|\P_{i, j} \leq |\B_{i, j}|$.
    So, we keep the following 2 bounds handy: 
    \begin{equation}\label{eqn:Bij-Pij-bounds}
        |\B_{i, j}| = O\paren{|\Y_{i, j}|} \ ; |\P_{i, j}| = O\paren{\mmin{|\Y_{i, j}|, n^{1-a} }}    
    \end{equation}

    We now proceed to bound the time taken by the iteration. It is easy to observe that the \linesref{2-APASP-faster-main-Pij-initialise}{2-APASP-faster-main-shift-values} run in $\widetilde{O}\paren{|\Y_{i, j}|n}$ time.

    Furthermore, the {\em for} loop starting at \linesref{2-APASP-faster-main-updateEstimateAfterShift-start} has $O\paren{|\Y_{i, j}|^2}$ iterations and each iteration takes $O\paren{1}$ time. 
    So, the loop overall takes $O(|\Y_{i, j}|^2)$ time. Note that $|\Y_{i, j}| \leq n$, and \obvref{omega-inequality-1} implies that 
    $n \leq n^{\omega\paren{1-a, 1, 1-a} + a-1}$. Hence, the runtime of the {\em for} loop starting at 
    \linesref{2-APASP-faster-main-updateEstimateAfterShift-start} is bounded by the following expression: 
    $O\paren{|\Y_{i, j}| n^{\omega\paren{1-a, 1, 1-a} + a-1 + o\paren{1}}}$.  

    Due to the discussion so far, \linesref{2-APASP-faster-main-shifted-ApproxMinPlus} is the only line whose runtime analysis remains. For sake of convenience, we use in this proof the operator $M_1 \boxast M_2$ to denote the offset-shifted $\paren{1+\varepsilon}$-approximate \minplusname-matrix product of matrices $M_1$ and $M_2$ (we do not write the $\varepsilon$, offset values in this notation to avoid clutter). As remarked earlier in \obvref{shifted-ApproxMinPlus-blocks}, the product $M_1 \boxast M_2$ can be computed by dividing $M_1, M_2$ into blocks and then combining the blocks later. Now, consider two cases:
    \begin{itemize}
        \item {\bf Case 1} - $|\Y_{i, j}| \geq n^{1-a}$: In this case, one can view the matrix $B_{i, j}$ as collection of $k = \ceil{\frac{|\Y_{i, j}|}{n^{1-a}} }$ matrices $C_1, C_2, \dots, C_k$, each of dimension $n \times n^{1-a}$. 
        \begin{align*}
            B_{i, j} &= \begin{pmatrix} C_1 & C_2 & \dots & C_k  \end{pmatrix}    \\
            \implies A_{i, j} \boxast B_{i, j} &= \begin{pmatrix} A_{i, j} \boxast C_1 & A_{i, j} \boxast C_2 & \dots & A_{i, j} \boxast C_k  \end{pmatrix}
        \end{align*}
        
        Recall that $A_{i, j}$ has dimensions $|\P_{i, j}| \times |V|$ and $|\P_{i, j}| = O\paren{n^{1-a}}$ (\eqnref{Bij-Pij-bounds}). Hence, the product $A_{i, j} \boxast C_1$ can be computed in $\widetilde{O}\paren{n^{\omega\paren{1-a, 1, 1-a} + o\paren{1}} \log W_{\max}}$ time (\lemref{offset-shifted-ApproxminPlus-runtime}). Since $k = O\paren{\frac{|\Y_{i, j}|}{n^{1-a}}}$, the total runtime is bounded by the following expression:
        \begin{align*}
        \widetilde{O}\paren{k \cdot n^{\omega\paren{1-a, 1, 1-a} + o\paren{1}} \log W_{\max}}  &= \widetilde{O}\paren{\frac{|\Y_{i, j}|}{n^{1-a}} \cdot n^{\omega\paren{1-a, 1, 1-a} + o\paren{1}} \log W_{\max}}\\
        &= \widetilde{O}\paren{|\Y_{i, j}| \cdot n^{\omega\paren{1-a, 1, 1-a} + a-1 + o\paren{1}} \log W_{\max}}
        \end{align*}

        So, the claimed runtime follows in this case.

        \item {\bf Case 2} - $|\Y_{i, j}| \leq n^{1-a}$: Here, we assume for sake of analysis that dimensions of $A_{i, j}$ are $|\Y_{i, j}| \times n$. This can be achieved by adding some padding rows to $A_{i, j}$ (this holds because $|\P_{i, j}|\leq |\Y_{i, j}|$ by \eqnref{Bij-Pij-bounds}. We write the matrix $A_{i, j}$ as collection of $k = \ceil{\frac{n}{|\Y_{i, j}|^{\frac{1}{1-a}}} }$ matrices $C_1, C_2, \dots, C_k$, each of dimension $|\Y_{i, j}| \times |\Y_{i, j}|^{\frac{1}{1-a}}$. Similarly, $B_{i, j}$ is written as collection of matrices $D_1, D_2, \dots, D_k$, each of dimension $|\Y_{i, j}|^{\frac{1}{1-a}} \times |\Y_{i, j}|$. That is,
        
        \begin{align*}
            A_{i, j} \boxast B_{i, j} &= \begin{pmatrix} C_1 & C_2 & \dots & C_k  \end{pmatrix} \boxast \begin{pmatrix} D_1 \\ D_2 \\ \dots \\D_k \end{pmatrix} \\ 
        \end{align*}
        
        Since our matrix multiplication is of the form \minplusname, the above expression can be written as:
        
        \begin{align*}
            A_{i, j} \boxast B_{i, j} &= \mmin{C_1 \boxast D_1, C_2 \boxast D_2, \dots, C_k \boxast D_k} 
        \end{align*}
        where $\mmin{\cdot}$ means entrywise minimum of the $k$ matrix products. Using \lemref{offset-shifted-ApproxminPlus-runtime}, we get that computing $C_1 \boxast D_1$ takes $\widetilde{O}\paren{\paren{|\Y_{i, j}|^{\frac{1}{1-a}}}^{\omega\paren{1-a, 1, 1-a} + o\paren{1}} \log W_{\max}}$ time. Thus, repeating this for computing all $C_i \boxast D_i, \ i \in \bracke{k}$ results in the following total time:
        
        \begin{align*}
            \widetilde{O}\paren{k \cdot \paren{|\Y_{i, j}|^{\frac{\omega\paren{1-a, 1, 1-a}}{1-a} + o\paren{1}}} \log W_{\max}}  
            &= \widetilde{O}\paren{\frac{n}{|\Y_{i, j}|^{\frac{1}{1-a}}} \cdot \paren{|\Y_{i, j}|^{\frac{\omega\paren{1-a, 1, 1-a}}{1-a} + o\paren{1}}} \log W_{\max}}     \\
            &= \widetilde{O}\paren{n \cdot {|\Y_{i, j}|^{\frac{\omega\paren{1-a, 1, 1-a} - 1}{1-a} + o\paren{1}}} \log W_{\max}} \\
            &= \widetilde{O}\paren{n \cdot {|\Y_{i, j}|^{\frac{\omega\paren{1-a, 1, 1-a} - 1}{1-a} - 1 + o\paren{1}}} |\Y_{i, j}| \log W_{\max}} \\
            &= \widetilde{O}\paren{n \cdot {|\Y_{i, j}|^{\frac{\omega\paren{1-a, 1, 1-a} + a - 2}{1-a} + o\paren{1}}} |\Y_{i, j}| \log W_{\max}}
        \end{align*}

        As implied by \obvref{omega-inequality-1}, $\omega\paren{1-a, 1, 1-a} + a - 2 \geq 0$. Furthermore, we are in the case $|\Y_{i, j}| \leq n^{1-a}$. So, the expression $|\Y_{i, j}|^{\frac{\omega\paren{1-a, 1, 1-a} + a - 2}{1-a} } \leq \paren{n^{1-a}}^{\frac{\omega\paren{1-a, 1, 1-a} + a - 2}{1-a} }$. Hence, the runtime is bounded by the following expression:

        \begin{align*}
            \widetilde{O}\paren{n \cdot {\paren{n^{1-a}}^{\frac{\omega\paren{1-a, 1, 1-a} - 1}{1-a} - 1 + o\paren{1}}} |\Y_{i, j}| \log W_{\max}}
        \end{align*} 

        The above expression can further be simplified as:
        \begin{align*} 
            \widetilde{O}\paren{{{n}^{1 + {\omega\paren{1-a, 1, 1-a} - 1} - 1 + a + o\paren{1}}} |\Y_{i, j}| \log W_{\max}}
            = \widetilde{O}\paren{{{n}^{{\omega\paren{1-a, 1, 1-a}} + a-1 + o\paren{1}}} |\Y_{i, j}| \log W_{\max}} 
        \end{align*}
        
        Hence, the claimed runtime follows in this case as well.

    \end{itemize}  

    So, each iteration of the {\bf for} loop starting at \linesref{2-APASP-faster-main-bucketIter-start} of \algref{2-APASP-faster-main} runs in total time bounded by $\widetilde{O} \paren{|\Y_{i, j}| n^{\omega\paren{1-a, 1, 1-a} + a-1 + o\paren{1}} \log W_{\max}}$.
\end{proof}

\begin{observation}\label{obv:Bj-partition}
    The family $\B_j$ of buckets (defined in \linesref{2-APASP-faster-main-pivotDistance-bucketing}) forms a partition of the vertex set $V$.
\end{observation}

The \obvref{Bj-partition} has the following simple corollary:

\begin{corollary}\label{cor:Yij-bounded}
    For a given family $\B_j$ of buckets, the following claim holds:

    $$\sum_{\B_{j}\bracke{i} \in \B_j} |\B_j[i] \cup \B_j[i+1]| \leq 2n$$

    Alternatively, the above bound can be written as following:

    $$\sum_{\substack{\B_{j}\bracke{i} \in \B_j \\ \Y_{i, j} = \B_j\bracke{i} \cup \B_j\bracke{i+1}}} |\Y_{i, j}| \leq 2n$$
\end{corollary}

\begin{lemma}\label{lem:2-APASP-faster-main-bucketSizeLoop-iter-runtime}
    Each iteration of the {\bf for} loop starting at \linesref{2-APASP-faster-main-bucketSizeLoop-start} of \algref{2-APASP-faster-main} runs in the following time:
    $$ \widetilde{O}\paren{n^{a + \omega\paren{1-a, 1, 1-a} + o\paren{1}} \log W_{\max}}$$ 
\end{lemma}
\begin{proof}
    By \lemref{PivotDistance-bucketing-runtime}, \linesref{2-APASP-faster-main-pivotDistance-bucketing} runs in $O(n)$ time, which lies within the claimed time bounds.
    
    Furthermore, as claimed by \lemref{2-APASP-faster-main-bucketIter-iter-runtime}, each iteration of the {\bf for} loop starting at \linesref{2-APASP-faster-main-bucketIter-start} runs in time $ \widetilde{O} \paren{|\Y_{i, j}| n^{\omega\paren{1-a, 1, 1-a} + a-1 + o\paren{1}} \log W_{\max}}$. Hence, the total runtime over all such iterations is given by the following expression:
    
    \begin{align*}
       \sum_{\substack{\B_{j}\bracke{i} \in \B_j \\ \Y_{i, j} = B_j\bracke{i} \cup \B_j\bracke{i+1}}} \widetilde{O} \paren{|\Y_{i, j}| n^{\omega\paren{1-a, 1, 1-a} + a-1 + o\paren{1}} \log W_{\max}} 
       &= \widetilde{O} \paren{\sum_{\substack{\B_{j}\bracke{i} \in \B_j \\ \Y_{i, j} = B_j\bracke{i} \cup \B_j\bracke{i+1}}} |\Y_{i, j}| n^{\omega\paren{1-a, 1, 1-a} + a-1 + o\paren{1}} \log W_{\max}}  \\
       &= \widetilde{O} \paren{n \cdot n^{\omega\paren{1-a, 1, 1-a} + a-1 + o\paren{1}} \log W_{\max}} \ \paren{\text{By } \corref{Yij-bounded}} \\
       &= \widetilde{O} \paren{n^{\omega\paren{1-a, 1, 1-a} + a + o\paren{1}} \log W_{\max}}
    \end{align*}
    
    Hence, the claimed runtime follows.
\end{proof}

\begin{lemma}\label{lem:faster-main-runtime}
    Assuming $\abs{E} = O\paren{n^2}$ and $q_1 = n^{-a}$ for some $a \in \bracke{0,1}$, the runtime of \algref{2-APASP-faster-main} is:

    $$
    \widetilde{O}\paren{n^{2+a} + n^{a+\omega\paren{1-a, 1, 1-a} + o\paren{1}} \log^2 W_{\max}  }
    $$

    For $a=\frac{\omega-2}{\omega-1}$,
the runtime is at most:
$$
\widetilde O\paren{
n^{\omega-\frac{\paren{\omega-2}^{2}}{\omega-1}
+o\paren{1}}
\cdot\log^2 W_{\max}
}
$$

 Using the previous FMM bounds
\cite{DupEisKozMehRuiSeeZhoAlmVasBal2026,VasXuXuZho2024,AlmDuaVasXuXuZho2024},
the runtime is
$\widetilde O\paren{
n^{2.24186294}\cdot\log^2 W_{\max}
}$.
Using the new FMM bounds
\cite{OpenAI2026,Uznanski2026,AmaBerZui2026}
gives
$\widetilde O\paren{
n^{2.17157288}\cdot\log^2 W_{\max}
}$
runtime.

\end{lemma}
\begin{proof}
    To bound the total running time, we analyse the time complexity of \algref{2-APASP-faster-main} piece by piece. Since $q_1 = n^{-a}$, it will be helpful to keep in mind that $|S_1| = n^{1-a}, |E_1| = n^{1+a}$:
    
    \begin{itemize}
        \item {\bf \linesref{2-APASP-faster-main-initialiseHierarchy}}: From \obvref{pivotsdistance}, this takes $\widetilde{O}\paren{|E|+n}$ total time, which can be written as $\widetilde{O}\paren{n^2}$.
        
        \item {\bf \linesref{2-APASP-faster-main-distanceUpdate-start}{2-APASP-faster-main-distanceUpdate-end}}: Since each line of the {\em for} loop takes only $O\paren{1}$ time, hence the loop overall takes $O\paren{n^2}$ time.

        \item {\bf \linesref{2-APASP-faster-main-estimateAPSP}}: Here,  we invoke the \textsc{Estimate\_augmented\_MSSP} procedure twice, with set of sources as the whole vertex set, i.e., $S = V$. From \lemref{estimateMSSP-runtime}, this takes time $\widetilde{O}\paren{|E_1| n + n^2}$, which simplifies to $\widetilde{O}\paren{n^{2+a}}$.  

        \item {\bf \linesref{2-APASP-faster-main-approxMSSP}}: From \obvref{ApproxMSSP-pivot-runtime}, this takes $\widetilde{O}\paren{m^{1+o\paren{1}} + n^{\omega\paren{1, 1-a, 1} + o\paren{1}}}$ time. Since $m = O\paren{n^2}$, the runtime can instead be written as $\widetilde{O}\paren{n^{2+o\paren{1}} + n^{\omega\paren{1, 1-a, 1} + o\paren{1}}}$.

        \item {\bf \linesref{2-APASP-faster-main-pivot-approxMinPlus}}: Since $\varepsilon$ is a fixed constant, using \thmref{ApproxMinPlus}, this line takes $\widetilde{O}\paren{n^{\omega\paren{1, 1-a, 1} + o\paren{1}} \log W_{\max}}$ time.
        
        \item {\bf {\em for} loop of \linesref{2-APASP-faster-main-bucketSizeLoop-start}}: From \lemref{2-APASP-faster-main-bucketSizeLoop-iter-runtime}, each iteration of the loop takes $\widetilde{O}\paren{n^{a + \omega\paren{1-a, 1, 1-a} + o\paren{1}} \log W_{\max}}$ time. Since the loop runs for $O\paren{\log nW_{\max}}$ iterations, we get the total runtime over all iterations as $\widetilde{O}\paren{n^{a + \omega\paren{1-a, 1, 1-a} + o\paren{1}} \log^2 W_{\max}}$.

        \item {\bf \linesref{2-APASP-faster-main-estimateAPSP-again}}: As discussed previously, this takes time $\widetilde{O}\paren{n^{2+a}}$.  
        
    \end{itemize}

    Hence, \algref{2-APASP-faster-main} runs in time: $\widetilde{O}\paren{n^{2+a} + n^{\omega\paren{1, 1-a, 1} + o\paren{1}}\cdot \log W_{\max} + n^{a + \omega\paren{1-a, 1, 1-a} + o\paren{1}} \cdot \log^2 W_{\max} }$.

    Using \obvref{omega-inequality-2} and \thmref{omega-cyclicShift}, we get that the runtime can be simplified to the following:

    $$ \widetilde{O}\paren{n^{2+a} + n^{a + \omega\paren{1-a, 1, 1-a} + o\paren{1}} \cdot \log^2 W_{\max} } $$

    This proves the desired bounds.
    To provide numerical exponents, we apply the rectangular FMM bounds
\cite{Uznanski2026,AmaBerZui2026}, for $0\leq a<1$:
$$
\begin{aligned}
a+\omega\paren{1-a,1,1-a}
&=
a+\paren{1-a}\cdot
\omega\paren{1,\frac{1}{1-a},1}\\
&\leq
a+\paren{1-a}\cdot
\paren{1+\frac{1}{1-a}+\frac{1-a}{4}}\\
&=
2+\frac{\paren{1-a}^{2}}{4}
\end{aligned}
$$
Balancing $2+a=2+\frac{\paren{1-a}^{2}}{4}$
gives $a=3-2\sqrt{2}$ and
$\widetilde O\paren{
n^{5-2\sqrt{2}+o\paren{1}}\cdot\log^2 W_{\max}
}$
runtime, which gives us 
$\widetilde O\paren{
n^{2.17157288}\cdot\log^2 W_{\max}
}$.
Using the earlier rectangular FMM bounds
\cite{VasXuXuZho2024,AlmDuaVasXuXuZho2024}
instead, choosing $a=0.24186294$ gives
$\widetilde O\paren{
n^{2.24186294}\cdot\log^2 W_{\max}
}$
runtime.

Finally, partitioning the inner dimension
into $O\paren{n^a}$ blocks reduces the product
to $O\paren{n^a}$ square products of dimension
$O\paren{n^{1-a}}$. Hence:
$$
\omega\paren{1-a,1,1-a}
\leq a+\paren{1-a}\omega
$$
Choosing $a=\frac{\omega-2}{\omega-1}$ gives:
$$
2+a
=
2a+\paren{1-a}\omega
=
\omega-\frac{\paren{\omega-2}^{2}}{\omega-1}
$$
Therefore, the runtime is at most
$\widetilde O\paren{
n^{\omega-\frac{\paren{\omega-2}^{2}}{\omega-1}
+o\paren{1}}\cdot\log^2 W_{\max}
}$,
whose exponent of $n$ is strictly below
$\omega$ if $\omega>2$.

\end{proof}

\section{Fully Rectangular $\paren{\max,\min}$-Matrix Multiplication}\label{rect_(max,min)}

Duan and Pettie \cite{DuaPet2009} showed that the \maxminname-product of two square $n\times n$ matrices can be computed in $\widetilde O\paren{n^{\frac{3+\omega}{2}+o\paren{1}}}$ time. Their algorithm combines fast matrix multiplication with sparse dominance products and row balancing. Grandoni, Italiano, {\L}ukasiewicz, Parotsidis and Uzna{\'n}ski \cite{GraItaLukParUzn2021} later observed that the same approach extends to the rectangular setting of an $n\times n^{b}$ matrix and an $n^{b}\times n$ matrix, obtaining a running time of $\widetilde O\paren{n^{\frac{2+b+\omega\paren{b}}{2}+o\paren{1}}}$. In this section, we fully prove the claim for fully rectangular matrices. Given $A\in\mathbb{R}^{n^{a}\times n^{b}}$ and $B\in\mathbb{R}^{n^{b}\times n^{c}}$, we show how to compute $A\maxmin B$ in $\widetilde O\paren{n^{\frac{a+b+c+\omega\paren{a,b,c}}{2}+o\paren{1}}}$ time.

For the \maxminname-product, we may assume that all entries of
$A$ and $B$ are distinct. Sort the entries of both matrices
together by value, placing entries of $A$ before entries of $B$
when their values are equal and breaking ties within each matrix
arbitrarily. Represent each entry by a tuple of value and rank in this order, and compare
these pairs lexicographically.  The original output values are
recovered simply by taking the first coordinate. We begin with:

\begin{observation}\label{obv:maxmin_options}
For every $i\in\bracke{n^{a}}$ and $j\in\bracke{n^{c}}$,
the value $\paren{A\maxmin B}\bracke{i,j}$ is the larger of
the following two values:
\begin{enumerate}
    \item The largest $A\bracke{i,k}$ satisfying
    $A\bracke{i,k}\leq B\bracke{k,j}$.
    \item The largest $B\bracke{k,j}$ satisfying
    $B\bracke{k,j}\leq A\bracke{i,k}$.
\end{enumerate}
In both cases $k\in\bracke{n^{b}}$ and if a set is empty we assume set its maximum to be $-\infty$.
\end{observation}

We first develop a fully rectangular sparse \domname-product algorithm for matrices of dimensions $n^{a}\times n^{b}$ and $n^{b}\times n^{c}$. We then partition the entries of the two matrices into $O\paren{t}$ common consecutive value ranges, chosen such that every range contains at most $O\paren{\frac{n^{a+b}}{t}}$ entries of $A$ and at most $O\paren{\frac{n^{b+c}}{t}}$ entries of $B$. 
For the first of the two quantities above, these ranges allow us to identify the largest range containing a feasible entry $A\bracke{i,k}$. Within this range, we combine the row balancing procedure from \appref{balance} with additional \domname-products to identify a single row part containing the largest such entry. Each such part contains at most $O\paren{\frac{n^{b}}{t}}$ entries, which we then inspect.
The second quantity is computed by applying the same argument after exchanging the roles of $A$ and $B$. Altogether, the resulting running time is
$\widetilde O\paren{t\cdot n^{\omega\paren{a,b,c}+o\paren{1}}+\frac{n^{a+b+c}}{t}}$,
which is minimized by selecting $t=n^{\frac{a+b+c-\omega\paren{a,b,c}}{2}}$, yielding
$\widetilde O\paren{n^{\frac{a+b+c+\omega\paren{a,b,c}}{2}+o\paren{1}}}$ time.

We first require a sparse version of the fully rectangular \domname-product. Duan and Pettie \cite{DuaPet2009} established the corresponding bound for square matrices. The following lemma extends this bound to matrices of dimensions $n^{a}\times n^{b}$ and $n^{b}\times n^{c}$. As in \appref{balance}, the sparsity of a matrix is measured by the number of its finite entries.

\begin{lemma}\label{lem:rect_dom_sparse}
Let
$A\in\paren{\mathbb{R}\cup\bracce{+\infty}}^{n^{a}\times n^{b}}$
and
$B\in\paren{\mathbb{R}\cup\bracce{-\infty}}^{n^{b}\times n^{c}}$,
and let $m_{A}$ and $m_{B}$ denote their respective numbers of finite entries. Then $A\dom B$ can be computed in
$
\widetilde O\paren{
n^{\omega\paren{a,b,c}+o\paren{1}}
+
\frac{m_{A}\cdot m_{B}}{n^{b}}
}
$ time.
\end{lemma}

\begin{proof}
If $m_{A}=0$ the claim follows trivially as $A\dom B$ is the zero matrix. We therefore assume $m_{A}>0$. Let $
\cbal{A}=\paren{A',A''}$. 
Recall that $
k_{A}^{\mathrm{col}}
=
\ceil{\frac{m_{A}}{n^{b}}}$
and that every non-final part of a column of $A$ is assigned by
$\rho_{A}^{\mathrm{col}}$ to a distinct column of $A''$. 
For every $\paren{k,q}\in\NFC{A}$, let
$J_{k}^{\paren{q}}$ denote the corresponding set of original
row indices and define:
$$\mu_{k,q}
=
\mmax{A\bracke{\ell,k}}{\ell\in J_{k}^{\paren{q}}}
$$
This value is obtained from the last entry of the sorted part.
In particular, all entries in this part are at most
$B\bracke{k,j}$  when $\mu_{k,q}\leq B\bracke{k,j}$.

Construct the Boolean matrices
$\hat{A}\in\bracce{0,1}^{n^{a}\times n^{b}}$ and
$\hat{B}\in\bracce{0,1}^{n^{b}\times n^{c}}$ by setting:
$$
\hat{A}\bracke{i,h}
=
\begin{cases}
1 &  A''\bracke{i,h}<+\infty\\
0 & \text{otherwise}
\end{cases}
\qquad \textnormal{and}\qquad  
\hat{B}\bracke{h,j}
=
\begin{cases}
1
& \begin{array}{l}
\text{if $h=\rho_{A}^{\mathrm{col}}\paren{k,q}$ for some}\\
\text{$\paren{k,q}\in\NFC{A}$ and $\mu_{k,q}\leq B\bracke{k,j}$}
\end{array}\\
0 & \text{otherwise}
\end{cases}
$$
We compute the \mmname-product
$\hat{C}=\hat{A}\cdot\hat{B}$ in
$ O\paren{n^{\omega\paren{a,b,c}+o\paren{1}}}$ time. For every $i\in\bracke{n^{a}}$ and $j\in\bracke{n^{c}}$ the entry $\hat{C}\bracke{i,j}$ counts the non-final parts
$J_{k}^{\paren{q}}$ containing $i$ with
$\mu_{k,q}\leq B\bracke{k,j}$.
Thus it counts exactly the comparisons
$A\bracke{i,k}\leq B\bracke{k,j}$ whose entry from $A$ belongs
to a completely dominated non-final part.
No comparison is counted twice, since every finite entry of $A$
belongs to exactly one part.

Initialize $C^{\star}$ to be a matrix full of zeros. 
For every finite entry $B\bracke{k,j}$, skip it if column $k$ of $A$
has no finite entries. Otherwise, use binary search on the values
$\mu_{k,q}$ to find the first non-final part $J_{k}^{\paren{q}}$
with $\mu_{k,q}>B\bracke{k,j}$.
If such a part exists, inspect its entries. Otherwise, inspect
the final part of column $k$.
For every inspected entry satisfying
$A\bracke{i,k}\leq B\bracke{k,j}$, increase
$C^\star\bracke{i,j}$ by one.
Return $C=\hat{C}+C^\star$.

For each finite entry $B\bracke{k,j}$, all parts preceding the
inspected part are non-final and are  completely dominated, so their
contributions are already included in $\hat{C}$.
The inspected part is either final or satisfies
$\mu_{k,q}>B\bracke{k,j}$, so in any case, its contributions for this entry of
$B$ were not counted by $\hat{C}$.   

If this part is non-final, every entry in a later part is at least
$\mu_{k,q}>B\bracke{k,j}$, since the parts follow a sorted order.
If it is final, there are no later parts.
Thus every successful comparison is counted exactly once in
$\hat{C}+C^\star$ and in $C=A\dom B$.

For each $k$, the values $\mu_{k,q}$ are non-decreasing in $q$,
so each binary search takes $O\paren{\log n} = \widetilde O \paren{1}$ time. Each inspected part contains at most $k_{A}^{\mathrm{col}}$ entries,
overall
$
\widetilde O\paren{
m_{B}
+
m_{B}\cdot k_{A}^{\mathrm{col}}
}
=
\widetilde O\paren{
m_{B}
+
\frac{m_{A}\cdot m_{B}}{n^{b}}
}
$
runtime.

Balancing $A$, constructing the Boolean matrices, and initializing
and adding the output matrices takes
$\widetilde O\paren{n^{a+b}+n^{b+c}+n^{a+c}}$ time,
which is covered by $\widetilde O\paren{n^{\omega\paren{a,b,c}}}$. Since $m_{B}\leq n^{b+c}\leq n^{\omega\paren{a,b,c}}$,
the total running time is indeed:
$$
\widetilde O\paren{
n^{\omega\paren{a,b,c}+o\paren{1}}
+
\frac{m_{A}\cdot m_{B}}{n^{b}}
}
$$

\end{proof}

\begin{example2}\label{exm:rect_dom_sparse}
Consider
$
A\in
\paren{\mathbb{R}\cup\bracce{+\infty}}^{4\times 6}
$
and
$
B\in
\paren{\mathbb{R}\cup\bracce{-\infty}}^{6\times 3}
$
given by:
$$
A=
\begin{pmatrix}
9&2&+\infty&+\infty&+\infty&+\infty\\
1&+\infty&6&+\infty&+\infty&+\infty\\
7&8&+\infty&11&+\infty&+\infty\\
3&5&4&+\infty&+\infty&+\infty
\end{pmatrix}
\qquad\text{and}\qquad
B=
\begin{pmatrix}
2&8&10\\
4&5&9\\
5&3&7\\
12&10&11\\
-\infty&-\infty&-\infty\\
-\infty&-\infty&-\infty
\end{pmatrix}
$$

The matrix $A$ contains $m_{A}=10$ finite entries. Since $A$ has $6$ columns, we have $k_{A}^{\mathrm{col}}=\ceil{\frac{10}{6}}=2$. We now compute $\cbal{A}=\paren{A',A''}$. The column parts of $A$ are:
$$
\begin{aligned}
&I_{1}^{\paren{1}}=\bracce{2,4},
&&I_{1}^{\paren{2}}=\bracce{3,1},\\
&I_{2}^{\paren{1}}=\bracce{1,4},
&&I_{2}^{\paren{2}}=\bracce{3},\\
&I_{3}^{\paren{1}}=\bracce{4,2},&&\\
&I_{4}^{\paren{1}}=\bracce{3}.&&
\end{aligned}
$$

Thus:
$
\NFC{A}=\bracce{\paren{1,1},\paren{2,1}}
$,
$
\mu_{1,1}=3
$
and
$
\mu_{2,1}=5
$. Choosing
$
\rho_{A}^{\mathrm{col}}\paren{1,1}=2
$
and
$
\rho_{A}^{\mathrm{col}}\paren{2,1}=5
$ we get:

$$
A'
=
\begin{pmatrix}
9&+\infty&+\infty&+\infty&+\infty&+\infty\\
+\infty&+\infty&6&+\infty&+\infty&+\infty\\
7&8&+\infty&11&+\infty&+\infty\\
+\infty&+\infty&4&+\infty&+\infty&+\infty
\end{pmatrix}
\qquad\text{and}\qquad
A''
=
\begin{pmatrix}
+\infty&+\infty&+\infty&+\infty&2&+\infty\\
+\infty&1&+\infty&+\infty&+\infty&+\infty\\
+\infty&+\infty&+\infty&+\infty&+\infty&+\infty\\
+\infty&3&+\infty&+\infty&5&+\infty
\end{pmatrix}
$$

Following the proof of \lemref{rect_dom_sparse}, the Boolean matrix
$\hat{A}$ is:
$$
\hat{A}
=
\begin{pmatrix}
0&0&0&0&1&0\\
0&1&0&0&0&0\\
0&0&0&0&0&0\\
0&1&0&0&1&0
\end{pmatrix}
$$

Because
$
\rho_{A}^{\mathrm{col}}\paren{1,1}=2
$
and
$
\rho_{A}^{\mathrm{col}}\paren{2,1}=5
$,  row $h=2$ of $\hat{B}$ is constructed using row
$k=1$ of $B$ and the value $\mu_{1,1}=3$, while row $h=5$ of $\hat{B}$
is constructed using row $k=2$ of $B$ and the value $\mu_{2,1}=5$. Since:
$$
B\bracke{1,\dontcare}
=
\begin{pmatrix}
2&8&10
\end{pmatrix}
\qquad\text{and}\qquad
B\bracke{2,\dontcare}
=
\begin{pmatrix}
4&5&9
\end{pmatrix}
$$

We obtain:
$$
\hat{B}\bracke{2,\dontcare}
=
\begin{pmatrix}
0&1&1
\end{pmatrix}
\qquad\text{and}\qquad
\hat{B}\bracke{5,\dontcare}
=
\begin{pmatrix}
0&1&1
\end{pmatrix}
$$

All other rows of $\hat{B}$ are zero, since no non-final part is assigned to
columns $1,3,4$ or $6$ of $A''$. Therefore:
$$
\hat{B}
=
\begin{pmatrix}
0&0&0\\
0&1&1\\
0&0&0\\
0&0&0\\
0&1&1\\
0&0&0
\end{pmatrix}
$$

Ergo, computing $\hat{C}
=
\hat{A}\cdot\hat{B}$, we get:

$$
\hat{C}
=\begin{pmatrix}
0&0&0&0&1&0\\
0&1&0&0&0&0\\
0&0&0&0&0&0\\
0&1&0&0&1&0
\end{pmatrix} \cdot \begin{pmatrix}
0&0&0\\
0&1&1\\
0&0&0\\
0&0&0\\
0&1&1\\
0&0&0
\end{pmatrix} = 
\begin{pmatrix}
0&1&1\\
0&1&1\\
0&0&0\\
0&2&2
\end{pmatrix}
$$

The entry $\hat{C}\bracke{i,j}$ counts the non-final parts
$I_{k}^{\paren{q}}$ such that $i\in I_{k}^{\paren{q}}$ and
$\mu_{k,q}\leq B\bracke{k,j}$. 
For example, consider the entry
$
\hat{C}\bracke{4,2}=2
$.
The first contribution corresponds to
$
h=2
$.
Since
$\rho_{A}^{\mathrm{col}}\paren{1,1}=2$, column $2$ of $A''$
stores the non-final part
$I_{1}^{\paren{1}}=\bracce{2,4}$ from column $1$ of $A$.
Because $4\in I_{1}^{\paren{1}}$, we have
$\hat{A}\bracke{4,2}=1$. Moreover,
$\mu_{1,1}=3\leq B\bracke{1,2}=8$, and therefore
$\hat{B}\bracke{2,2}=1$. The second contribution corresponds to $h=5$. Since
$\rho_{A}^{\mathrm{col}}\paren{2,1}=5$, column $5$ of $A''$
stores the non-final part
$I_{2}^{\paren{1}}=\bracce{1,4}$ from column $2$ of $A$.
Because $4\in I_{2}^{\paren{1}}$, we have
$\hat{A}\bracke{4,5}=1$. Moreover,
$\mu_{2,1}=5\leq B\bracke{2,2}=5$, and therefore
$\hat{B}\bracke{5,2}=1$. Overall, the two terms corresponding to $h=2$ and $h=5$ contribute
to $\hat{C}\bracke{4,2}$, and hence
$\hat{C}\bracke{4,2}=2$.

$\hat{C}$ does not count comparisons where the relevant
non-final part is not completely dominated. 
We now define $C^\star$ as the matrix containing the additional
contributions obtained from the explicit inspections. We initialize
$
C^\star
$
to be full of zeros.

Consider $B\bracke{1,1}=2$. Since $\mu_{1,1}=3>2$, the non-final part
$I_{1}^{\paren{1}}=\bracce{2,4}$ is not counted by $\hat{C}$ and must
be inspected explicitly. The inspection gives
$A\bracke{2,1}=1\leq 2$ and $A\bracke{4,1}=3>2$.
Therefore, $C^\star\bracke{2,1}$ is increased by one, while
$C^\star\bracke{4,1}$ remains unchanged. The following part
$I_{1}^{\paren{2}}=\bracce{3,1}$
is the final part of column $1$. It does not contribute in this case,
since all its entries are larger than $B\bracke{1,1}=2$.

As another case, consider $B\bracke{1,2}=8$. The non-final part
$I_{1}^{\paren{1}}$ is already completely dominated, since
$\mu_{1,1}=3\leq B\bracke{1,2}=8$, and its contribution is already
included in $\hat{C}$. Since there is no additional non-final part in
column $1$, we inspect the final part
$I_{1}^{\paren{2}}=\bracce{3,1}$.
The inspection gives
$A\bracke{3,1}=7\leq 8$ and
$A\bracke{1,1}=9>8$. Hence, $C^\star\bracke{3,2}$ is increased by one.

We repeat this 
inspection procedure for every finite entry of $B$, resulting in the matrix:

$$
C^\star
=
\begin{pmatrix}
1&0&1\\
1&0&1\\
1&1&3\\
1&0&1
\end{pmatrix}
$$

By construction, every comparison
$A\bracke{i,k}\leq B\bracke{k,j}$ is counted exactly once:
comparisons belonging to completely dominated non-final parts are
counted in $\hat{C}$, while all remaining comparisons are counted in
$C^\star$. Therefore, $C=\hat{C}+C^\star$. Hence:

$$
\begin{aligned}
\hat{C}+C^\star
&=
\begin{pmatrix}
0&1&1\\
0&1&1\\
0&0&0\\
0&2&2
\end{pmatrix}
+
\begin{pmatrix}
1&0&1\\
1&0&1\\
1&1&3\\
1&0&1
\end{pmatrix}
=
\begin{pmatrix}
1&1&2\\
1&1&2\\
1&1&3\\
1&2&3
\end{pmatrix}
\\
&=
\begin{pmatrix}
9&2&+\infty&+\infty&+\infty&+\infty\\
1&+\infty&6&+\infty&+\infty&+\infty\\
7&8&+\infty&11&+\infty&+\infty\\
3&5&4&+\infty&+\infty&+\infty
\end{pmatrix}
\dom
\begin{pmatrix}
2&8&10\\
4&5&9\\
5&3&7\\
12&10&11\\
-\infty&-\infty&-\infty\\
-\infty&-\infty&-\infty
\end{pmatrix}
\\
&=
A\dom B=C
\end{aligned}
$$

\end{example2}

The algorithm in \lemref{rect_dom_sparse} runs in
$\widetilde O\paren{n^{\omega\paren{a,b,c}+o\paren{1}}}$ time when the matrices are relatively sparse, that is 
$m_{A}\cdot m_{B}\leq n^{b+\omega\paren{a,b,c}}$. For denser matrices, we extend another result of Duan and Pettie
\cite{DuaPet2009} to fully rectangular matrices.
We partition the finite entries of $A$ into groups, using
\mmname-products for completely dominated groups and
\lemref{rect_dom_sparse} for the remaining comparisons.

\begin{lemma}\label{lem:rect_dom_dense}
Let $A\in\paren{\mathbb{R}\cup\bracce{+\infty}}^{n^{a}\times n^{b}}$
and $B\in\paren{\mathbb{R}\cup\bracce{-\infty}}^{n^{b}\times n^{c}}$,
and let $m_{A}$ and $m_{B}$ denote their respective numbers of finite
entries. Assume that $m_{A}\cdot m_{B}\geq n^{b+\omega\paren{a,b,c}}$. Then 
$A\dom B$ can be computed within
$\widetilde O\paren{
\sqrt{m_{A}\cdot m_{B}}\cdot
n^{\frac{\omega\paren{a,b,c}-b}{2}+o\paren{1}}
}$ runtime.
\end{lemma}

\begin{proof}
Fix $t\in \bracke{ m_{A}}$ to be determined later.
Sort the finite entries of $A$ in non-decreasing order, breaking
ties arbitrarily, and divide the sorted list into $t$ nonempty
consecutive groups of at most $\ceil{\frac{m_{A}}{t}}$ entries.
For each $r\in\bracke{t}$, let $L_{r}$ denote the $r^{\textnormal{th}}$ group
of finite entries of $A$, and let $\ell_{r}$ and $\ess_{r}$
be the largest and smallest values in $L_{r}$, respectively.
Define:
$$
A_{r}\bracke{i,k}
=
\begin{cases}
A\bracke{i,k}
& \begin{array}{l}
A\bracke{i,k}\in L_{r}
\end{array}\\
+\infty & \text{otherwise}
\end{cases}
\qquad\textnormal{and}\qquad
B_{r}\bracke{k,j}
=
\begin{cases}
B\bracke{k,j} &  \ess_{r}\leq B\bracke{k,j}<\ell_{r}\\
-\infty & \text{otherwise}
\end{cases}
$$

In addition, we construct the Boolean matrices $\hat{A}_{r}$ and $\hat{B}_{r}$:
$$
\hat{A}_{r}\bracke{i,k}
=
\begin{cases}
1 & A_{r}\bracke{i,k}<+\infty\\
0 & \text{otherwise}
\end{cases}
\qquad\textnormal{and}\qquad
\hat{B}_{r}\bracke{k,j}
=
\begin{cases}
1 & B\bracke{k,j}\geq \ell_{r}\\
0 & \text{otherwise}
\end{cases}
$$

For every $r\in\bracke{t}$, compute:
$
\hat{C}_{r}=\hat{A}_{r}\cdot\hat{B}_{r}
$ and $
C_{r}=A_{r}\dom B_{r}
$ and 
return:
$$
C=\msum{\paren{\hat{C}_{r}+C_{r}}}{r=1}{t}
$$

Consider a finite entry $A\bracke{i,k}$ belonging to $L_{r}$
and satisfying $A\bracke{i,k}\leq B\bracke{k,j}$.
If $B\bracke{k,j}\geq\ell_{r}$, this comparison is counted
in $\hat{C}_{r}\bracke{i,j}$.
Otherwise,
$\ess_{r}\leq A\bracke{i,k}\leq B\bracke{k,j}<\ell_{r}$,
so it is counted in $C_{r}\bracke{i,j}$.
Both matrices count only successful comparisons, and these
two cases are disjoint.
Since every finite entry of $A$ belongs to exactly one group,
every successful comparison is counted exactly once.
Hence precisely   $C=A\dom B$.

We now consider the runtime. Let $m_{B_{r}}$ denote the number of finite entries of $B_{r}$.
Since $\ell_{r}\leq\ess_{r+1}$, each finite entry of $B$ appears
in at most one $B_{r}$ and if $\ess_{r}=\ell_{r}$ then $B_{r}$ has no finite entries. Thus $\msum{m_{B_{r}}}{r=1}{t}\leq m_{B}$. Computing all $\hat{C}_{r}$ takes
$\widetilde O\paren{t\cdot n^{\omega\paren{a,b,c}+o\paren{1}}}$ time.
Applying \lemref{rect_dom_sparse} to compute all the matrices $C_{r}$ takes:
$$
\widetilde O\paren{
\msum{
\paren{
n^{\omega\paren{a,b,c}+o\paren{1}}
+
\frac{\ceil{\frac{m_{A}}{t}}\cdot m_{B_{r}}}{n^{b}}
}
}{r=1}{t}
}
$$
runtime. Sorting the entries, constructing the matrices and adding their
products can be done within a runtime of $\widetilde O \paren{t\cdot \paren{n^{a+b}+n^{b+c}+n^{a+c}}} \subseteq \widetilde O \paren{t\cdot n^{\omega\paren{a,b,c}+o\paren{1}}}$. Combined with the runtime to compute $\hat{C}_r$, the total runtime remains within:

$$
\widetilde O\paren{
t\cdot n^{\omega\paren{a,b,c}+o\paren{1}}
+
\msum{
\frac{\ceil{\frac{m_{A}}{t}}\cdot m_{B_{r}}}{n^{b}}
}{r=1}{t}
}
=
\widetilde O\paren{
t\cdot n^{\omega\paren{a,b,c}+o\paren{1}}
+\frac{m_{A}\cdot m_{B}}{t\cdot n^{b}}
}
$$

All that remains is balancing the terms. We choose:
$t=\ceil{\sqrt{
\frac{m_{A}\cdot m_{B}}{n^{b+\omega\paren{a,b,c}}}
}}$.
The assumption $m_{A}\cdot m_{B}\geq n^{b+\omega\paren{a,b,c}}$
ensures that the expression inside the ceiling is at least $1$.
Since $m_{B}\leq n^{b+c}\leq n^{b+\omega\paren{a,b,c}}$,
we also have: $t\leq \ceil{\sqrt{m_{A}}}\leq m_{A}$. Therefore $t\in\bracke{ m_{A}}$ and substituting this $t$ yields a total runtime of:

$$\widetilde O\paren{
\sqrt{m_{A}\cdot m_{B}}\cdot
n^{\frac{\omega\paren{a,b,c}-b}{2}+o\paren{1}}
}$$

\end{proof}

We now use \lemref{rect_dom_sparse} and row balancing to extend
the \maxminname-product algorithm of Duan and Pettie
\cite{DuaPet2009} to fully rectangular matrices.
For each pair $i,j$ we use \domname-products to identify the last range
containing an entry $A\bracke{i,k}$ satisfying
$A\bracke{i,k}\leq B\bracke{k,j}$.
Within this range, row balancing and further \domname-products
identify the last part of the $i^{\textnormal{th}}$ row containing
such an entry.
We inspect the $O\paren{\frac{n^{b}}{t}}$ entries of this part
one by one and select the largest $A\bracke{i,k}$ satisfying
$A\bracke{i,k}\leq B\bracke{k,j}$.

\begin{lemma}\label{lem:rect_maxmin}
Let $A\in\mathbb{R}^{n^{a}\times n^{b}}$ and
$B\in\mathbb{R}^{n^{b}\times n^{c}}$.
For every  $ t\in\bracke{ n^{b}}$ we can compute the product $A\maxmin B$
 within
$\widetilde O\paren{
t\cdot n^{\omega\paren{a,b,c}+o\paren{1}}
+\frac{n^{a+b+c}}{t}
}$ runtime.
\end{lemma}

\begin{proof}
To break ties, we replace the entries by their ranks in the order fixed at the
beginning of this section. We continue to denote the resulting
matrices by $A$ and $B$, and recover the original values by the end. By \obvref{maxmin_options}, it suffices to compute the matrices
$\overset{\scriptscriptstyle\leq}{C}$ and
$\overset{\scriptscriptstyle\geq}{C}$ defined by:
$$
\overset{\scriptscriptstyle\leq}{C}\bracke{i,j}
=
\mmax{
A\bracke{i,k}
\midline
A\bracke{i,k}\leq B\bracke{k,j}
}{
k\in\bracke{n^{b}}
}
\qquad\textnormal{and}\qquad
\overset{\scriptscriptstyle\geq}{C}\bracke{i,j}
=
\mmax{
B\bracke{k,j}
\midline
A\bracke{i,k}\geq B\bracke{k,j}
}{
k\in\bracke{n^{b}}
}
$$
We first compute $\overset{\scriptscriptstyle\leq}{C}$.  Let
$t_{A}=\ceil{\frac{n^{a+b}}{t}}$ and
$t_{B}=\ceil{\frac{n^{b+c}}{t}}$. Sort all entries of $A$ and $B$ together in  an increasing order.
For each $r\in \bracke{2\ceil{\mmax{
\frac{n^{a+b}}{t_{A}},
\frac{n^{b+c}}{t_{B}}}
}}$, take the next entries in this order until either
$t_{A}$ entries from $A$ or $t_{B}$ entries from $B$ have been taken,
or the list ends.
Store the selected entries of $A$ in $A_{r}$ and those of $B$ in
$B_{r}$, preserving their original row and column indices.
The matrices $A_{r},B_{r}$ have the same dimensions as $A,B$,
respectively, with all remaining entries set to $+\infty$ in $A_{r}$
and $-\infty$ in $B_{r}$.  Lastly, let $L_{r}$ be the set of finite values stored in $A_{r}$ and $B_{r}$.

Let $\rbal{A_{r}}=\paren{A'_{r},A''_{r}}$.
Since $m_{A_{r}}\leq t_{A}$, each row part of $A_{r}$ has at most
$k_{A_{r}}^{\mathrm{row}}
=\ceil{\frac{m_{A_{r}}}{n^{a}}}
\leq\ceil{\frac{n^{b}}{t}}
\in O\paren{\frac{n^{b}}{t}}$
entries. Let $J_{r,i}^{\paren{q}}$ denote the set of column indices
in the $q^{\textnormal{th}}$ part of row $i$ in $A_{r}$.
The final part remains in row $i$ of $A'_{r}$, while each non-final
part $J_{r,i}^{\paren{q}}$ is stored  in row
$\rho_{A_{r}}^{\mathrm{row}}\paren{i,q}$ of $A''_{r}$.

For every $r$ we compute the matrices $\overset{\scriptscriptstyle\leq}{C}_{r}=A_{r}\dom B$,
$\overset{\scriptscriptstyle\leq}{C'}_{r}=A'_{r}\dom B$, and
$\overset{\scriptscriptstyle\leq}{C''}_{r}=A''_{r}\dom B$.
For each pair $i,j$, we use
$\overset{\scriptscriptstyle\leq}{C}_{r}\bracke{i,j}$
to identify the last range containing an entry $A\bracke{i,k}$
satisfying $A\bracke{i,k}\leq B\bracke{k,j}$.
Within this range,
$\overset{\scriptscriptstyle\leq}{C'}_{r}\bracke{i,j}$
counts these entries in the final part of row $i$.
If this count is zero, we use
$\overset{\scriptscriptstyle\leq}{C''}_{r}
\bracke{\rho_{A_{r}}^{\mathrm{row}}\paren{i,q},j}$
to identify the last non-final part containing an entry
satisfying the inequality.

To compute these matrices, we separate comparisons with entries of
$B$ in $L_{r}$ from comparisons with entries in later ranges.
We construct the Boolean matrices
$\hat{A}_{r},\hat{A}'_{r},\hat{A}''_{r}
\in\bracce{0,1}^{n^{a}\times n^{b}}$ and
$\hat{B}_{r}\in\bracce{0,1}^{n^{b}\times n^{c}}$ by setting:

$$
\hat{A}_{r}\bracke{i,k}
=
\begin{cases}
1 & A_{r}\bracke{i,k}<+\infty\\
0 & \text{otherwise}
\end{cases}
\qquad\textnormal{and}\qquad
\hat{B}_{r}\bracke{k,j}
=
\begin{cases}
1 & B\bracke{k,j}\in\mcup{L_{h}}{h>r}{}\\
0 & \text{otherwise}
\end{cases}
$$
$$
\hat{A}'_{r}\bracke{i,k}
=
\begin{cases}
1 & A'_{r}\bracke{i,k}<+\infty\\
0 & \text{otherwise}
\end{cases}
\qquad\textnormal{and}\qquad
\hat{A}''_{r}\bracke{i,k}
=
\begin{cases}
1 & A''_{r}\bracke{i,k}<+\infty\\
0 & \text{otherwise}
\end{cases}
$$
We now compute the three matrices using:
$$
\begin{aligned}
\overset{\scriptscriptstyle\leq}{C}_{r}
&=A_{r}\dom B_{r}+\hat{A}_{r}\cdot\hat{B}_{r}\\
\overset{\scriptscriptstyle\leq}{C'}_{r}
&=A'_{r}\dom B_{r}+\hat{A}'_{r}\cdot\hat{B}_{r}\\
\overset{\scriptscriptstyle\leq}{C''}_{r}
&=A''_{r}\dom B_{r}+\hat{A}''_{r}\cdot\hat{B}_{r}
\end{aligned}
$$

In each identity, the \domname-product with $B_{r}$ counts the
comparisons with entries of $B$ in $L_{r}$.
Every entry of $B$ in a later range is larger than every finite
entry of $A_{r},A'_{r},A''_{r}$, so the corresponding
\mmname-product counts all comparisons with later ranges.
Entries of $B$ in earlier ranges contribute zero.
Thus each identity gives the required count against the full
matrix $B$.

Each of $A_{r},A'_{r},A''_{r}$ has at most $t_{A}$ finite entries,
and $B_{r}$ has at most $t_{B}$ finite entries.
Applying \lemref{rect_dom_sparse} to the three \domname-products
and computing the three \mmname-products therefore takes:
$$
\widetilde O\paren{
n^{\omega\paren{a,b,c}+o\paren{1}}+\frac{t_{A}\cdot t_{B}}{n^{b}}
}
\subseteq
\widetilde O\paren{
n^{\omega\paren{a,b,c}+o\paren{1}}+\frac{n^{a+b+c}}{t^{2}}
}
$$
time per range. 

For each fixed pair $i,j$, scan all range indices $r$ and select
the largest one satisfying
$\overset{\scriptscriptstyle\leq}{C}_{r}\bracke{i,j}>0$. If no such $r$ exists, set
$\overset{\scriptscriptstyle\leq}{C}\bracke{i,j}=-\infty$
and continue to the next pair.

For the selected range $r$, we now locate the last part of the
$i^{\textnormal{th}}$ row of $A_{r}$ containing an entry
$A\bracke{i,k}$ satisfying $A\bracke{i,k}\leq B\bracke{k,j}$.
If $\overset{\scriptscriptstyle\leq}{C'}_{r}\bracke{i,j}>0$,
inspect the final part of the $i^{\textnormal{th}}$ row.
This part contains an entry satisfying the inequality, and its
values are larger than those in all earlier parts.
Otherwise, the final part contains no entry satisfying the
inequality. Since
$\overset{\scriptscriptstyle\leq}{C}_{r}\bracke{i,j}>0$,
the required entry must lie in a non-final part.
Scan the non-final parts of the $i^{\textnormal{th}}$ row and
select the largest $q$ satisfying
$\overset{\scriptscriptstyle\leq}{C''}_{r}
\bracke{\rho_{A_{r}}^{\mathrm{row}}\paren{i,q},j}>0$.

Now, inspect the entries whose column indices belong to
$J_{r,i}^{\paren{q}}$.
Among the inspected entries, select the largest
$A\bracke{i,k}$ satisfying
$A\bracke{i,k}\leq B\bracke{k,j}$ and store its value in
$\overset{\scriptscriptstyle\leq}{C}\bracke{i,j}$. By the choice of $r$, no later range contains an entry of the
$i^{\textnormal{th}}$ row satisfying
$A\bracke{i,k}\leq B\bracke{k,j}$.
Within range $r$, the inspected part is the last part of this
row containing an entry satisfying the inequality.
Since the ranges and the row parts are ordered by value,
the inspected part contains the largest $A\bracke{i,k}$
satisfying $A\bracke{i,k}\leq B\bracke{k,j}$.
Thus we obtain $\overset{\scriptscriptstyle\leq}{C}$.

We now consider the runtime.
For each pair $i,j$, selecting $r$ takes $O\paren{t}$ time,
since there are $O\paren{t}$ ranges.
Over all pairs, this takes $O\paren{t\cdot n^{a+c}}$ time.

To bound the time spent selecting $q$, recall that each non-final
part of $A_{r}$ is stored in a distinct row of $A''_{r}$.
Since $A''_{r}$ has $n^{a}$ rows, there are at most $n^{a}$
non-final parts in each $A_{r}$.
Each part is tested for at most $n^{c}$ values of $j$,
so scanning these parts over all ranges takes
$O\paren{t\cdot n^{a+c}}$ time.
Each inspected part contains at most
$\ceil{\frac{n^{b}}{t}}$ entries.
Hence inspecting the selected parts over all pairs takes
$O\paren{\frac{n^{a+b+c}}{t}}$ time.

Additionally, sorting the entries, balancing the rows and constructing the
matrices requires an additional runtime of $
\widetilde O\paren{
t\cdot\paren{n^{a+b}+n^{b+c}+n^{a+c}}
}
\subseteq
\widetilde O\paren{t\cdot n^{\omega\paren{a,b,c}+o\paren{1}}}$.
Overall, we compute
$\overset{\scriptscriptstyle\leq}{C}$ within a runtime of 
$\widetilde O\paren{
t\cdot n^{\omega\paren{a,b,c}+o\paren{1}}
+\frac{n^{a+b+c}}{t}
}
$. 

A symmetrical argument follows for computing   $\overset{\scriptscriptstyle\geq}{C}$.  Instead of $A$ and $B$ consider $B^{\mathsf{T}}$ and $A^{\mathsf{T}}$.  This exchanges $a$ and $c$, so the runtime remains the same
since $\omega\paren{c,b,a}=\omega\paren{a,b,c}$.

Finally, we iterate over each pair $i,j$, setting: $C\bracke{i,j}
=
\mmax{
\overset{\scriptscriptstyle\leq}{C}\bracke{i,j},
\overset{\scriptscriptstyle\geq}{C}\bracke{i,j}
}$. 
Replace each entry of $C$ by its corresponding original input value before the tie breaking.
Since this replacement preserves minimum and maximum,
\obvref{maxmin_options} implies that the resulting matrix is the
\maxminname-product of the original inputs.

This final step simply takes $O\paren{n^{a+c}}$ time, so the total
runtime remains
$\widetilde O\paren{
t\cdot n^{\omega\paren{a,b,c}+o\paren{1}}
+\frac{n^{a+b+c}}{t}
}$, as required.

\end{proof}

The following example shows how the three  matrices determine the last range and row part to inspect:

\begin{example2}\label{exm:rect_maxmin}
Consider $A\in\mathbb{R}^{3 \times 5}$ and $B\in\mathbb{R}^{5 \times 4}$ given by: 

$$
A=
\begin{pmatrix}
9&2&7&4&5\\
28&20&26&22&24\\
18&30&12&33&14
\end{pmatrix}
\qquad\textnormal{and}\qquad
B=
\begin{pmatrix}
1&29&8&10\\
31&11&13&15\\
3&16&17&19\\
23&21&25&35\\
6&27&32&34
\end{pmatrix}
$$

To make things simpler, all entries are distinct. Take $t=3$. Then
$t_{A}=\ceil{\frac{15}{3}}=5$ and
$t_{B}=\ceil{\frac{20}{3}}=7$.
The nonempty ranges are:

$$
\begin{array}{c|c|c}
r & \textnormal{finite values in $A_{r}$}
  & \textnormal{finite values in $B_{r}$}\\
\hline
1 & \bracce{2,4,5,7,9} & \bracce{1,3,6,8}\\
2 & \bracce{12,14,18} & \bracce{10,11,13,15,16,17,19}\\
3 & \bracce{20,22,24,26,28} & \bracce{21,23,25,27}\\
4 & \bracce{30,33} & \bracce{29,31,32,34,35}
\end{array}
$$

The first and third ranges end at $9$ and $28$, respectively, upon
taking $5$ entries from $A$. The second ends at $19$, upon taking
$7$ entries from $B$, and the fourth ends when the list is exhausted. 
We follow the construction in the proof of \lemref{rect_maxmin},
first computing $\overset{\scriptscriptstyle\leq}{C}$.
For $r=1$, we have:
$$
A_{1}=
\begin{pmatrix}
9&2&7&4&5\\
+\infty&+\infty&+\infty&+\infty&+\infty\\
+\infty&+\infty&+\infty&+\infty&+\infty
\end{pmatrix}
\qquad\textnormal{and}\qquad
B_{1}=
\begin{pmatrix}
1&-\infty&8&-\infty\\
-\infty&-\infty&-\infty&-\infty\\
3&-\infty&-\infty&-\infty\\
-\infty&-\infty&-\infty&-\infty\\
6&-\infty&-\infty&-\infty
\end{pmatrix}
$$

Since $m_{A_{1}}=5$, we have
$k_{A_{1}}^{\mathrm{row}}=\ceil{\frac{5}{3}}=2$.
The parts of the first row are:
$$
J_{1,1}^{\paren{1}}=\bracce{2,4},
\qquad
J_{1,1}^{\paren{2}}=\bracce{5,3},
\qquad
J_{1,1}^{\paren{3}}=\bracce{1}
$$

The first part contain the values (and coincidentally, the indices) $2,4$; the second part contains the values $5,7$;
 the final part contains only $9$.
Choosing $\rho_{A_{1}}^{\mathrm{row}}\paren{1,1}=2$ and
$\rho_{A_{1}}^{\mathrm{row}}\paren{1,2}=3$ gives
$\rbal{A_{1}}=\paren{A'_{1},A''_{1}}$, where:
$$
A'_{1}=
\begin{pmatrix}
9&+\infty&+\infty&+\infty&+\infty\\
+\infty&+\infty&+\infty&+\infty&+\infty\\
+\infty&+\infty&+\infty&+\infty&+\infty
\end{pmatrix}
\qquad\textnormal{and}\qquad
A''_{1}=
\begin{pmatrix}
+\infty&+\infty&+\infty&+\infty&+\infty\\
+\infty&2&+\infty&4&+\infty\\
+\infty&+\infty&7&+\infty&5
\end{pmatrix}
$$

The Boolean matrices $\hat{A}_{1}$ and $\hat{B}_{1}$ are:
$$
\hat{A}_{1}=
\begin{pmatrix}
1&1&1&1&1\\
0&0&0&0&0\\
0&0&0&0&0
\end{pmatrix}
\qquad\textnormal{and}\qquad
\hat{B}_{1}=
\begin{pmatrix}
0&1&0&1\\
1&1&1&1\\
0&1&1&1\\
1&1&1&1\\
0&1&1&1
\end{pmatrix}
$$

Recall that $\hat{B}_{1}\bracke{k,j}=1$ precisely when $B\bracke{k,j}$
belongs to a range after $L_{1}$. The matrices $\hat{A}'_{1}$ and $\hat{A}''_{1}$ are obtained from
$A'_{1}$ and $A''_{1}$ by replacing finite entries by $1$ and all
other entries by $0$:

$$
\hat{A}'_{1}=
\begin{pmatrix}
1&0&0&0&0\\
0&0&0&0&0\\
0&0&0&0&0
\end{pmatrix}
\qquad\textnormal{and}\qquad
\hat{A}''_{1}=
\begin{pmatrix}
0&0&0&0&0\\
0&1&0&1&0\\
0&0&1&0&1
\end{pmatrix}
$$

Ergo, the three matrices are computed as follows:
$$
\begin{aligned}
\overset{\scriptscriptstyle\leq}{C}_{1}
&=A_{1}\dom B_{1}+\hat{A}_{1}\cdot\hat{B}_{1}\\
&=
\begin{pmatrix}
1&0&0&0\\
0&0&0&0\\
0&0&0&0
\end{pmatrix}
+
\begin{pmatrix}
2&5&4&5\\
0&0&0&0\\
0&0&0&0
\end{pmatrix}
=
\begin{pmatrix}
3&5&4&5\\
0&0&0&0\\
0&0&0&0
\end{pmatrix}
\end{aligned}
$$
$$
\begin{aligned}
\overset{\scriptscriptstyle\leq}{C'}_{1}
&=A'_{1}\dom B_{1}+\hat{A}'_{1}\cdot\hat{B}_{1}\\
&=
\begin{pmatrix}
0&0&0&0\\
0&0&0&0\\
0&0&0&0
\end{pmatrix}
+
\begin{pmatrix}
0&1&0&1\\
0&0&0&0\\
0&0&0&0
\end{pmatrix}
=
\begin{pmatrix}
0&1&0&1\\
0&0&0&0\\
0&0&0&0
\end{pmatrix}
\end{aligned}
$$
$$
\begin{aligned}
\overset{\scriptscriptstyle\leq}{C''}_{1}
&=A''_{1}\dom B_{1}+\hat{A}''_{1}\cdot\hat{B}_{1}\\
&=
\begin{pmatrix}
0&0&0&0\\
0&0&0&0\\
1&0&0&0
\end{pmatrix}
+
\begin{pmatrix}
0&0&0&0\\
2&2&2&2\\
0&2&2&2
\end{pmatrix}
=
\begin{pmatrix}
0&0&0&0\\
2&2&2&2\\
1&2&2&2
\end{pmatrix}
\end{aligned}
$$

In each equality, the \domname-product counts comparisons with
entries of $B$ in $L_{1}$, while the \mmname-product counts
comparisons with entries in $L_{2}$ to $L_{4}$.

For example, consider
$\overset{\scriptscriptstyle\leq}{C}_{1}\bracke{1,1}=3$.
The product $A_{1}\dom B_{1}$ counts
$A\bracke{1,5}=5\leq B\bracke{5,1}=6$, with both entries in $L_{1}$.
The product $\hat{A}_{1}\cdot\hat{B}_{1}$ counts
$A\bracke{1,2}=2\leq B\bracke{2,1}=31$ and
$A\bracke{1,4}=4\leq B\bracke{4,1}=23$, whose entries from $B$
belong to later ranges. Accumulating these counts gives $1+2=3$.

The matrices for the remaining nonempty ranges are:
$$
A_{2}=
\begin{pmatrix}
+\infty&+\infty&+\infty&+\infty&+\infty\\
+\infty&+\infty&+\infty&+\infty&+\infty\\
18&+\infty&12&+\infty&14
\end{pmatrix}
\qquad\textnormal{and}\qquad
B_{2}=
\begin{pmatrix}
-\infty&-\infty&-\infty&10\\
-\infty&11&13&15\\
-\infty&16&17&19\\
-\infty&-\infty&-\infty&-\infty\\
-\infty&-\infty&-\infty&-\infty
\end{pmatrix}
$$
$$
A_{3}=
\begin{pmatrix}
+\infty&+\infty&+\infty&+\infty&+\infty\\
28&20&26&22&24\\
+\infty&+\infty&+\infty&+\infty&+\infty
\end{pmatrix}
\qquad\textnormal{and}\qquad
B_{3}=
\begin{pmatrix}
-\infty&-\infty&-\infty&-\infty\\
-\infty&-\infty&-\infty&-\infty\\
-\infty&-\infty&-\infty&-\infty\\
23&21&25&-\infty\\
-\infty&27&-\infty&-\infty
\end{pmatrix}
$$
$$
A_{4}=
\begin{pmatrix}
+\infty&+\infty&+\infty&+\infty&+\infty\\
+\infty&+\infty&+\infty&+\infty&+\infty\\
+\infty&30&+\infty&33&+\infty
\end{pmatrix}
\qquad\textnormal{and}\qquad
B_{4}=
\begin{pmatrix}
-\infty&29&-\infty&-\infty\\
31&-\infty&-\infty&-\infty\\
-\infty&-\infty&-\infty&-\infty\\
-\infty&-\infty&-\infty&35\\
-\infty&-\infty&32&34
\end{pmatrix}
$$
Each matrix retains the entries belonging to its range at their original positions. We now balance the rows of $A_{2},A_{3},A_{4}$. In each case, the final part stays in its original row of $A'_{r}$, while each non-final part is stored in its assigned row of $A''_{r}$. The column indices remain unchanged.

For $r=2$, we have $m_{A_{2}}=3$ and $k_{A_{2}}^{\mathrm{row}}=\ceil{\frac{3}{3}}=1$. All three finite entries belong to the third row. Sorting them gives the column indices $3,5,1$, so the parts are:
$$
J_{2,3}^{\paren{1}}=\bracce{3},
\qquad
J_{2,3}^{\paren{2}}=\bracce{5},
\qquad
J_{2,3}^{\paren{3}}=\bracce{1}
$$
These parts contain the values $12$, $14$ and $18$, respectively. We choose $\rho_{A_{2}}^{\mathrm{row}}\paren{3,1}=1$ and $\rho_{A_{2}}^{\mathrm{row}}\paren{3,2}=2$. Thus the non-final parts containing $12$ and $14$ are stored in rows $1$ and $2$ of $A''_{2}$, respectively, while the final part containing $18$ remains in row $3$ of $A'_{2}$. We obtain:
$$
A'_{2}=
\begin{pmatrix}
+\infty&+\infty&+\infty&+\infty&+\infty\\
+\infty&+\infty&+\infty&+\infty&+\infty\\
18&+\infty&+\infty&+\infty&+\infty
\end{pmatrix}
\qquad\textnormal{and}\qquad
A''_{2}=
\begin{pmatrix}
+\infty&+\infty&12&+\infty&+\infty\\
+\infty&+\infty&+\infty&+\infty&14\\
+\infty&+\infty&+\infty&+\infty&+\infty
\end{pmatrix}
$$

For $r=3$, we have $m_{A_{3}}=5$ and $k_{A_{3}}^{\mathrm{row}}=\ceil{\frac{5}{3}}=2$. The finite entries belong to the second row, and their column indices in increasing order of value are $2,4,5,3,1$. The parts are:
$$
J_{3,2}^{\paren{1}}=\bracce{2,4},
\qquad
J_{3,2}^{\paren{2}}=\bracce{5,3},
\qquad
J_{3,2}^{\paren{3}}=\bracce{1}
$$
The two non-final parts contain the values $20,22$ and $24,26$, respectively, while the final part contains $28$. Choosing $\rho_{A_{3}}^{\mathrm{row}}\paren{2,1}=1$ and $\rho_{A_{3}}^{\mathrm{row}}\paren{2,2}=2$ stores the first two parts in rows $1$ and $2$ of $A''_{3}$. The final part remains in row $2$ of $A'_{3}$. Hence:
$$
A'_{3}=
\begin{pmatrix}
+\infty&+\infty&+\infty&+\infty&+\infty\\
28&+\infty&+\infty&+\infty&+\infty\\
+\infty&+\infty&+\infty&+\infty&+\infty
\end{pmatrix}
\qquad\textnormal{and}\qquad
A''_{3}=
\begin{pmatrix}
+\infty&20&+\infty&22&+\infty\\
+\infty&+\infty&26&+\infty&24\\
+\infty&+\infty&+\infty&+\infty&+\infty
\end{pmatrix}
$$

For $r=4$, we have $m_{A_{4}}=2$ and $k_{A_{4}}^{\mathrm{row}}=\ceil{\frac{2}{3}}=1$. The third row has two parts:
$$
J_{4,3}^{\paren{1}}=\bracce{2}
\qquad\textnormal{and}\qquad
J_{4,3}^{\paren{2}}=\bracce{4}
$$
The first part contains $30$ and is non-final, while the final part contains $33$. We choose $\rho_{A_{4}}^{\mathrm{row}}\paren{3,1}=1$, so $30$ is stored in row $1$ of $A''_{4}$, in its original column $2$. The value $33$ remains in row $3$, column $4$ of $A'_{4}$. Thus:
$$
A'_{4}=
\begin{pmatrix}
+\infty&+\infty&+\infty&+\infty&+\infty\\
+\infty&+\infty&+\infty&+\infty&+\infty\\
+\infty&+\infty&+\infty&33&+\infty
\end{pmatrix}
\qquad\textnormal{and}\qquad
A''_{4}=
\begin{pmatrix}
+\infty&30&+\infty&+\infty&+\infty\\
+\infty&+\infty&+\infty&+\infty&+\infty\\
+\infty&+\infty&+\infty&+\infty&+\infty
\end{pmatrix}
$$

For each $r\in\bracce{2,3,4}$, we use the same three calculations:
$$
\left\{
\begin{aligned}
\overset{\scriptscriptstyle\leq}{C}_{r}
&=A_{r}\dom B_{r}+\hat{A}_{r}\cdot\hat{B}_{r}\\[0.4em]
\overset{\scriptscriptstyle\leq}{C'}_{r}
&=A'_{r}\dom B_{r}+\hat{A}'_{r}\cdot\hat{B}_{r}\\[0.4em]
\overset{\scriptscriptstyle\leq}{C''}_{r}
&=A''_{r}\dom B_{r}+\hat{A}''_{r}\cdot\hat{B}_{r}
\end{aligned}
\right.
$$

Each time, the \domname-products with $B_{r}$ count comparisons within $L_{r}$ while the \mmname-products count comparisons with entries of $B$ in later ranges. For $r=4$, there is no later range, so $\hat{B}_{4}$ is the zero matrix. Overall, the resulting matrices count comparisons against the full matrix $B$:

$$
\begin{alignedat}{3}
\overset{\scriptscriptstyle\leq}{C}_{2}
&=
\begin{pmatrix}
0&0&0&0\\
0&0&0&0\\
0&3&2&2
\end{pmatrix}
\qquad&
\overset{\scriptscriptstyle\leq}{C'}_{2}
&=
\begin{pmatrix}
0&0&0&0\\
0&0&0&0\\
0&1&0&0
\end{pmatrix}
\qquad&
\overset{\scriptscriptstyle\leq}{C''}_{2}
&=
\begin{pmatrix}
0&1&1&1\\
0&1&1&1\\
0&0&0&0
\end{pmatrix}
\\[1em]
\overset{\scriptscriptstyle\leq}{C}_{3}
&=
\begin{pmatrix}
0&0&0&0\\
2&2&2&2\\
0&0&0&0
\end{pmatrix}
\qquad&
\overset{\scriptscriptstyle\leq}{C'}_{3}
&=
\begin{pmatrix}
0&0&0&0\\
0&1&0&0\\
0&0&0&0
\end{pmatrix}
\qquad&
\overset{\scriptscriptstyle\leq}{C''}_{3}
&=
\begin{pmatrix}
2&0&1&1\\
0&1&1&1\\
0&0&0&0
\end{pmatrix}
\\[1em]
\overset{\scriptscriptstyle\leq}{C}_{4}
&=
\begin{pmatrix}
0&0&0&0\\
0&0&0&0\\
1&0&0&1
\end{pmatrix}
\qquad&
\overset{\scriptscriptstyle\leq}{C'}_{4}
&=
\begin{pmatrix}
0&0&0&0\\
0&0&0&0\\
0&0&0&1
\end{pmatrix}
\qquad&
\overset{\scriptscriptstyle\leq}{C''}_{4}
&=
\begin{pmatrix}
1&0&0&0\\
0&0&0&0\\
0&0&0&0
\end{pmatrix}
\end{alignedat}
$$

We now use these count matrices as in the proof of \lemref{rect_maxmin}. For each pair $i,j$, we first select the last range $L_{r}$ for which $\overset{\scriptscriptstyle\leq}{C}_{r}\bracke{i,j}>0$. Within this range, we check whether the final part of row $i$ contains an entry satisfying $A\bracke{i,k}\leq B\bracke{k,j}$. If it does, we inspect that part. Otherwise, we inspect the last non-final part with a positive count. If no range has a positive count, we set $\overset{\scriptscriptstyle\leq}{C}\bracke{i,j}=-\infty$.

Start with  $i=j=1$. We have $\overset{\scriptscriptstyle\leq}{C}_{1}\bracke{1,1}=3$, $\overset{\scriptscriptstyle\leq}{C}_{2}\bracke{1,1}=0$, $\overset{\scriptscriptstyle\leq}{C}_{3}\bracke{1,1}=0$ and $\overset{\scriptscriptstyle\leq}{C}_{4}\bracke{1,1}=0$.
Thus only $L_{1}$ contains entries of the first row satisfying $A\bracke{1,k}\leq B\bracke{k,1}$. Ergo, we select $r=1$.   

As $\overset{\scriptscriptstyle\leq}{C'}_{1}\bracke{1,1}=0$, the final part contains no entry satisfying this inequality. Because $\rho_{A_{1}}^{\mathrm{row}}\paren{1,1}=2$ and $\rho_{A_{1}}^{\mathrm{row}}\paren{1,2}=3$, the first and second non-final parts are stored in rows $2$ and $3$ of $A''_{1}$, respectively. We have:
$\overset{\scriptscriptstyle\leq}{C''}_{1}\bracke{2,1}=2$ and $\overset{\scriptscriptstyle\leq}{C''}_{1}\bracke{3,1}=1$. As both are positive  we select $q=2$, which is the second non-final part $J_{1,1}^{\paren{2}}=\bracce{5,3}$. This is the last part with a positive count, even though its count is smaller than that of the first part. The indices $5,3$ are the columns of the entries to inspect in the original first row.

Inspecting these two entries gives $A\bracke{1,5}=5\leq B\bracke{5,1}=6$ and $A\bracke{1,3}=7>B\bracke{3,1}=3$. Only the value $5$ satisfies the inequality in this part. The preceding part contains the smaller values $2,4$, and the final part contains no entry satisfying the inequality. Hence the maximal entry value that satisfies this is  $\overset{\scriptscriptstyle\leq}{C}\bracke{1,1}=5$.

For $i=1,j=2$, we have $\overset{\scriptscriptstyle\leq}{C}_{1}\bracke{1,2}=5$, $\overset{\scriptscriptstyle\leq}{C}_{2}\bracke{1,2}=0$, $\overset{\scriptscriptstyle\leq}{C}_{3}\bracke{1,2}=0$ and $\overset{\scriptscriptstyle\leq}{C}_{4}\bracke{1,2}=0$. We again select $r=1$, but now $\overset{\scriptscriptstyle\leq}{C'}_{1}\bracke{1,2}=1$. We therefore inspect the final part $J_{1,1}^{\paren{3}}=\bracce{1}$. Since $A\bracke{1,1}=9\leq B\bracke{1,2}=29$, we obtain $\overset{\scriptscriptstyle\leq}{C}\bracke{1,2}=9$. There is no need to inspect either non-final part, since all their values are smaller than $9$.

For $i=3,j=4$, we have $\overset{\scriptscriptstyle\leq}{C}_{1}\bracke{3,4}=0$, $\overset{\scriptscriptstyle\leq}{C}_{2}\bracke{3,4}=2$, $\overset{\scriptscriptstyle\leq}{C}_{3}\bracke{3,4}=0$ and $\overset{\scriptscriptstyle\leq}{C}_{4}\bracke{3,4}=1$. Both $L_{2}$ and $L_{4}$ contain entries of the third row satisfying $A\bracke{3,k}\leq B\bracke{k,4}$. We select $r=4$, since it is the last range with a positive count. Although its count is smaller, every value in $L_{4}$ is larger than every value in $L_{2}$. Since $\overset{\scriptscriptstyle\leq}{C'}_{4}\bracke{3,4}=1$, we inspect the final part $J_{4,3}^{\paren{2}}=\bracce{4}$. Its only entry satisfies $A\bracke{3,4}=33\leq B\bracke{4,4}=35$, therefore $\overset{\scriptscriptstyle\leq}{C}\bracke{3,4}=33$.

Repeating this procedure for every pair  of indices $i,j$  results in:
$$
\overset{\scriptscriptstyle\leq}{C}=
\begin{pmatrix}
5&9&7&9\\
22&28&24&24\\
30&18&14&33
\end{pmatrix}
$$ 

Fix a pair $i,j$. We use the entries $\overset{\scriptscriptstyle\leq}{C}_{r}\bracke{i,j}$ for $r\in\bracke{4}$ to select the last range $L_{r}$ with a positive count. For the selected $r$, $\overset{\scriptscriptstyle\leq}{C'}_{r}\bracke{i,j}$ determines whether to inspect the final part, and the entries $\overset{\scriptscriptstyle\leq}{C''}_{r}\bracke{\rho_{A_{r}}^{\mathrm{row}}\paren{i,q},j}$ identify the last non-final part with a positive count if the final-part count is zero. The largest entry satisfying $A\bracke{i,k}\leq B\bracke{k,j}$ in the selected part of row $i$ of $A_{r}$ gives $\overset{\scriptscriptstyle\leq}{C}\bracke{i,j}$.

To compute $\overset{\scriptscriptstyle\geq}{C}$, we use the procedure that computed $\overset{\scriptscriptstyle\leq}{C}$, with $B^{\mathsf{T}}\in\mathbb{R}^{4\times 5}$ as its first input and $A^{\mathsf{T}}\in\mathbb{R}^{5\times 3}$ as its second. The values  $1,\ldots,35$ remain the same. 
We keep $t=3$. Each range now ends after $\ceil{\frac{20}{3}}=7$ entries from $B^{\mathsf{T}}$ or $\ceil{\frac{15}{3}}=5$ entries from $A^{\mathsf{T}}$, or when the sorted list ends. Thus the ranges $L_{1},\ldots,L_{4}$ remain unchanged.

For indices $j,i$, the procedure selects the largest $\paren{B^{\mathsf{T}}}\bracke{j,k}=B\bracke{k,j}$ satisfying $\paren{B^{\mathsf{T}}}\bracke{j,k}\leq \paren{A^{\mathsf{T}}}\bracke{k,i}$, which is exactly the condition $A\bracke{i,k}\geq B\bracke{k,j}$. Transposing the matrix, the returned value is therefore $\overset{\scriptscriptstyle\geq}{C}\bracke{i,j}$. 

Consider $i=1,j=3$. The procedure now uses row $3$ of $B^{\mathsf{T}}$ and column $1$ of $A^{\mathsf{T}}$. For $k=1$, we have $B\bracke{1,3}=8\leq A\bracke{1,1}=9$, while for $k=2,3,4,5$ the comparisons are $13>2$, $17>7$, $25>4$ and $32>5$, respectively. Thus only $k=1$ satisfies the inequality. The value $8$ belongs to $L_{1}$, so we select $r=1$. Within this range, row $3$ of $B^{\mathsf{T}}$ contains only the entry $8$, which forms its final part. Inspecting it returns $8$ at position $\paren{3,1}$, giving $\overset{\scriptscriptstyle\geq}{C}\bracke{1,3}=8$ after transposing.

Repeating this procedure for every pair of indices $j,i$ and transposing the result gives:
$$
\overset{\scriptscriptstyle\geq}{C}
=
\begin{pmatrix}
3&6&23\\
-\infty&21&21\\
8&17&25\\
-\infty&19&15
\end{pmatrix}^{\mathsf{T}}
=
\begin{pmatrix}
3&-\infty&8&-\infty\\
6&21&17&19\\
23&21&25&15
\end{pmatrix}
$$

where $-\infty$ arise when no $k$ satisfies $A\bracke{i,k}\geq B\bracke{k,j}$. Finally, for every pair of indices $i,j$, we set $C\bracke{i,j}=\mmax{\overset{\scriptscriptstyle\leq}{C}\bracke{i,j},\overset{\scriptscriptstyle\geq}{C}\bracke{i,j}}$. We get:
$$
C=A\maxmin B=
\begin{pmatrix}
5&9&8&9\\
22&28&24&24\\
30&21&25&33
\end{pmatrix}
$$

In particular, $\overset{\scriptscriptstyle\leq}{C}\bracke{1,3}=7$ and $\overset{\scriptscriptstyle\geq}{C}\bracke{1,3}=8$, so $C\bracke{1,3}=\mmax{7,8}=8$. Following the definition, $\paren{A\maxmin B}\bracke{1,3}=\mmax{8,2,7,4,5}=8$ indeed.

\end{example2}

Going back to \lemref{rect_maxmin}, we balance the two terms in its runtime to obtain the following bound:

\begin{conclusion}\label{cnc:rect_maxmin}
Let $a,b,c\in\bracke{0,1}$, $A\in\mathbb{R}^{n^{a}\times n^{b}}$ and $B\in\mathbb{R}^{n^{b}\times n^{c}}$. Then $A\maxmin B$ can be computed within $\widetilde O\paren{n^{\frac{a+b+c+\omega\paren{a,b,c}}{2}+o\paren{1}}}$ time.
\end{conclusion}

\begin{proof}
Choose $t=\floor{n^{\frac{a+b+c-\omega\paren{a,b,c}}{2}}}$. Since $a+c\leq \omega\paren{a,b,c}\leq a+b+c$, we have $1\leq t\leq n^{\frac{b}{2}} \leq n^{b}$, so \lemref{rect_maxmin} applies. Overall, $t\in\Theta\paren{n^{\frac{a+b+c-\omega\paren{a,b,c}}{2}}}$, hence the runtime becomes: 
$$
\widetilde O\paren{
t\cdot n^{\omega\paren{a,b,c}+o\paren{1}}
+\frac{n^{a+b+c}}{t}
}
=
\widetilde O\paren{
n^{\frac{a+b+c+\omega\paren{a,b,c}}{2}+o\paren{1}}
}
$$

\end{proof}

For square matrices and for rectangular products with equal outer dimensions, this observation recovers the following bounds of Duan and Pettie \cite{DuaPet2009} and Grandoni, Italiano, {\L}ukasiewicz, Parotsidis and Uzna{\'n}ski \cite{GraItaLukParUzn2021}, respectively.

\begin{observation}\label{obv:square_maxmin}
If $a=b=c=1$ we can compute $A\maxmin B$ in $\widetilde O\paren{n^{\frac{3+\omega}{2}+o\paren{1}}} $ runtime. If only $a=c=1$ we can compute $A\maxmin B$ in $\widetilde O\paren{n^{\frac{2+b+\omega\paren{b}}{2}+o\paren{1}}}$ runtime. 
\end{observation}

Finally, multiplying the entries by $-1$ exchanges minimum and maximum without changing the matrices dimensions. Thus the same runtime is available for the \minmaxname-product.

\begin{corollary}\label{cor:rect_minmax}
Let $a,b,c\in\bracke{0,1}$, $A\in\mathbb{R}^{n^{a}\times n^{b}}$ and $B\in\mathbb{R}^{n^{b}\times n^{c}}$. Computing $A\maxmin B$ and computing $A\minmax B$ reduces to one another. Each product can be computed within $\widetilde O\paren{n^{\frac{a+b+c+\omega\paren{a,b,c}}{2}+o\paren{1}}}$ time.
\end{corollary}

\begin{proof}
For every $i\in\bracke{n^{a}}$ and $j\in\bracke{n^{c}}$, negation gives:
$$
\begin{aligned}
\paren{A\minmax B}\bracke{i,j}
&=
\mmin{
\max\bracce{A\bracke{i,k},B\bracke{k,j}}
}{k\in\bracke{n^{b}}}\\
&=
-\mmax{
\min\bracce{-A\bracke{i,k},-B\bracke{k,j}}
}{k\in\bracke{n^{b}}}\\
&=
-\paren{\paren{-A}\maxmin\paren{-B}}\bracke{i,j}
\end{aligned}
$$
Hence we compute $A\minmax B$ by negating both input matrices, computing $\paren{-A}\maxmin\paren{-B}$ using \cncref{rect_maxmin}, and lastly negating the output. The reverse reduction follows from the identity:
$$
A\maxmin B=-\paren{\paren{-A}\minmax\paren{-B}}
$$
In either direction, negating the inputs and the output takes $O\paren{n^{a+b}+n^{b+c}+n^{a+c}}$ time, which is neglectable, considering the runtime of \cncref{rect_maxmin}, so both  the \maxminname-product and  the \minmaxname-product require the same runtime.

\end{proof}

We use \corref{rect_minmax} in
\algref{MinMax-update-through-pivot} of
\secref{2-slower}.
Consider the case $\sizeindex{u\squiggly v}{2}=1$,
and let
$\paren{x,y}=\rightabovearrow{u\squiggly v}{2}$.
If $u\notin\ball{x}{1}$, the preceding computations
give:
$$
\mmax{
d\bracke{u,p_{1}\paren{x}},
d\bracke{p_{1}\paren{x},v}}
\leq\delta\paren{u,v}
$$
The \minmaxname-product update therefore gives
$d\bracke{u,v}\leq 2\delta\paren{u,v}$.
The case $v\notin\ball{y}{1}$ is symmetric,
while the remaining case is already exact.
In \appref{2,W2}, replacing this update by
\algref{combinatorial-update-through-pivots}
gives a combinatorial $\paren{2,W_{2}}$-APASP.
Thus, the \minmaxname-product eliminates the
additive $W_{2}$ term in this branch.

\section{Conclusion and Open Problems}\label{conclusion}

We present the first $2$-APASP algorithm for
weighted undirected graphs 
whose runtime exponent is strictly
 $<\omega$ whenever $\omega>2$. The algorithm in \secref{2-faster} therefore answers
\queref{exact2} affirmatively. 
Dor, Halperin and Zwick~\cite{DorHalZwi2000}
show that achieving any fixed multiplicative
stretch below $2$ is at least as hard as BMM.
Consequently, assuming that BMM admits no algorithm
with a runtime exponent below $\omega$, then 
$\alpha=2$ is the smallest multiplicative stretch
for which $\alpha$-APASP can achieve such an exponent.

Although this settles the multiplicative stretch,
the running time can potentially be improved.
For every fixed $\varepsilon>0$, Dory, Forster,
Kirkpatrick, Nazari, Vassilevska Williams and
de Vos~\cite{DorForKirNazVasVos2023}
compute a $\paren{2+\varepsilon}$-APASP in:
$$
\widetilde O\paren{
n^{2+a}
+
n^{\omega\paren{1,1-a,1}+o\paren{1}}
\cdot\log W_{\max}
}
$$
time, for any $a\in\bracke{0,1}$.
Our $2$-APASP algorithm in \secref{2-faster}
instead takes:
$$
\widetilde O\paren{
n^{2+a}
+
n^{a+\omega\paren{1-a,1,1-a}+o\paren{1}}
\cdot\log^{O\paren{1}}W_{\max}
}
$$
runtime.
By \thmref{omega-cyclicShift} and
\obvref{omega-inequality-2}, we have:
$\omega\paren{1,1-a,1}
=
\omega\paren{1,1,1-a}
\leq
a+\omega\paren{1-a,1,1-a}$. 
This leaves the following question.

\begin{question}\label{que:2and2+eps}
For any $a\in\bracke{0,1}$, can $2$-APASP
for weighted undirected graphs  be computed within a runtime of $\widetilde O\paren{
n^{2+a}
+
n^{\omega\paren{1,1-a,1}+o\paren{1}}
\cdot\log^{O\paren{1}}W_{\max}
}$, matching the dependence on $n$ of the
$\paren{2+\varepsilon}$-APASP algorithm of
Dory, Forster, Kirkpatrick, Nazari,
Vassilevska Williams and de Vos
\cite{DorForKirNazVasVos2023}?
\end{question}

A second direction concerns combinatorial algorithms.
For weighted undirected graphs, the best known
combinatorial runtime for $2$-APASP remains
$\widetilde O\paren{n^{2}+m\cdot n^{\frac{1}{2}}}$,
due to Baswana and Kavitha~\cite{BasKav2010}. This runtime bound has remained
unimproved for two decades.

\begin{question}\label{que:combinatorial2}
Can a combinatorial algorithm compute $2$-APASP
for weighted undirected graphs with a polynomial
improvement over
$\widetilde O\paren{n^{2}+m\cdot n^{\frac{1}{2}}}$
runtime for some range of graph densities?
In particular, can it achieve expected
$\widetilde O\paren{n^{\frac{5}{2}-\eta}}$
runtime for some constant $\eta>0$?
\end{question}

As a step towards \queref{combinatorial2},
\appref{2,W2} presents a combinatorial branch
of the algorithm in \secref{2-slower}.
Replacing its final \minmaxname-product update
by additional SSSP computations gives a
$\paren{2,W_{2}}$-APASP in 
$\widetilde O\paren{n^{\frac{7}{3}}}$ runtime.
Although this guarantee already follows from
Cohen and Zwick~\cite{CohZwi1997}, the construction
continues the analysis of \secref{2-slower}
and identifies the case in which an additive
$W_{2}$ term remains, which we mark as a \qoute{weak spot}.

We suggest further examining this case to obtain
a $\paren{2,cW_{2}}$-APASP for some constant
$c\in\paren{0,1}$ within the same $\widetilde O\paren{n^{\frac{7}{3}}}$ runtime.
Reducing the additive term would be an intermediate
step towards eliminating it entirely, therefore aiming towards
answering \queref{combinatorial2}.

\newpage
\clearpage
\renewcommand*{\bibfont}{\small}
\begingroup
\phantomsection\printbibliography[heading=bibintoc]
\endgroup

\appendix


\section{Row and Column Balancing}\label{balance}

In this appendix, we present the row and column
balancing procedures used in the fully rectangular
matrix multiplication algorithms of
\secref{rect_(max,min)}.

We consider a matrix
$A\in\paren{\mathbb{R}\cup\bracce{+\infty}}^{n^{a}\times n^{b}}$
and denote by $m_{A}$ the number of its finite entries. Our matrix
multiplication algorithms exploit the sparsity of $A$ by avoiding
operations involving entries equal to $+\infty$.
 However, the finite
entries of $A$ may be highly concentrated in a small number of rows or
columns. Row and column balancing redistribute\footnote{We assume $m_{A}>0$. If $m_{A}=0$, we
simply set $\rbal{A}=\paren{A,A}$.} these entries more
\qoute{uniformly}, while preserving their relevant column or row
indices, respectively.

\begin{definition}\label{def:rbal}
Let $k_{A}^{\mathrm{row}}=\ceil{\frac{m_{A}}{n^{a}}}$. 
For every row $i\in\bracke{n^{a}}$, sort its finite entries in an 
increasing order, breaking ties according to their column indices.
Partition the
the corresponding column indices of the resulting
order into consecutive sets
$J_{i}^{(1)},J_{i}^{(2)},\ldots,J_{i}^{(s_{i})}$, such that every non-final
set $J_{i}^{\paren{q}}$ for $1\leq q < s_{i}$ contains exactly $k_{A}^{\mathrm{row}}$ indices and the final set $J_{i}^{(s_{i})}$ contains at
most $k_{A}^{\mathrm{row}}$ indices\footnote{Rows with no finite values at all have no sets.}. 
For every $q\in\bracke{s_{i}}$, let
$
\widetilde{A}_{i}^{(q)}
\in
\paren{\mathbb{R}\cup\bracce{+\infty}}^{n^{b}}
$
denote the padded row vector corresponding to $J_{i}^{\paren{q}}$, defined
by:
$$
\widetilde{A}_{i}^{(q)}\bracke{j}
=
\begin{cases}
A\bracke{i,j},
    & \text{if $j\in J_{i}^{\paren{q}}$},\\
+\infty,
    & \text{otherwise}
\end{cases}
$$

Let
$\NFR{A}
=
\bracce{
\paren{i,q}
\midline
i\in\bracke{n^{a}}
\text{ and }
1\leq q<s_{i}
}$
denote the set of non-final sets and fix an injective mapping
$\rho^{\mathrm{row}}_{A}:\NFR{A}\rightarrow\bracke{n^{a}}$.
The \emph{row balancing} of $A$ is the pair
$\rbal{A}=\paren{A',A''}$, where
$A',A''\in
\paren{\mathbb{R}\cup\bracce{+\infty}}^{n^{a}\times n^{b}}$.
The matrix $A'$ retains the final set of every row and is defined by:
$$
A'\bracke{i,j}
=
\begin{cases}
\widetilde{A}_{i}^{(s_{i})}\bracke{j}
    & \text{if $s_{i}>0$}\\
+\infty
    & \text{otherwise}
\end{cases}
$$

For any 
$i\in\bracke{n^{a}}$ and $1\leq q<s_{i}$
the matrix $A''$ stores the non-final set  $J_{i}^{\paren{q}}$ $1\leq q <s_{i}$ from $A$ in its assigned 
row  $\rho^{\mathrm{row}}_{A}\paren{i,q} \in \bracke{n^{a}}$ in $A''$. It is defined by:
$$
A''\bracke{h,j}
=
\begin{cases}
\widetilde{A}_{i}^{(q)}\bracke{j}
    & \text{if $h=\rho^{\mathrm{row}}_{A}\paren{i,q}$ for some
      $\paren{i,q}\in\NFR{A}$}\\
+\infty
    & \text{otherwise}
\end{cases}
$$
\end{definition}

Every finite entry of $A$ appears exactly once in either $A'$ or
$A''$ and retains its original column index. Every row of $A'$
contains at most $k_{A}^{\mathrm{row}}$ finite entries, while every row of $A''$ in
the image of $\rho^{\mathrm{row}}_{A}$ contains exactly $k_{A}^{\mathrm{row}}$ finite entries; all other
rows of $A''$ contain only $+\infty$. Moreover,
$\paren{\rho_{A}^{\mathrm{row}}}^{-1}\paren{h}=\paren{i,q}$ identifies the original row $i$ and the indices from $J_{i}^{\paren{q}}$ whose corresponding entries from $A$ are
stored in the row $h$ of $A''$.

The injective mapping $\rho^{\mathrm{row}}_{A}$ exists since the non-final sets are
pairwise disjoint and each contains exactly $k_{A}^{\mathrm{row}}$ finite entries.
Therefore,
$k_{A}^{\mathrm{row}}\cdot\abs{\NFR{A}}\leq m_{A}$, which implies
$\abs{\NFR{A}}\leq\frac{m_{A}}{k_{A}^{\mathrm{row}}}\leq n^{a}$. Hence, the
$n^{a}$ rows of $A''$ suffice to store all non-final sets.
Similarly, we define the column-balancing of $A$:

\begin{example2}\label{exm:rbal}
Consider:
$$
A=
\begin{pmatrix}
4&1&3&2\\
+\infty&5&+\infty&+\infty\\
+\infty&+\infty&+\infty&6
\end{pmatrix}
$$

Here $m_{A}=6$ and
$k_{A}^{\mathrm{row}}=\ceil{\frac{6}{3}}=2$.
Sorting the finite entries of the first row gives the column
indices $2,4,3,1$, partitioned into
$J_{1}^{\paren{1}}=\bracce{2,4}$ and
$J_{1}^{\paren{2}}=\bracce{3,1}$.
Each remaining row has only a final part. Thus $\NFR{A}=\bracce{\paren{1,1}}$.
Choosing $\rho_{A}^{\mathrm{row}}\paren{1,1}=3$ gives
$\rbal{A}=\paren{A',A''}$, where
$$
A'=
\begin{pmatrix}
4&+\infty&3&+\infty\\
+\infty&5&+\infty&+\infty\\
+\infty&+\infty&+\infty&6
\end{pmatrix},
\qquad
A''=
\begin{pmatrix}
+\infty&+\infty&+\infty&+\infty\\
+\infty&+\infty&+\infty&+\infty\\
+\infty&1&+\infty&2
\end{pmatrix}.
$$

The final parts remain in their original rows in $A'$, while
the non-final part of the first row moves to the third row of
$A''$. In both cases, the original column indices are preserved.
\end{example2}

\begin{definition}\label{def:cbal}
Suppose that
$\rbal{A^{\mathsf{T}}}=\paren{\paren{A^{\mathsf{T}}}',\paren{A^{\mathsf{T}}}''}$.
The \emph{column balancing} of $A$ is the pair:
$$
\cbal{A}
=
\paren{
\paren{\paren{A^{\mathsf{T}}}'}^{\mathsf{T}},
\paren{\paren{A^{\mathsf{T}}}''}^{\mathsf{T}}
}
$$
\end{definition}

By applying \defref{rbal} to $A^{\mathsf{T}}$ we set
$k_{A}^{\mathrm{col}}=\ceil{\frac{m_{A}}{n^{b}}}$,
$\NFC{A}=\NFR{A^{\mathsf{T}}}$ and obtain an injective mapping
$\rho^{\mathrm{col}}_{A}:\NFC{A}\rightarrow\bracke{n^{b}}$.
Every finite entry of $A$ retains its original row index and appears exactly once in one of the two matrices in
 $\cbal{A}$. Every column
of the first matrix contains at most $k_{A}^{\mathrm{col}}$ finite
entries, while every column of the second matrix in the image of
$\rho^{\mathrm{col}}_{A}$ contains exactly $k_{A}^{\mathrm{col}}$ finite
entries; all remaining columns contain only $+\infty$. Moreover,
$\paren{\rho_{A}^{\mathrm{col}}}^{-1}\paren{h}=\paren{j,q}$ identifies the
original column $j$ and the corresponding non-final set stored in column $h$.

\section{Quadratic $2$-APASP for Unweighted Graphs}
\label{unweighted2}
Contrary to weighted graphs, computing $2$-APASP for unweighted graphs can be perfomed in $\widetilde{O} \paren{n^{2+o\paren{1}}}$ runtime. We give a simple $2$-APASP algorithm for an unweighted
undirected graph $G=\paren{V,E}$. This algorithm is essentially a  simplified version of the
 generic $2$-APASP algorithm of Dory, Forster, Kirkpatrick, Nazari, Vassilevska Williams,
and de Vos for unweighted graphs~\cite{DorForKirNazVasVos2023}. We recall the classic result by Baswana and Kavitha and use it as black box:

\begin{theorem}\label{thm:BasKav-2Approx}(Theorem 7.6 in \cite{BasKav2010})
    Given an undirected weighted graph $G$ with $n$ vertices and $m$ edges, there exists an algorithm to compute the 2-Approximate All-Pair Shortest Paths in $G$, and the algorithm runs in $\widetilde{O}\paren{m \sqrt{n} + n^2}$ time with high probability.\footnote{Though \cite{BasKav2010} state their result in expected running time, this can be converted to high probability using chernoff bounds. Discussion on the same can be found in \cite{DorForKirNazVasVos2023} (Theorem 3.3). }
\end{theorem}

The idea behind our approach is to 
combine the $2$-APASP of Baswana
and Kavitha~\cite{BasKav2010} on a sparse subgraph with the
$\paren{1+\varepsilon}$-MSASP algorithm of Elkin and
Neiman~\cite{ElkNei2020} and the
$\paren{1+\varepsilon}$-approximate \minplusname-product algorithm of
Zwick~\cite{Zwick2002}. For both calls, we choose
$\varepsilon$ such that $\paren{1+\varepsilon}^{2}<\frac{5}{4}$, say  $\varepsilon=\frac{1}{9}$.

\algref{unweighted2} samples each vertex from $V$ with probability $n^{-\frac{1}{2}}$.
We assume $w\paren{e}=1$ for any $e\in E$ and construct $E_{S}=\bracce{\paren{u,v} \midline \delta\paren{u,S}>1=w\paren{u,v}}$, as defined in \secref{nearest}, and set
$G_{1}=\paren{V,E_{1}}$.
It computes the pivots $p_{S}\paren{x}$ and their
distances using \obvref{pivotsdistance}, implemented
by a single BFS invocation as the SSSP.
It then invokes the $2$-APASP algorithm of Baswana and Kavitha \cite{BasKav2010} on $G_{1}$
and updates $d$ with its estimates. 

It proceeds to invoke the $\paren{1+\varepsilon}$-MSASP of Elkin and Neiman \cite{ElkNei2020} from $S$ over $G$ with $\varepsilon = \frac{1}{9}$. Having achieved estimated between members of $S$ and the entire vertex set $V$, it computes estimates between any two members of $V$ by a $\paren{1+\varepsilon}$-approximate \minplusname-product, for the same $\varepsilon=\frac{1}{9}$, over the matrices $V\times S$ and $S\times V$, whose entries are defined by $d$. Lastly, as the graph is unweighted, we verify the entries are integers by rounding them down.

\medskip

\begin{algorithm}[H]\label{alg:unweighted2}
\setstretch{1.15}
\DontPrintSemicolon
\caption{\textsc{Unweighted-2-APASP}$\paren{G}$}
\textbf{Input:} An unweighted undirected graph
$G=\paren{V,E}$.

\textbf{Output:} A symmetric table $d\bracke{\cdot,\cdot}$
satisfying
$\delta\paren{u,v}\leq d\bracke{u,v}
\leq 2\delta\paren{u,v}$
for all $\paren{u,v}\in V\times V$.

\medskip

Initialize $d\bracke{u,v}\gets\infty$ for all
$\paren{u,v}\in V\times V$, set $d\bracke{u,v}\gets 1$  for $\paren{u,v}\in E$

$S\gets\textsc{Sample}\paren{V,n^{-\frac{1}{2}}}$

Compute the pivots $p_{S}\paren{u}$ and distances
$\delta\paren{u,S}$ by a multi-source BFS from $S$
(\obvref{pivotsdistance})

Construct $E_{1}\gets E_{S}$ and
$G_{1}\gets\paren{V,E_{1}}$

Compute a $2$-APASP in $G_{1}$ and update $d$
accordingly (\thmref{BasKav-2Approx})
\label{algline:unweighted2-bk}

Set $\varepsilon\gets\frac{1}{9}$

\textsc{Approximate\_MSSP}$\paren{G,S,d,\varepsilon}$
to compute a $\paren{1+\varepsilon}$-MSASP and update $d$

Update $d$ by a $\paren{1+\varepsilon}$-Approximation of
$d\bracke{V,S}\minplus d\bracke{S,V}$
(\thmref{ApproxMinPlus})
\label{algline:unweighted2-ampmm}

\ForEach{$\paren{u,v}\in V\times V$}
{
    $d\bracke{u,v}\gets\floor{d\bracke{u,v}}$\;
    \label{algline:unweighted2-round}
}

\Return $d$
\end{algorithm}

\medskip

\begin{theorem}\label{thm:unweighted2}
\algref{unweighted2} computes 
$2$-APASP  for unweighted
undirected graphs in
$\widetilde O\paren{
n^{\omega\paren{1,\frac{1}{2},1}+o\paren{1}}
}$ runtime. With current $\omega$ values ~\cite{AmaBerZui2026,Uznanski2026}, the runtime becomes $\widetilde O \paren{n^{2+o\paren{1}}}$. 
\end{theorem}

\begin{proof}
The algorithms of Baswana and Kavitha and Elkin and
Neiman do not underestimate distances in $G$, nor does
the approximate \minplusname-product invocation. Since distances in $G$ are
integers, rounding down preserves
$d\bracke{u,v}\geq\delta\paren{u,v}$.

Let $u,v\in V$. If $\paren{u,v}\in E$ then $d\bracke{u,v}=\delta\paren{u,v}$. Fix a shortest path $u\squiggly v$. It remains to consider $\delta\paren{u,v}\geq 2$.  If $u\squiggly v\subseteq E_{1}$, then
$\delta_{G_{1}}\paren{u,v}=\delta\paren{u,v}$.
The invocation of Baswana and Kavitha on $G_{1}$
therefore gives:
$$
d\bracke{u,v}
\leq 2\delta_{G_{1}}\paren{u,v}
=2\delta\paren{u,v}
$$

Otherwise, let 
$
\paren{x,y}=\rightabovearrow{u\squiggly v}{1}$. By the definition of $E_{S}$: $
\delta\paren{x,p_{S}\paren{x}}\leq 1$. Consequently:
$ \delta\paren{u,p_{S}\paren{x}}
+\delta\paren{p_{S}\paren{x},v}\leq
\delta\paren{u,x}
+2\delta\paren{x,p_{S}\paren{x}}
+\delta\paren{x,v}
\leq \delta\paren{u,v}+2 $.  Both the  MSASP and the approximate \minplusname-product invocations estimates the distances from both
pivots within a factor of $1+\varepsilon=\frac{10}{9}$. Thus, after rounding down:
$$
d\bracke{u,v}
\leq
\floor{
\paren{1+\varepsilon}^{2}\cdot 
\paren{\delta\paren{u,v}+2}
}
=
\floor{
\frac{100}{81}\cdot \paren{\delta\paren{u,v}+2}
}
$$

If $\delta\paren{u,v}=2$ we get $d\bracke{u,v} \leq \floor{\frac{400}{81}} = 4 = 2\cdot \delta\paren{u,v}$. Assuming $\delta\paren{u,v}\geq 3$ we get: $\frac{100}{81}\cdot \paren{\delta\paren{u,v}+2} = 2\cdot \delta\paren{u,v}+\frac{200-62\cdot\delta\paren{u,v}}{81}\leq 2\delta\paren{u,v} +\frac{14}{81} < 2\delta\paren{u,v}+1$. Consequently: $d\bracke{u,v} \leq \floor{\frac{100}{81}\cdot \paren{\delta\paren{u,v}+2}}\leq \floor{2\delta\paren{u,v}+\frac{14}{81}}= 2\delta\paren{u,v}$, where the last equality follows as $G$ is unweighted and therefore distances in $G$ are integers.

As for the runtime, by \obvref{hssize} and
\obvref{edgessize}, with high probability that 
$
\abs{S}\in\widetilde O\paren{n^{\frac{1}{2}}}$ and $
\abs{E_{1}}\in\widetilde O\paren{n^{\frac{3}{2}}}
$. Computing the pivots and constructing $E_{1}$ takes
$O\paren{m+n}$ time.
The invocation of Baswana and Kavitha on $G_{1}$
takes
$
\widetilde O\paren{
n^{2}
+n^{\frac{1}{2}}\cdot \abs{E_{1}}
}
=
\widetilde O\paren{n^{2}}
$ runtime. The MSASP invocation of Elkin and
Neiman takes:
$
\widetilde O\paren{
m^{1+o\paren{1}}
+n^{\omega\paren{\frac{1}{2},1,1}+o\paren{1}}
}
$. The approximate \minplusname-product invocation multiplies matrices of dimensions
$n\times O\paren{n^{\frac{1}{2}}}$ and
$O\paren{n^{\frac{1}{2}}}\times n$. For our constant $\varepsilon = \frac{1}{9}$ , this takes: $\widetilde O\paren{
n^{\omega\paren{1,\frac{1}{2},1}+o\paren{1}}
}$ runtime. As $\omega\paren{\frac{1}{2},1,1}
=\omega\paren{1,\frac{1}{2},1}$ this remains within the same runtime costs. Initializing $d$, updating its entries and rounding
down take $O\paren{n^{2}}$ time. The total runtime is therefore:
$
\widetilde O\paren{
n^{\omega\paren{1,\frac{1}{2},1}+o\paren{1}}
}
$.

\end{proof}

The rectangular matrix multiplication bounds
of Uznanski \cite{Uznanski2026} or of 
Amanov, van den Berg and Zuiddam~\cite{AmaBerZui2026} 
give
$\omega\paren{1,\frac{1}{2},1}=2$.
Thus, \algref{unweighted2} computes a $2$-APASP for
unweighted undirected graphs in 
$\widetilde O\paren{n^{2+o\paren{1}}}$ time, therefore solving \queref{n^2} for the unweighted setting.

\section{$\paren{2,W_{2}}$-APASP}\label{2,W2}

A natural goal is to obtain a combinatorial
$2$-APASP faster than the
$\widetilde O\paren{n^{2}+m\cdot n^{\frac{1}{2}}}$
algorithm of Baswana and Kavitha~\cite{BasKav2010}.
Setting $b=c=\frac{1}{3}$ in
\algref{2-APASP-slower-main} makes all its SSSP
computations take $\widetilde O\paren{n^{\frac{7}{3}}}$ time,
which improves upon this runtime for dense graphs.
The remaining step uses a \minmaxname-product.
We replace this step by additional SSSP computations,
obtaining a $\paren{2,W_{2}}$-APASP within the
same runtime.
We include this variant for completeness.
Its guarantee is already implied by the $+\mmin{2W_{1},4W_{2}}$-APASP of Cohen and
Zwick~\cite{CohZwi1997}. Indeed, $+\mmin{2W_{1},4W_{2}} \leq + W_{1}+2W_{2} \leq\paren{2,W_{2}} $. 

Let $S_{1}$ and $S_{2}$ be the same as in \algref{2-APASP-slower-main}.  We retain all computations preceding the final
\minmaxname-product update.
$d\bracke{s,v}=\delta\paren{s,v}$
for every $s\in S_{2}$ and $v\in V$, and
$d\bracke{x,p_{1}\paren{x}}
=\delta\paren{x,S_{1}}$
for every $x\in V$.
We replace the call to
\algref{MinMax-update-through-pivot}
by \algref{combinatorial-update-through-pivots}.
All preceding computations of the main algorithm
remain unchanged.

Consider the remaining case
$\sizeindex{u\squiggly v}{2}=1$, and let
$
\paren{x,y}
=
\rightabovearrow{u\squiggly v}{2}
=
\leftabovearrow{u\squiggly v}{2}
$. If
$\sizeindex{u\squiggly x}{1}
=\sizeindex{y\squiggly v}{1}=0$,
then $\sizeindex{u\squiggly v}{1}=1$,
so the initial two SSSP rounds already compute
the exact distance.
Otherwise, w.l.o.g.\footnote{ Reverse the path if necessary.} 
$\sizeindex{u\squiggly x}{1}\geq 1$. Let $
\paren{z,w}
=
\rightabovearrow{u\squiggly x}{1} $ and $
\paren{r,t}
=
\leftabovearrow{u\squiggly x}{1}
$. Thus,
$\sizeindex{u\squiggly z}{1}
=\sizeindex{t\squiggly x}{1}=0$
and $\sizeindex{z\squiggly t}{2}=0$.
If $\sizeindex{u\squiggly x}{1}=1$, then
$\paren{z,w}=\paren{r,t}$. We do not distinguish whether $\paren{z,w}=\paren{r,t}$ or not.

\medskip

\begin{algorithm}[H]
\setstretch{1.15}
\DontPrintSemicolon
\caption{\textsc{Combinatorial\_Update\_through\_pivots}$
\paren{G_{1},S_{1},S_{2},p_{2}\paren{\cdot},d}$}
\label{alg:combinatorial-update-through-pivots}

\textbf{Input:} The graph $G_{1}=\paren{V,E_{1},w}$,
the sampled sets $S_{1},S_{2}$, the pivots
$p_{2}\paren{\cdot}$ and the symmetric distance
estimates $d$ obtained by the preceding computations
of \algref{2-APASP-slower-main}.

\textbf{Output:} Symmetrically updates $d$ in-place.

\medskip

$F\gets
\bracce{
\paren{u,s}\in V\times S_{1}
\midline
d\bracke{u,s}
<
d\bracke{u,p_{2}\paren{u}}
}$

Construct the undirected graph
$H\gets\paren{
V,E_{1}\cup F\cup\paren{S_{2}\times V}
}$

Retain the original weights on $E_{1}$ and assign
weight $d\bracke{u,v}$ to every auxiliary edge
$\paren{u,v}$.
For parallel edges, retain the smaller weight.

\textsc{Estimate\_augmented\_MSSP}$\paren{H,V,d}$
(\algref{Estimate-augmented-MSSP})

\Return $Null$
\end{algorithm}

\medskip

We therefore look at the path $u\squiggly z \straight w\squiggly r \straight t  \squiggly x \straight y \squiggly v$. To connect between $u$ and $x$ we consider the bunch edges. 
The edges of $F$ correspond precisely to the bunch edges 
as defined in \secref{hierarchy}:
$F=
\bracce{
\paren{u,s}
\midline
u\in V,\ s\in\ball{u}{1}{2}
}$. Indeed, if $s\in\ball{u}{1}{2}$, then
$\delta\paren{u,s}<\delta\paren{u,S_{2}}$.
By \obvref{pathinball}, the shortest path
$u\squiggly s\subseteq E_{2}$. The preceding SSSP invocation from $s\in S_{1}$
over $G_{2}$ therefore computes
$d\bracke{u,s}=\delta\paren{u,s}$.
Conversely, since $d$ does not underestimate
distances and
$d\bracke{u,p_{2}\paren{u}}=\delta\paren{u,S_{2}}$,
the condition used to construct $F$ implies
$s\in\ball{u}{1}{2}$.
Thus, every edge of $F$ has its exact distance
as its weight, and constructing $F$ requires
no additional distance computations.

\begin{theorem}\label{thm:2,W2}
\algref{2-APASP-slower-main},  replacing its
final call to \algref{MinMax-update-through-pivot}
by \algref{combinatorial-update-through-pivots},
is a combinatorial $\paren{2,W_{2}}$-APASP with  $\widetilde O\paren{n^{\frac{7}{3}}}$ runtime.
\end{theorem}

\begin{proof}
The preceding computations do not underestimate
distances, by \lemref{lowerBound-proof}.
Every auxiliary edge added to $H$ has its exact
distance as its weight.
The additional SSSP computations therefore
preserve $d\bracke{u,v}\geq\delta\paren{u,v}$.

Fix a shortest path $u\squiggly v$.
If $\sizeindex{u\squiggly v}{1}\leq 1$,
the initial two SSSP rounds already compute
the exact distance.
The cases
$\sizeindex{u\squiggly v}{2}=0$ and
$\sizeindex{u\squiggly v}{2}\geq 2$
are covered as shown in the proof of \lemref{slower-mainAlgo-proof}.
We therefore consider the remaining case and
use the vertices $x,y,z,w,r,t$ defined above.

Since $\paren{x,y}\notin E_{2}$, we have
$\delta\paren{x,p_{2}\paren{x}}\leq w\paren{x,y}$.
The exact auxiliary edges through
$p_{2}\paren{x}\in S_{2}$ give:
$$
\begin{aligned}
d\bracke{u,v}
&\leq
\delta\paren{u,p_{2}\paren{x}}
+\delta\paren{p_{2}\paren{x},v}\\
&\leq
\delta\paren{u,v}
+2\delta\paren{x,p_{2}\paren{x}}\\
&\leq
\delta\paren{u,v}+2w\paren{x,y}
\end{aligned}
$$

If $2w\paren{x,y}\leq\delta\paren{u,v}$,
this already gives
$d\bracke{u,v}\leq 2\delta\paren{u,v}$.
Otherwise, $\paren{x,y}$ is the unique heaviest
edge of $u\squiggly v$.
Since $\paren{r,t}$ is a different edge and
$\paren{r,t}\notin E_{1}$, we obtain:
$\delta\paren{t,p_{1}\paren{t}}
\leq w\paren{r,t}
\leq W_{2}$. We observe that while $\paren{x,y}$ is a different edge, it might be possible that $\paren{z,w} = \paren{r,t}$. Hence, $w\paren{z,w},w\paren{r,t}\leq W_{2}$, but we cannot bound with $W_{3}$. 

We now  show which estimate from
$p_{1}\paren{t}$ is available before the
additional SSSP invocation.
Since $\sizeindex{t\squiggly x}{1}=0$,
the path $t\squiggly y$ contains exactly one
edge outside $E_{1}$, namely $\paren{x,y}$.
The initial two SSSP rounds therefore compute
$d\bracke{t,y}=\delta\paren{t,y}$.
The update in  \linesref{2-APASP-slower-main-distanceUpdate-start}
{2-APASP-slower-main-distanceUpdate-end}
then gives: $d\bracke{p_{1}\paren{t},y}
\leq
\delta\paren{p_{1}\paren{t},t}
+\delta\paren{t,y}$.
Furthermore, $\sizeindex{y\squiggly v}{2}=0$. After invoking  SSSP  from $p_{1}\paren{t} \in S_{1}$ over $G_{2}$ in
\linesref{2-APASP-slower-main-pivotMSSP-again}, we get: $d\bracke{p_{1}\paren{t},v}
\leq
\delta\paren{p_{1}\paren{t},t}
+\delta\paren{t,v}$.

We now distinguish between two cases according to whether
$p_{1}\paren{t}$ belongs to $\ball{z}{1}{2}$. 

\begin{enumerate}
\item \dashuline{$p_{1}\paren{t}\in\ball{z}{1}{2}$}\\
Hence $\paren{z,p_{1}\paren{t}}\in F$ and has weight
$\delta\paren{z,p_{1}\paren{t}}$.
The additional SSSP from $v$ uses the
auxiliary star edge  $\paren{v,p_{1}\paren{t}}$,
the edge from $p_{1}\paren{t}$ to $z$,
and the path $z\squiggly u\subseteq E_{1}$. Ergo:
$$
\begin{aligned}
d\bracke{u,v}
&\leq
\delta\paren{t,v}
+\delta\paren{t,p_{1}\paren{t}}
+\delta\paren{p_{1}\paren{t},z}
+\delta\paren{z,u}\\
&\leq
\delta\paren{t,v}
+2\delta\paren{t,p_{1}\paren{t}}
+\delta\paren{t,z}
+\delta\paren{z,u}\\
&=
\delta\paren{u,v}
+2\delta\paren{t,p_{1}\paren{t}}\\
&\leq
\delta\paren{u,v}+2W_{2}\\
&\leq
2\delta\paren{u,v}
\end{aligned}
$$

\item \dashuline{$p_{1}\paren{t}\notin\ball{z}{1}{2}$}\\
By definition: $\delta\paren{z,p_{2}\paren{z}}
\leq
\delta\paren{z,p_{1}\paren{t}}
\leq
\delta\paren{z,t}
+\delta\paren{t,p_{1}\paren{t}}$. 
Using the  auxiliary edges through
$p_{2}\paren{z}\in S_{2}$ gives: 
$$
\begin{aligned}
d\bracke{u,v}
&\leq
\delta\paren{u,p_{2}\paren{z}}
+\delta\paren{p_{2}\paren{z},v}\\
&\leq
\delta\paren{u,v}
+2\delta\paren{z,p_{2}\paren{z}}\\
&\leq
\delta\paren{u,v}
+2\delta\paren{z,t}
+2\delta\paren{t,p_{1}\paren{t}}
\end{aligned}
$$
Together with
$d\bracke{u,v}\leq\delta\paren{u,v}+2w\paren{x,y}$,
we obtain:
$$
\begin{aligned}
d\bracke{u,v}
&\leq
\delta\paren{u,v}
+\mmin{
2w\paren{x,y},
2\delta\paren{z,t}
+2\delta\paren{t,p_{1}\paren{t}}
}\\
&\leq
\delta\paren{u,v}
+w\paren{x,y}
+\delta\paren{z,t}
+\delta\paren{t,p_{1}\paren{t}}\\
&\leq
\delta\paren{u,v}+\delta\paren{z,y}+w\paren{r,t}\\
&\leq
2\delta\paren{u,v}+W_{2}
\end{aligned}
$$
\end{enumerate}

As for the runtime, by the analysis of
\lemref{mainAlgo-runtime}, the computations preceding
the omitted \minmaxname-product require: $\widetilde O\paren{
n^{3-b-c}+n^{2+b}+n^{2+c}
}
=
\widetilde O\paren{n^{\frac{7}{3}}}$
for $b=c=\frac{1}{3}$. By \obvref{bunchsize2}, \obvref{edgessize}
and \obvref{hssize2}, respectively: 

$$
\begin{aligned}
\abs{F}
&\leq
\msum{
\abs{\ball{u}{1}{2}}}
{u\in V}
\in\widetilde O\paren{n^{\frac{4}{3}}}\\
\mathbb E\bracke{\abs{E_{1}}}
&\in\widetilde O\paren{n^{\frac{4}{3}}}\\
n\cdot\abs{S_{2}}
&\in\widetilde O\paren{n^{\frac{4}{3}}}
\end{aligned}
$$
Thus, $\abs{H}\in\widetilde O\paren{n^{\frac{4}{3}}}$. We observe that in order to construct $F$ we simply check the bunch condition for all pairs in
$V\times S_{1}$, which takes a neglectful additional $O\paren{n^{\frac{5}{3}}}$ runtime. 
By \lemref{estimateMSSP-runtime}, the additional
SSSP invocation from all vertices require $\widetilde O\paren{
n\cdot\paren{n^{\frac{4}{3}}+n}
}
=
\widetilde O\paren{n^{\frac{7}{3}}}$ runtime. Hence, the total  runtime remains
$\widetilde O\paren{n^{\frac{7}{3}}}$.
\end{proof}

\section{$\paren{2+\varepsilon,W_{2}}$-APASP}\label{2+eps,W2}

\medskip

\begin{algorithm}[H]
\setstretch{1.15}
\DontPrintSemicolon
\caption{$\paren{2+\varepsilon,W_{2}}$-APASP}
\label{alg:2+eps,W2}

\textbf{Input:} A weighted undirected graph
$G=\paren{V,E,w}$ with integer weights in
$\bracce{0}\cup\bracke{W_{\max}}\cup\bracce{\infty}$,
parameters $b,c\geq 0$ with $b+c\leq 1$
and $0<\varepsilon\leq 1$.

\textbf{Output:} A symmetric table $d$ satisfying
$\delta\paren{u,v}\leq d\bracke{u,v}
\leq\paren{2+\varepsilon}\cdot\delta\paren{u,v}+W_{2}$
for every $u,v\in V$.

\medskip

Set $\varepsilon'\gets\frac{\varepsilon}{6}$

Run \algref{2-APASP-slower-main} with sampling
probabilities $n^{-b}$ and $n^{-c}$,
omitting its final call to
\algref{MinMax-update-through-pivot} and replacing
$\textsc{Estimate\_augmented\_MSSP}\paren{G,S_{2},d}$
by
$\textsc{Approximate\_MSSP}
\paren{G,S_{2},d,\varepsilon'}$
(\algref{ApproxMSSP-pivot})

$F\gets
\bracce{
\paren{u,s}\in V\times S_{1}
\midline
d\bracke{u,s}<d\bracke{u,p_{2}\paren{u}}
}$

Construct $H\gets\paren{V,E_{1}\cup F}$,
retaining the original weights on $E_{1}$
and assigning weight $d\bracke{u,s}$ to
every $\paren{u,s}\in F$.
For parallel edges, retain the smaller weight.

$\textsc{Estimate\_augmented\_MSSP}\paren{H,V,d}$
(\algref{Estimate-augmented-MSSP})

$A\gets d\bracke{V,S_{2}}$

Compute a $\paren{1+\varepsilon'}$-approximation
$C$ of $A\minplus A^{\mathsf T}$
(\thmref{ApproxMinPlus})

\ForEach{$\paren{u,v}\in V\times V$}{
    $d\bracke{u,v}\gets
    \mmin{d\bracke{u,v},C\bracke{u,v},C\bracke{v,u}}$

    $d\bracke{v,u}\gets d\bracke{u,v}$
}

\Return $d$
\end{algorithm}

\medskip

As a step towards \queref{2and2+eps}, we extend
the construction of \appref{2,W2} using FMM algorithms.
For every fixed $\varepsilon>0$ and
$a\in\bracke{0,\frac{1}{2}}$, we replace the runtime
expression $\widetilde O\paren{
n^{2+a}
+n^{\omega\paren{1,1-a,1}+o\paren{1}}
\cdot\log W_{\max}
}$ 
of the $\paren{2+\varepsilon}$-APASP algorithm of
Dory, Forster, Kirkpatrick, Nazari,
Vassilevska Williams and de Vos
\cite{DorForKirNazVasVos2023} by
$\widetilde O\paren{
n^{2+a}
+n^{\omega\paren{1,1-2a,1}+o\paren{1}}
\cdot\log W_{\max}}$ at the cost of an additive stretch of $W_{2}$.

A related result is the $\paren{1+\varepsilon,2W_{1}}$-APASP of
Saha and Ye~\cite{SahYe2023} within the same runtime, which was improved to $\paren{1+\varepsilon,\mmin{2W_{1},4W_{2}}}$-APASP
by Roditty and Sapir~\cite{RodSap2025}. Indeed, it holds that $\paren{1+\varepsilon,\mmin{2W_{1},4W_{2}}} \leq \paren{1+\varepsilon,W_{1}+2W_{2}} \leq\paren{2+\varepsilon,W_{2}}$. However, we believe that by considering our $\paren{2+\varepsilon,W_{2}}$-APASP algorithm, one can get closer to \queref{2and2+eps}, by identifying the hard caseour approach, similarly to the combintorial \appref{2,W2}.

As we use the $\paren{1+\varepsilon}$-approximate \minplusname-product, we assume integer edge weights in $\bracce{0}\cup\bracke{W_{\max}}\cup\bracce{\infty}$. Let $0<\varepsilon\leq 1$ and
$b,c\geq 0$ with $b+c\leq 1$. We retain the hierarchy with sampling probabilities
$n^{-b}$ and $n^{-c}$ of \algref{2-APASP-slower-main} and set $\varepsilon'=\frac{\varepsilon}{6}$.
We replace the MSSP invocation from $S_{2}$ over $G$
by the $\paren{1+\varepsilon'}$-MSASP algorithm
of Elkin and Neiman~\cite{ElkNei2020}.
We also omit the auxiliary edges $S_{2}\times V$
from the graph used by the final SSSP invocation.
Instead, we compute the estimates through $S_{2}$
using a $\paren{1+\varepsilon'}$-approximate
\minplusname-product of Zwick~\cite{Zwick2002}.
All other computations remain unchanged.

\begin{theorem}\label{thm:2+eps,W2}
\algref{2+eps,W2} computes a
$\paren{2+\varepsilon,W_{2}}$-APASP in 
$\widetilde O\paren{
n^{2+a}+
\paren{\frac{1}{\varepsilon}}^{O\paren{1}}
\cdot n^{\omega\paren{1,1-2a,1}+o\paren{1}}
\cdot\log W_{\max}
}$
runtime.
\end{theorem}
\begin{proof}
The exact pivot distances and the SSSP computations
over $E_{1}$ and $E_{2}$ remain unchanged.
Thus, as in \appref{2,W2}:
$$
F=
\bracce{
\paren{u,s}
\midline
u\in V,\ s\in\ball{u}{1}{2}
}
$$
Every edge of $F$ has its exact distance as its weight.
All previous updates preserve
$d\bracke{u,v}\geq\delta\paren{u,v}$.
The $\paren{1+\varepsilon'}$-MSASP invocation gives
$\delta\paren{u,s}\leq d\bracke{u,s}
\leq\paren{1+\varepsilon'}\delta\paren{u,s}$
for every $u\in V$ and $s\in S_{2}$.
Hence, the approximate \minplusname-product gives:
$$
\begin{aligned}
C\bracke{u,v}
&\leq
\paren{1+\varepsilon'}\cdot 
\mmin{
d\bracke{u,s}+d\bracke{s,v}
}{s\in S_{2}}\\
&\leq
\paren{1+\varepsilon'}^{2}\cdot  
\mmin{
\delta\paren{u,s}+\delta\paren{s,v}
}{s\in S_{2}}
\end{aligned}
$$
The proof of \thmref{2,W2} uses either estimates
obtained from original edges, exact pivot distances
and edges of $F$, or estimates through a member
of $S_{2}$.
The former remain available, while the latter
are now approximated within a factor of
$\paren{1+\varepsilon'}^{2}$.
Consequently:
$$
d\bracke{u,v}
\leq
\paren{1+\varepsilon'}^{2}
\cdot\paren{2\delta\paren{u,v}+W_{2}}
$$

Since $0<\varepsilon\leq 1$ and $\varepsilon'=\frac{\varepsilon}{6}<\frac{1}{6}<\frac{2}{5}$, we obtain:
$$
\begin{aligned}
d\bracke{u,v}
&\leq
2\delta\paren{u,v}+W_{2}
+\paren{2\varepsilon'+\varepsilon'^{2}}
\cdot\paren{2\delta\paren{u,v}+W_{2}}\\
& \leq 2\delta\paren{u,v}+W_2
+\paren{2\varepsilon'+\varepsilon'^2}\cdot 
\paren{2\delta\paren{u,v}+\frac{\delta\paren{u,v}}{2}}\\
&=
\paren{
2+5\varepsilon'
+\frac{5}{2}\varepsilon'^{2}
}\cdot \delta\paren{u,v}+W_{2}\\
&<
\paren{
2+6\varepsilon'
}\cdot \delta\paren{u,v}+W_{2}\\
&\leq
\paren{2+\varepsilon}\cdot\delta\paren{u,v}+W_{2}
\end{aligned}
$$

For completeness, we observe that if $\sizeindex{u\squiggly v}{1}\leq 1$,
the initial two SSSP rounds still compute
the exact distance.

As for the runtime, the retained SSSP computations
over $E_{1}$ and $E_{2}$ require
$\widetilde O\paren{n^{2+b}+n^{2+c}}$
 runtime.
By \obvref{bunchsize2} and \obvref{edgessize}:
$$
\begin{aligned}
\abs{F}
&\leq
\msum{
\abs{\ball{u}{1}{2}}}
{u\in V}
\in\widetilde O\paren{n^{1+c}}\\
\mathbb E\bracke{\abs{E_{1}}}
&\in\widetilde O\paren{n^{1+b}}
\end{aligned}
$$
To  construct $F$ we simply check the bunch condition
for all pairs in $V\times S_{1}$, which takes
$\widetilde O\paren{n^{2-b}}$ runtime.  By \lemref{estimateMSSP-runtime}, the additional
SSSP invocation from all vertices over
$H=\paren{V,E_{1}\cup F}$ therefore requires
$\widetilde O\paren{n^{2+b}+n^{2+c}}$ runtime. 

The sampling bounds for $S_{2}$, together with
\thmref{ApproxMSSP} and \thmref{ApproxMinPlus},
give $\widetilde O\paren{
\paren{\frac{1}{\varepsilon}}^{O\paren{1}}
\cdot n^{\omega\paren{1,1-b-c,1}+o\paren{1}}
\cdot\log W_{\max}
}$ runtime for the MSASP invocation and
the approximate \minplusname-product.
Hence, the  overall runtime is:

$$
\widetilde O\paren{
n^{2+b}+n^{2+c}
+\paren{\frac{1}{\varepsilon}}^{O\paren{1}}
\cdot n^{\omega\paren{1,1-b-c,1}+o\paren{1}}
\cdot\log W_{\max}
}
$$
Setting $b=c=a$, for any
$a\in\paren{0,\frac{1}{2}}$, yields:
$$
\widetilde O\paren{
n^{2+a}
+\paren{\frac{1}{\varepsilon}}^{O\paren{1}}
\cdot n^{\omega\paren{1,1-2a,1}+o\paren{1}}
\cdot\log W_{\max}
}
$$
\end{proof}

Current rectangular FMM bounds
\cite{Uznanski2026,AmaBerZui2026} give
$\omega\paren{1,1-2a,1}
\leq 2-2a+\frac{1}{4\paren{1-2a}}$
for $a\in\bracke{0,\frac{1}{4}}$.
Balancing this expression with $2+a$ gives:
$$
2+a=2-2a+\frac{1}{4\cdot\paren{1-2a}}
\quad\Longleftrightarrow\quad
24a^{2}-12a+1=0
$$
Choosing $a=\frac{3-\sqrt{3}}{12}$ therefore gives:
$$
\widetilde O\paren{
\paren{\frac{1}{\varepsilon}}^{O\paren{1}}
\cdot n^{\frac{27-\sqrt{3}}{12}+o\paren{1}}
\cdot\log W_{\max}
}
\subseteq
\widetilde O\paren{
\paren{\frac{1}{\varepsilon}}^{O\paren{1}}
\cdot n^{2.10566244}
\cdot\log W_{\max}
}
$$
 runtime.

Reducing the additive stretch to $c W_{2}$
for some constant $c\in\paren{0,1}$, while retaining
this runtime, would already be a step towards
\queref{2and2+eps}.
Eliminating the additive term entirely would give
a faster $\paren{2+\varepsilon}$-APASP algorithm.
Eliminating both the additive term and the
additional multiplicative $\varepsilon$
would answer \queref{2and2+eps}.

\section{Runtimes Improvements for Known APASP Results}\label{app:allResults}

For completeness, we summarize known APASP results
for weighted and unweighted undirected graphs, with
an emphasis on updating the algebraic runtimes.
We revisit the analyses of the existing algorithms
and substitute OpenAI's bound
$\omega\leq \frac{9}{4}$~\cite{OpenAI2026},
together with the rectangular FMM bounds of
Uzna{\'n}ski~\cite{Uznanski2026} or of
Amanov, van den Berg and Zuiddam~\cite{AmaBerZui2026}.
\tabref{allWeighted} and \tabref{allUnweighted}
summarize the approximation guarantees, updated
runtimes and additional assumptions for weighted
and unweighted graphs, respectively.

Some runtimes in the following tables depend on parameters
whose values are determined by the current rectangular FMM bounds.
We define these parameters as follows:

\begin{enumerate}
 \item $F_{k}\paren{t}=
    \frac{
    \paren{3-2t\cdot \paren{3k+1}+\paren{3k^{2}+2k-2}\cdot t^{2}}^{2}
    }{
    4\paren{1-t\cdot \paren{k-1}}\cdot \paren{1-kt}\cdot \paren{1-t\cdot \paren{k+2}}
    }$
    is the function used in the root equations below, where 
    $0<t<\frac{1}{k+2+\sqrt{6}}$.

    \item $\tau_{k}$ determines the algebraic exponent
    $2+\tau_{k}$ for unweighted
    $+2\cdot\paren{k+1}$-APASP.
    Set $\tau_{0}=0.20799054$ and $\tau_{1}=\frac{1}{7}$.
    For $k\geq 2$, $\tau_{k}$ satisfies
    $F_{k}\paren{\tau_{k}}
    =1+2\cdot\paren{k+2}\cdot\tau_{k}$.
    The cases $k=0,1$ use
    \cite{JinKirStaVas2026}, while $k\geq 2$ uses
    \cite{SahYe2023}.

    \item $\sigma_{k}$ determines the algebraic exponent
    $2+\sigma_{k}$ for unweighted
    $\paren{1+\varepsilon,2\cdot\paren{k+1}}$-APASP
    and weighted
    $\paren{1+\varepsilon,2\cdot\msum{W_{i}}{i=1}{k+1}}$-APASP.
    Set $\sigma_{0}=\frac{3-\sqrt{3}}{12}$.
    For $k\geq 1$, $\sigma_{k}$ satisfies
    $F_{k}\paren{\sigma_{k}}=2+\sigma_{k}$.

    \item $\gamma_{\ell}
    =\frac{1}{2\cdot \paren{\ell+2+\sqrt{\ell+2}}}$
    determines the algebraic exponent $2+\gamma_{\ell}$
    for weighted
    $\paren{\frac{3\ell+4}{\ell+2}+\varepsilon}$-APASP.
    In particular, $\gamma_{1}=\sigma_{0}$
    and $\gamma_{2}=\frac{1}{12}$.
\end{enumerate}

\begingroup
\footnotesize
\setlength{\tabcolsep}{2.5pt}
\setlength{\LTcapwidth}{\textwidth}
\setlength{\cellspacetoplimit}{3pt}
\setlength{\cellspacebottomlimit}{3pt}
\renewcommand{\arraystretch}{1.3}
\captionsetup{position=bottom}

\begin{longtable}{
|>{\centering\arraybackslash}S{m{.21\textwidth}}
|>{\centering\arraybackslash}S{m{.32\textwidth}}
|>{\centering\arraybackslash}S{m{.29\textwidth}}
|>{\centering\arraybackslash}S{m{.126\textwidth}}|
}

\hline
\multirow{2}{*}{\textbf{Stretch}}
& \multirow{2}{*}{
    \shortstack{\textbf{Combinatorial}\\\textbf{Runtime}}
}
& \multicolumn{2}{c|}{\textbf{Algebraic}}
\\ \cline{3-4}
& & \textbf{Runtime}
& \textbf{Remarks}
\\ \hhline{|=|=|=|=|}
\endfirsthead

\hline
\multirow{2}{*}{\textbf{Stretch}}
& \multirow{2}{*}{
    \shortstack{\textbf{Combinatorial}\\\textbf{Runtime}}
}
& \multicolumn{2}{c|}{\textbf{Algebraic}}
\\ \cline{3-4}
& & \textbf{Runtime}
& \textbf{Remarks}
\\ \hhline{|=|=|=|=|}
\endhead

\endfoot
\endlastfoot

$+2W_{2}$
& $n^{2}+n^{\frac{3}{2}}\cdot m^{\frac{1}{2}}$
  \mbox{\cite{CohZwi1997}}
& ---
& ---
\\ \hline

$+\mmin{2W_{1},4W_{2}}$
& $n^{\frac{7}{3}}$
  \mbox{\cite{CohZwi1997}}
& ---
& ---
\\ \hline

$+2\cdot\msum{W_{i}}{i=1}{k+1}$
& {\scriptsize
  $\mmin{\!n^{2+\frac{1}{3k+2}},
  \!n^{2}+\!n^{2-\frac{1}{k+2}}\cdot m^{\frac{1}{k+2}}\!}$
  }
  \mbox{\cite{CohZwi1997,RodSap2025}}
& ---
& ---
\\ \hline

$1+\varepsilon$
& ---
& $\frac{1}{\varepsilon}\cdot
  n^{\frac{9}{4}+o\paren{1}}$
  \mbox{\cite{Zwick2002,BriKunWun2019}}
& $\varepsilon>0$
\\ \hline

$\paren{1+\varepsilon,\mmin{2W_{1},4W_{2}}}$
& ---
& $\paren{\frac{1}{\varepsilon}}^{O\paren{1}}\cdot
  n^{2.10566244}\cdot\log W_{\max}$
  \mbox{\cite{RodSap2025}}
& $\varepsilon>0$\par
  Integer weights
\\ \hline

$\paren{1+\varepsilon,2W_{1}}$
& ---
& \multicolumn{1}{@{}c@{}|}{%
  \begin{tabular}[c]{
  >{\centering\arraybackslash}S{m{.29\textwidth}}
  }
  $\frac{1}{\varepsilon}\cdot
  n^{2.10566244}\cdot\log W_{\max}$
  \mbox{\cite{SahYe2023}}
  \\ \hline
  $\paren{\frac{1}{\varepsilon}}^{3}\cdot
  n^{2.14644661}$
  \mbox{\cite{BerKas2007}}
  \\
  \end{tabular}%
}
& $\varepsilon>0$
\\ \hline

$\paren{1+\varepsilon,2\cdot\msum{W_{i}}{i=1}{k+1}}$
& ---
& $\frac{1}{\varepsilon}\cdot
  n^{2+\sigma_{k}+o\paren{1}}\cdot\log W_{\max}$
  \mbox{\cite{SahYe2023}}
& $\varepsilon>0$
\\ \hline

$\paren{\frac{5}{3}+\varepsilon,\mmin{W_{1},2W_{2}}}$
& ---
& $\paren{\frac{1}{\varepsilon}}^{O\paren{1}}\cdot
  n^{2.10566244}\cdot\log W_{\max}$
  \mbox{\cite{RodSap2025}}
& $\varepsilon>0$\par
  Integer weights
\\ \hline

$\paren{\frac{6k+3}{3k+2},\msum{W_{i}}{i=1}{k+1}}$
& $n^{2+\frac{1}{3k+2}}$
  \mbox{\cite{RodSap2025}}
& ---
& ---
\\ \hline

$2$
& $n^{2}+m\cdot n^{\frac{1}{2}}$
  \mbox{\cite{BasKav2010}}
& \multicolumn{2}{@{}c@{}|}{%
  \begin{tabular}[c]{
  >{\centering\arraybackslash}S{m{.29\textwidth}}
  |>{\centering\arraybackslash}S{m{.126\textwidth}}
  }
  $n^{2.34549151}$
  \mbox{[\secref{2-slower}]}
  & ---
  \\ \hline
  $n^{2.17157288}\cdot\log^{O\paren{1}}W_{\max}$
  \mbox{[\secref{2-faster}]}
  & Integer weights
  \\
  \end{tabular}%
}
\\ \hline

$\paren{2,W_{2}}$
& \mbox{$n^{\frac{7}{3}}$~[\appref{2,W2}]}
& ---
& ---
\\ \hline

$\paren{2,W_{1}}$
& $n^{2}$
  \mbox{\cite{BasKav2010,BerKas2007}}
& ---
& ---
\\ \hline

$2+\varepsilon$
& ---
& $\paren{\frac{1}{\varepsilon}}^{O\paren{1}}\cdot
  n^{2.14644661}\cdot\log W_{\max}$
  \mbox{\cite{Kavitha2012,DorForKirNazVasVos2023}}
& $\varepsilon>0$
\\ \hline

$\paren{2+\varepsilon,W_{2}}$
& ---
& $\paren{\frac{1}{\varepsilon}}^{O\paren{1}}\cdot
  n^{2.10566244}\cdot\log W_{\max}$
  \mbox{[\appref{2+eps,W2}]}
& $\varepsilon>0$\par
  Integer weights
\\ \hline

$\frac{7}{3}$
& $n^{2}+n\cdot m^{\frac{2}{3}}$
  \mbox{\cite{BasKav2010,AkaRod2021}}
& ---
& ---
\\ \hline

$\frac{7}{3}+\varepsilon$
& ---
& $\paren{\frac{1}{\varepsilon}}^{O\paren{1}}\cdot
  n^{2.10566244}\cdot\log W_{\max}$
  \mbox{\cite{RodSap2025}}
& $\varepsilon>0$
\\ \hline

$\frac{5}{2}$
& $n^{2}+n^{\frac{5}{4}}\cdot m^{\frac{1}{2}}$
  \mbox{\cite{Kavitha2012,AkaRod2021}}
& ---
& ---
\\ \hline

$\frac{5}{2}+\varepsilon$
& ---
& $\paren{\frac{1}{\varepsilon}}^{O\paren{1}}\cdot
  n^{2.08333334}\cdot\log W_{\max}$
  \mbox{\cite{RodSap2025}}
& $\varepsilon>0$
\\ \hline

$\frac{3\ell+4}{\ell+2}$
& $n^{2}+n^{2-\frac{3}{\ell+2}}\cdot
  m^{\frac{2}{\ell+2}}$
  \mbox{\cite{AkaRod2021}}
& ---
& ---
\\ \hline

$\frac{3\ell+4}{\ell+2}+\varepsilon$
& ---
& $\paren{\frac{1}{\varepsilon}}^{O\paren{1}}\cdot
  n^{2+\gamma_{\ell}+o\paren{1}}\cdot\log W_{\max}$
  \mbox{\cite{RodSap2025}}
& $\varepsilon>0$
\\ \hline

$3$
& $n^{2}$
  \mbox{\cite{CohZwi1997,BasKav2010}}
& ---
& ---
\\ \hline

\caption{Known $\paren{\alpha,\beta}$-APASP algorithm for weighted undirected graphs with current updated runtimes.}
\label{tab:allWeighted}\\

\end{longtable}
\endgroup

\begingroup
\footnotesize
\setlength{\tabcolsep}{2.5pt}
\setlength{\LTcapwidth}{\textwidth}
\setlength{\cellspacetoplimit}{3pt}
\setlength{\cellspacebottomlimit}{3pt}
\renewcommand{\arraystretch}{1.3}
\captionsetup{position=bottom}

\begin{longtable}{
|>{\centering\arraybackslash}S{m{.21\textwidth}}
|>{\centering\arraybackslash}S{m{.32\textwidth}}
|>{\centering\arraybackslash}S{m{.29\textwidth}}
|>{\centering\arraybackslash}S{m{.126\textwidth}}|
}

\hline
\multirow{2}{*}{\textbf{Stretch}}
& \multirow{2}{*}{
    \shortstack{\textbf{Combinatorial}\\\textbf{Runtime}}
}
& \multicolumn{2}{c|}{\textbf{Algebraic}}
\\ \cline{3-4}
& & \textbf{Runtime}
& \textbf{Remarks}
\\ \hhline{|=|=|=|=|}
\endfirsthead

\hline
\multirow{2}{*}{\textbf{Stretch}}
& \multirow{2}{*}{
    \shortstack{\textbf{Combinatorial}\\\textbf{Runtime}}
}
& \multicolumn{2}{c|}{\textbf{Algebraic}}
\\ \cline{3-4}
& & \textbf{Runtime}
& \textbf{Remarks}
\\ \hhline{|=|=|=|=|}
\endhead

\endfoot
\endlastfoot

$+2$
& $\mmin{n^{\frac{7}{3}},
  n^{2}+n^{\frac{3}{2}}\cdot m^{\frac{1}{2}}}$
  \mbox{\cite{DorHalZwi2000}}
& $n^{2.20799054}$
  \mbox{\cite{JinKirStaVas2026,JinParSahXu2026}}
& ---
\\ \hline

$+4$
& $\mmin{n^{2+\frac{1}{5}},
  n^{2}+n^{2-\frac{1}{3}}\cdot m^{\frac{1}{3}}}$
  \mbox{\cite{DorHalZwi2000}}
& $n^{2.14285715}$
  \mbox{\cite{JinKirStaVas2026}}
& ---
\\ \hline

$+6$
& $\mmin{n^{2+\frac{1}{8}},
  n^{2}+n^{2-\frac{1}{4}}\cdot m^{\frac{1}{4}}}$
  \mbox{\cite{DorHalZwi2000}}
& $n^{2.09618925}$
  \mbox{\cite{SahYe2023,JinParSahXu2026}}
& ---
\\ \hline

$+8$
& $\mmin{n^{2+\frac{1}{11}},
  n^{2}+n^{2-\frac{1}{5}}\cdot m^{\frac{1}{5}}}$
  \mbox{\cite{DorHalZwi2000}}
& $n^{2.07219878}$
  \mbox{\cite{SahYe2023,JinParSahXu2026}}
& ---
\\ \hline

$+10$
& $\mmin{n^{2+\frac{1}{14}},
  n^{2}+n^{2-\frac{1}{6}}\cdot m^{\frac{1}{6}}}$
  \mbox{\cite{DorHalZwi2000}}
& $n^{2.05786455}$
  \mbox{\cite{SahYe2023,JinParSahXu2026}}
& ---
\\ \hline

$+2\cdot\paren{k+1}$
& {\scriptsize
  $\mmin{\!n^{2+\frac{1}{3k+2}},
  \!n^{2}+\!n^{2-\frac{1}{k+2}}\cdot m^{\frac{1}{k+2}}\!}$
  }
  \mbox{\cite{DorHalZwi2000}}
& $n^{2+\tau_{k}+o\paren{1}}$
  \mbox{\cite{JinKirStaVas2026,SahYe2023,JinParSahXu2026}}
& ---
\\ \hline

$1+\varepsilon$
& ---
& $n^{\frac{9}{4}+o\paren{1}}$
  \mbox{\cite{Seidel1995}}
& ---
\\ \hline

\multicolumn{3}{|@{}c@{}|}{%
  \begin{tabular}[c]{
  >{\centering\arraybackslash}S{m{.21\textwidth}}
  |>{\centering\arraybackslash}S{m{.32\textwidth}}
  |>{\centering\arraybackslash}S{m{.29\textwidth}}
  }

  $\paren{1+\varepsilon,2}$
  & ---
  & $\frac{1}{\varepsilon}\cdot n^{2.10566244}$
    \mbox{\cite{DorForKirNazVasVos2023,SahYe2023}}
  \\ \hline

  $\paren{1+\varepsilon,4}$
  & ---
  & $\frac{1}{\varepsilon}\cdot n^{2.06467117}$
    \mbox{\cite{SahYe2023}}
  \\ \hline

  $\paren{1+\varepsilon,6}$
  & ---
  & $\frac{1}{\varepsilon}\cdot n^{2.04053306}$
    \mbox{\cite{SahYe2023}}
  \\ \hline

  $\paren{1+\varepsilon,2\cdot\paren{k+1}}$
  & ---
  & $\frac{1}{\varepsilon}\cdot
    n^{2+\sigma_{k}+o\paren{1}}$
    \mbox{\cite{SahYe2023}}
  \\

  \end{tabular}%
}
& $\varepsilon>0$
\\ \hline

$2$
& $\mmin{n^{2.25},n^{2}+m\cdot n^{\frac{1}{2}}}$
  \mbox{\cite{Roditty2023,BasKav2010}}
& $n^{2+o\paren{1}}$
  \mbox{\cite[\appref{unweighted2}]{DorForKirNazVasVos2023,SahYe2023}}
& ---
\\ \hline

$\paren{2,1}$
& $n^{2}$
  \mbox{\cite{BasKav2010,BerKas2007}}
& ---
& ---
\\ \hline

$\frac{7}{3}$
& $n^{2}$
  \mbox{\cite{SahYe2023}}
& ---
& ---
\\ \hline

$3$
& $n^{2}$
  \mbox{\cite{CohZwi1997,BasKav2010,DorHalZwi2000}}
& ---
& ---
\\ \hline

\caption{Known $\paren{\alpha,\beta}$-APASP algorithms for unweighted undirected graphs with current updated runtimes.}
\label{tab:allUnweighted}\\

\end{longtable}
\endgroup


\end{document}